\documentclass[12pt,preprint]{aastex}

 \newcommand{\inc}{{\it i}}

 \newcommand{\be}{\begin{equation}}
 \newcommand{\ee}{\end{equation}}
 \newcommand{\ba}{\begin{eqnarray}}
 \newcommand{\ea}{\end{eqnarray}}
 \newcommand{\bs}{\begin{subequations}}
 \newcommand{\es}{\end{subequations}}

 \shorttitle{The tidal despinning of Mercury}
 \shortauthors{Noyelles et al.}

 \usepackage{ifpdf}
 \usepackage{epsfig}
 \usepackage[latin1]{inputenc}
 \usepackage{amsfonts}
 \usepackage{amsthm}
 \usepackage{amssymb,latexsym,amsmath}
 \usepackage{amsbsy}
 \usepackage{amstext}
 \usepackage{amsopn}
 \usepackage{amsgen}
 \usepackage[dvips]{color}
 \usepackage{graphicx}
 \usepackage{color}
 \usepackage{wasysym}

\begin{document}

 \title{Spin-orbit evolution of Mercury revisited}

 \author{Beno\^it Noyelles}
 \affil{Department of Mathematics and the Namur Centre for Complex Systems (naXys),\\
 University of Namur, 8 Rempart de la Vierge, Namur B-5000 Belgium\\
 \rm e-mail: ~~benoit.noyelles$\,$@$\,$unamur.be~,~\\~\\}

 \author{Julien Frouard}
 \affil{Department of Physics \& Astronomy and Center for Interdisciplinary Exploration and Research in Astrophysics (CIERA),\\
Northwestern University, Evanston IL 60208 USA\\
 \rm e-mail: ~~frouard$\,$@$\,$imcce.fr~,~\\~\\}

 \author{Valeri Makarov}
 \affil{US Naval Observatory, Washington DC 20392 USA\\
 \rm e-mail: ~~vvm$\,$@$\,$usno.navy.mil~\\~\\}

 \author{and\\~\\}

 \author{Michael Efroimsky}
 \affil{US Naval Observatory, Washington DC 20392 USA\\
 \rm e-mail: ~~michael.efroimsky$\,$@$\,$usno.navy.mil~\\~\\}

 \begin{abstract}


  Although it is accepted that the significant eccentricity of Mercury (0.206) favours entrapment into the 3:2 spin-orbit resonance, open are the questions of how exactly and when the capture took place. A recent work by \citet{m2012} has demonstrated that trapping into this resonance is certain if the eccentricity is larger than $\,0.2\,$, provided that we use a realistic tidal model, the one which is based on the Darwin-Kaula expansion of the tidal torque.
	
 While in {\it{Ibid}}. a Mercury-like planet had its eccentricity fixed, we take into account its evolution. To that end, a family of possible histories of the eccentricity is generated, based on synthetic time evolution consistent with the expected statistics of the distribution of eccentricity. We employ a model of tidal friction, which takes into account both the rheology and self-gravitation of the planet.

 As opposed to the commonly used constant time lag (CTL) and constant phase lag (CPL) models, the physics-based tidal model changes dramatically the statistics of the possible final spin states. First, we discover that after only one encounter with the spin-orbit 3:2 resonance this resonance becomes the most probable end-state. Second, if a capture into this (or any other) resonance takes place, the capture becomes final, several crossings of the same state being forbidden by our model. Third, within our model the trapping of Mercury happens much faster than previously believed: for most histories, 10 - 20 Myr are sufficient. Fourth, even a weak laminar friction between the solid mantle and a molten core would most likely result in a capture in the 2:1 or even higher resonance, which is confirmed both semi-analytically and by limited numerical simulations.

 So the principal novelty of our paper is that the 3:2 end-state is more ancient than the same end-state obtained when the constant time lag model is employed.
 The swift capture justifies our treatment of Mercury as a homogeneous, unstratified body whose liquid core had not yet formed by the time of trapping.

 We also provide a critical analysis of the hypothesis by \citet{wcllr2012} that the early Mercury might had been retrograde, whereafter it synchronised its spin and
 then accelerated it to the 3:2 resonance. Accurate processing of the available data on cratering does not support that hypothesis, while the employment of a realistic rheology invalidates a key element of the hypothesis, an intermediate pseudosynchronous state needed to spin-up to the 3:2 resonance.
 \end{abstract}


 \section{Motivation and Plan}

 Half a century ago, radar observations determined the Mercurian spin period to be $\approx58$ days \citep{pd1965}, which corresponds to a 3:2 spin-orbit resonance. A later study \citep{mpjsh2007} revealed that the orientation of Mercury's spin axis is consistent with the Cassini State 1 \citep{c1965}, the obliquity being $\,2.04\pm0.08\,'\,$  \citep{mpshgjygpc2012}.

 This raised the question: how had Mercury been trapped into this resonance? A consensus exists that the high eccentricity of
 Mercury (currently $e\approx0.206$) favours the trapping by widening the resonance. At the same time, there is a cleavage in opinion on whether the 3:2 resonance was the most likely end-state of Mercury's spin-orbit evolution.

 The first work on the despinning of Mercury was published by \citet{gp1966}. They obtained a $7\%$ probability for the capture into the 3:2 resonance, assuming that the eccentricity of Mercury has always had its present value, and that the tidal torque has obeyed the MacDonald (1964) model. Some forty years later, \citet{cl2004} explored a chaotic evolution of Mercury's eccentricity, showing that repetitive episodes of eccentricity increases could have boosted the probability of entrapment in the 3:2 resonance to $55\%$.

 A more complete study, though, should include a large fluid core. Keeping the eccentricity constant, \citet{pb1977} explained that the core-mantle friction could have significantly enhanced the 3:2 capture probability, provided that Mercury had not been trapped into the 2:1 resonance prior to that. \citet{cl2009} revisited the problem, with the core included, and suggested that the eccentricity of Mercury might have been very low in the past, lowering drastically the probability of an early capture into the 2:1 resonance, and making a later trapping into the 3:2 spin state almost certain. Both works by Correia \& Laskar (2004, 2009) were based on the CTL (constant time lag) tidal model.

 The crater counts from the Mariner data and the first two flybys of MESSENGER may be interpreted in favour of Mercury's spin being synchronous in the past.
 This motivated \citet{wcllr2012} to assume that Mercury had initially been retrograde, evolving later to synchronism. \citet{cl2012} achieved the synchronism from an initially prograde rotation. Then this resonance could have been destabilised by a huge impact driving Mercury into a state of stable pseudosynchronous rotation maintained by a significant eccentricity, before falling into the 3:2 resonance. The stability of pseudosynchronous spin depends on the applicability of the CTL model.

 Unfortunately, both the MacDonald and CTL models are incompatible with the physically plausible rheology of terrestrial planets \citep{e2012}. \footnote{~Besides this, the MacDonald model is inherently inconsistent \citep{we2012,em2013} and cannot be used even in theoretical studies.} A study based on a more realistic tidal response was undertaken recently by \citet{m2012} for a uniform Mercury analog, who demonstrated that entrapment in 3:2 resonance is inevitable at eccentricities between 0.2 and 0.41, without invoking a core-mantle friction or eccentricity variation.

 This motivated us to revisit the spin-orbit evolution of Mercury for a range of possible options suggested in the literature. In Section \ref{sec:overview}, we present an overview of the preceding studies. Sections \ref{triaxial} deals with the triaxiality-generated torque. Section \ref{tid.sec} presents our developments of the tidal torque. A special emphasise is made on how an essentially arbitrary rheology should enter the expression for the tidal torque.  A short Section \ref{despin.sec} elaborates
 on the despinning timescale. In Section \ref{sec:eccentricity}, we compute secular variations of Mercury's eccentricity. Section \ref{sec:correia} describes our numerical
 method. Before putting it to action, we test it by reproducing the results by \citep{cl2004}.

 In Section \ref{sec:scen1},  we apply our method to a homogeneous, originally prograde Mercury, and demonstrate that its likeliest end-state is 3:2. For a cold early Mercury (with the Maxwell time $\,\tau_{_M}=500\,$ yr), it is reached in less than 20 Myr. For a warmer early Mercury (i.e., for lower values of $\,\tau_{_M}\,$), it is reached in less than 10 Myr. This justifies the homogeneity assumption, because heating-up and differentiation takes up to a billion of years (see Appendix). Despite this, to make our study complete, in Section \ref{sec:scen2} we explore the despinning of an initially prograde Mercury with a liquid core. In this case, even a moderate friction in the core-mantle boundary boosts the probabilities of capture into the 3:2 and higher resonances. For example, with the present value of Mercury's eccentricity, entrapment into the 2:1 resonance becomes certain (while trapping in the 5:2 resonance becomes more probable than traversing it). Therefore, had Mercury been hot and differentiated soon after its creation, its despinning all the way to the 3:2 state would be highly unlikely. In Section \ref{sec:scen3}, we provide a critical analysis of the hypothesis by Wieczorek et al. (2012) that the early Mercury might had been retrograde, whereafter it synchronised and then accelerated its spin to the 3:2 resonance. Accurate processing of the available data on cratering does not support that hypothesis; while the employment of a realistic rheology invalidates a key element of that hypothesis, an intermediate pseudosynchronous state needed to spin-up to the 3:2 resonance. So that hypothesis has to be abandoned.
 Finally, in Section 10 we draw conclusions.

 This preprint is an extended version of our paper submitted to {\it{Icarus}}. Here we present numerous details and derivations which we had to omit in the journal version of our work.

 \section{Scenarios of Mercury's spin-orbit evolution. An overview\label{sec:overview}}

\begin{figure}[ht]
 \epsscale{0.92}
 \plotone{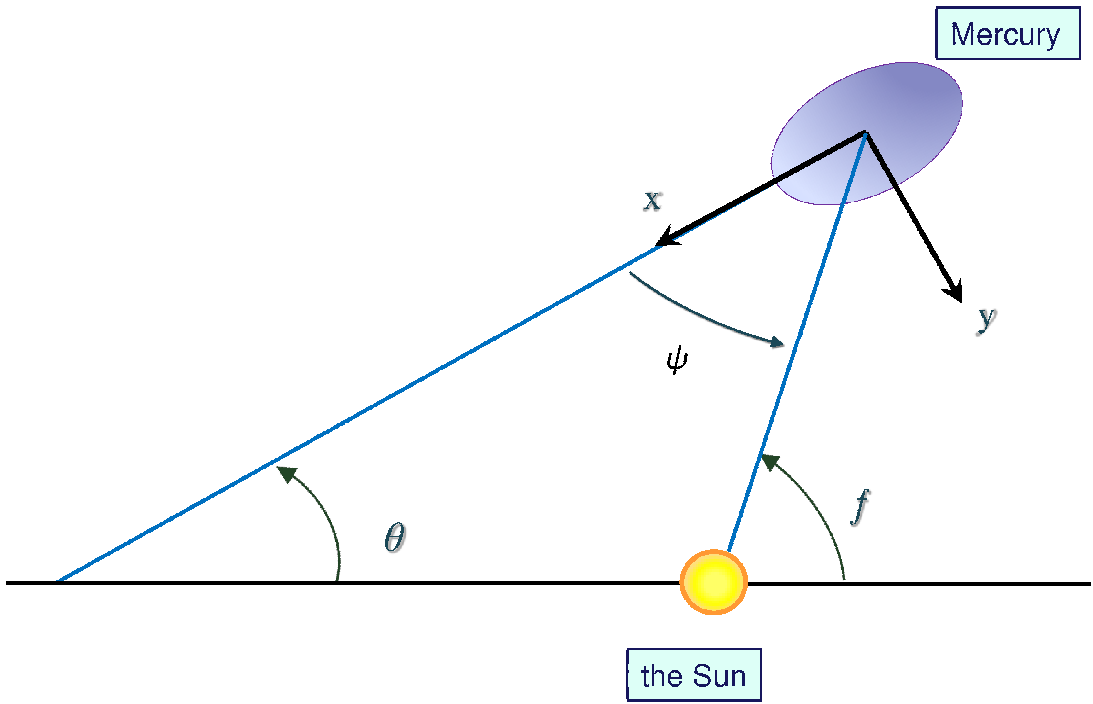}
 \caption{\small{The principal axes $\,x\,$ and $\,y\,$ of the planet correspond to the minimal and middle moments of inertia, respectively. The horizontal line is that of apsides,
 so $\,f\,$ is the true anomaly. In neglect of the apsidal precession, $\,\theta\,$ is the sidereal angle of the planet. The angle $\,\psi\,=\,f\,-\,\theta\,$ renders the
 separation between the planetocentric direction towards the star and the minimal-inertia axis $\,x\,$}
 \label{Figure}}
 \end{figure}

 Consider a planet of a mean radius $\,R\,$, mass $\,M_{planet}\,$ and the principal moments of inertia $\,A<B<C\,$.
 Assume that the spin of the planet is directed along its major-inertia axis $\,z\,$, the one related to the maximal moment of inertia $\,C\,$.
 The sidereal angle $\,\theta\,$ of the planet can be reckoned from the line of apsides to the largest-elongation axis $\,x\,$, the one related to the minimal moment of inertia $\,A\,$, as in Figure \ref{Figure}. The star exerts two torques on the planet. One, $\vec{\mathcal{T}}^{^{(TRI)}}$, is due to the planet's permanent triaxiality. Another, $\,\vec{\mathcal{T}}^{^{(TIDE)}}\,$, is caused by the tidal deformation of the planet. Rotation of the planet about its major-inertia axis $\,z\,$ is then governed by the torques' polar components. Expressing the maximal moment of inertia as $\,C\,=\,\xi\,M_{planet}\,R^{\,2}\,$, we write the equation of motion as
 \ba
 \stackrel{\bf\centerdot\,\centerdot}{\theta~}\,=~\frac{\,{\cal{T}}^{\rm{^{\,(TRI)}}}_z\,+~{\cal{T}}^{\rm{^{\,(TIDE)}}}_z}{C~~}~=~\frac{{\cal{T}}^{\rm{^{\,(TRI)}}}_z\,+~
 {\cal{T}}^{\rm{^{\,(TIDE)}}}_z}{\xi ~M_{planet}~R^{\,2}}\,~.
 \label{eq.eq}
 \label{eq:despinning}
 \ea
 The value of $\xi$ reflects the degree of inhomogeneity. While for a homogeneous sphere $\,\xi=0.4\,$, for Mercury it is known to be $\,0.346\,\pm\,0.014\,$ \citep{mpshgjygpc2012,NL}.
 The fact that $\,\xi\,<\,0.4\,$ indicates that the inner part of Mercury is denser than its outer part.

 This equation governs the spin history. The outcome also depends on the initial conditions and on disruptive events. Three options have been discussed in the literature hitherto.

 \subsection{Option 1: Spin-down of a homogeneous Mercury}

 Suppose Mercury was captured when it was yet homogeneous, and its liquid core was formed afterwards. Simplistic, this approach is convenient mathematically, wherefore it was used in the pioneer works. The first such investigation was undertaken by \citet{gp1966} who assumed that the initially spin was prograde and pretty fast.
 Their study produced a $7\%$ probability of capture into 3:2 resonance. Therefore, it was accepted that the current rotational state of Mercury was an outcome of a rather unlikely event.

 \citet{c1969} proposed that the eccentricity of Mercury could have varied under the influence of other planets, and that a larger value could enhance the probability of capture into the 3:2 resonance. However, he saw no reason why the eccentricity would have been bigger than $0.25$. The question remained open for a quarter of a century, until \citet{cl2004} ran 1,000 numerical simulations of the orbital evolution of Mercury over 4 Gyr. They then extracted 1,000 histories of Mercury's eccentricity and, among other things, found that it could have reached $0.45$. The authors integrated the equation (\ref{eq.eq}) with the following expression for the mean tidal torque (sometimes referred to as the `{\it{viscous model}}$\,$'):
 \ba
 \mathcal{T}_z^{^{(TIDE)}}~=~-~{3\,n^2\,M_{star}~k_{2}}{~\Delta t}~\,\frac{R^{\textstyle{^5}}}{a^3}\,~\left[\,\dot{\theta\,}~{\cal{A}}(e)~-~n~\mathcal{N}(e)\,\right]\,~,
 \label{eq:tidecl}
 \label{eq:macdonaldcorreia}
 \ea
 \ba
 {\cal{A}}(e)~=~\frac{\,1\,+\,3\,e^2\,+\,\frac{\textstyle 3}{\textstyle 8}~e^4\,}{\left(\,1~-~e^2\,\right)^{\textstyle{^{9/2}}}}\quad,\qquad
 \mathcal{N}(e)~=~\frac{\,1\,+\,\frac{\textstyle 15}{\textstyle 2}~e^2~+~\frac{\textstyle 45}{\textstyle 8}~e^4~+~\frac{\textstyle 5}{\textstyle 16}~e^6\,}{\left(\,1~-~e^2\,\right)^{\textstyle{^{\,6}}}}\,~,
 \label{eq:omege}
 \label{eq:fOmega}
 \label{eq:ne}
 \label{eq:fN}
 \ea
 $R\,$, $\,n\,$ and $\,e\,$ being Mercury's radius, mean motion and eccentricity; $\,M_{star}\,$ being the mass of the Sun;
 the Love number being $\,k_2=0.4\,$; and the time lag being chosen (pretty arbitrarily) as $\,n\,\Delta t=1/50\,$.
 Integrations began with a spin period of $\,20\,$ days, to see how many histories lead to the current period of $\,87.9\,$ days, i.e., to the 3:2 spin-orbit resonance.~\footnote{~In \citet{cl2004}, the
 product $\,n\,\Delta t\,$  was denoted with $\,Q\,$. The notation was misleading, as the so-defined $\,Q\,$ did not coincide with the quality factor (see Williams \& Efroimsky 2012, Section 4).}

 The formulae (\ref{eq:tidecl} - \ref{eq:fN}) were based on the assumption that the time lag $\,\Delta t\,$ bears no dependence upon frequency. Hence the name: the constant time lag (CTL) model. Detailed derivation of these formulae can be found in the Appendix to \citet{we2012}. In various forms, this result appeared much earlier in \citet{hut} and \citet{ekh1998}. In an implicit form, it was pioneered by \citet[Equation 24]{gp1966}.

 An important attribute of the CTL model is a pseudosynchronous equilibrium spin rate
 \begin{equation}
 \label{eq:pseudosynchronous}
 \dot{\theta}~=~n~+~6~n~e^2~+~\frac{3}{8}~n~e^4~+~\mathcal{O}(e^6)~~.
 \end{equation}
 For the present eccentricity of $\,\approx0.206\,$, this would be $\,\dot{\theta}\,\approx\,1.26\,n\,$. If the eccentricity reaches $0.285$, the pseudosynchronous equilibrium spin rate reaches the 3:2 spin-orbit resonance. This dramatically enhances the trapping, as the planet can get captured and can stay in this resonance even after the eccentricity decreases to a lower value. After many such crossings, the overall probability of capture into the 3:2 state must exceed 1/2.

 The physical validity of this scenario is questionable, because the existence of the pseudosynchronous spin state depends on the tidal model employed -- and because the CTL model is very different from the actual behaviour of solids and partial melts. An accurate description of tides must be based on a Fourier decomposition of the perturbing potential, with each term corresponding to an appropriate tidal mode. From these series, a similar one is derived for the torque. Then the tidal response at each mode should be entered. Mathematically, this means that, into each term, an appropriate value of $\,k_2/Q\,$ (or, more generally, of $\,k_l/Q_l\,$) must be inserted. Insertion of realistic frequency-dependencies of $\,k_l/Q_l\,$ renders pseudosynchronous rotation states unstable, as demonstrated in \citet{me2013}.

 Basing their study on the alleged existence of the pseudosynchronous state (\ref{eq:pseudosynchronous}), \citet{cl2004} obtained three types of capture:
  \begin{itemize}
  \item[1.] Mercury is trapped at the first crossing of a resonance (31 trajectories over 1,000).
  \item[2.] At the crossing of the resonance, Mercury is not trapped, but the pseudosynchronous spin rate (\ref{eq:pseudosynchronous}) turns out to be close enough to the resonance, to induce several crossings while the eccentricity oscillates; and the planet gets locked (168 trajectories).
  \item[3.] Mercury crosses the resonance and keeps despinning till reaching the pseudosynchronous rotation. Then secular variations of Mercury's eccentricity respin it, because increase in eccentricity entails increase in the pseudosynchronous spin rate. The planet gets into the 3:2 resonance or crosses it (to be despun back later, when the eccentricity becomes lower). Multiple crossings of this kind can result in a trapping (355 trajectories).
  \end{itemize}
 22 trajectories terminated in the 2:1 resonance, 554 in the 3:2 resonance, and 36 in the synchronous state. As the current 3:2 resonance appeared to be the most probable end state, the problem of Mercury's spin-orbit resonance seemed to have been solved. As we explained above, the presented solution critically depends on the choice of the tidal model.

 \subsection{Option 2: Spin-down of a differentiated Mercury}

 Based on the magnetic field measurements by Mariner 10 \citep{nblws1974}, Mercury was suspected to harbor a partially molten or completely liquid core \citep{nblw1975}.
 The core was detected by Earth-based radar observations of the longitudinal librations \citep{mpjsh2007}, which are significantly larger in amplitude than what should be expected from a uniformly solid planet.
 As a consequence, extra dissipation should be taking place at the core-mantle boundary (CMB). If the core had formed before Mercury was trapped into the 3:2 resonance, it must be taken into account in computation of the despinning. So the question becomes: when did the planet acquire its core? An answer to this question depends upon the
 scenario of formation of Mercury. Any such theory must be fit to explain why Mercury is poor in silicates, a fact ensuing from the planet's high density.

 One scenario is that of early volatilisation. Within this approach, the planet's composition may be explained if the removal process in Mercury's zone of the solar nebula was only slightly more effective for silicates than for iron. The fractionation required to produce an iron-rich planet is achieved through a combination of the gravitational and drag forces acting in the nebula during the formation of proto-Mercury \citep{w1978}. Another possibility is a slow volatilisation scenario which suggests that after the formation of Mercury the solar wind could have partially volatilised the mantle \citep{c1985,fc1987}. Within both these scenarios, differentiation of the planet was gradual. As explained in Appendix \ref{sec:protacalepsis}, core formation required hundreds of millions to a billion of years.

 A very different scenario is rapid differentiation through an impact energetic enough to have molten the originally chondritic proto-Mercury \citep{bsc1988,bahw2007}.
 The event would have resulted in a very warm tumbling Mercury wherein the CMB would have formed very quickly. The hypothesis of an early impact is not supported by the MESSENGER data. As was pointed out by \citep{pehmbgeghlmnsrsss2011}, the abundance of potassium $\,${\it{is inconsistent with those physical models of Mercury formation, which require extreme heating of the planet}}. This observation prompted \citep{wurm} to explore the action of photophoretic forces upon irradiated solids. The investigation has demonstrated that silicates are preferentially pushed into an optically thick disk. The subsequent planetesimal production at the
 outward-moving edge renders metal-rich planetesimals close to the star and metal-depleted ones farther out in the nebula. After this very early separation of metals and
 silicates, the further development of Mercury goes gradually, with a slow emergence of a core-mantle boundary.

 \citet{gp1967} originally introduced the core-mantle friction to explain the despinning of Venus. The friction was modeled as a torque linear in the relative angular velocity. \citet{c1969} introduced that torque into the theory of despinning of Mercury, and concluded that the core-mantle friction increased what he called the local probability of capture (the probability of capture into each resonance considered separately). The new torque did not change considerably the global probability, in the sense that Mercury was still very likely to be trapped into the 2:1 spin-orbit resonance and had very few chance to reach the current 3:2 rotation state. This result was confirmed and supplemented by \citet{pb1977} who concluded that the tidal quality factor $\,Q\,$ should be smaller than $\,100\,$, and that the kinematic viscosity of the molten core of Mercury with a laminar boundary layer should be comparable to that of water ($\,\approx\,0.01\,$~cm$^2$/s$\,$). The role of core-mantle friction was later revisited by \citet{cl2009} who stated that, owing to the chaotic evolution of the eccentricity of Mercury, the planet might have been trapped into the 2:1 resonance before a decrease of the eccentricity disrupted this configuration. In this last study, the 3:2 end-state has a probability of $\,26\%\,$, the 2:1 resonance being the most probable outcome. Anyway, the authors state that the probability of a final capture in the 3:2 resonance $\,${\it{can be increased up to $\,55\%\,$ or $\,73\%\,$ if the eccentricity of Mercury in the past has descended below the critical values $\,0.025\,$ or $\,0.005\,$, respectively}}. Such a low eccentricity can be reached through considering short-period effects omitted within a secular model \citep{l2008}. All of these studies have been conducted using the CTL tidal model which permits pseudosynchronism.
 However, contrarily to the rigid-Mercury case, this time the pseudosynchronous equilibrium state does not help to get trapped into the 3:2 resonance, since one crossing is usually enough.

 \subsection{Option 3: Post-synchronous evolution}\label{sce3.sec}

 The counts of craters on the surface of Mercury, from the Mariner 10 data ($\,\approx45\%$ coverage) and the first two flybys of MESSENGER, led \citet{wcllr2012} and \citet{cl2012} to suggest that Mercury's rotation was initially retrograde, and the tides (possibly helped by the core-mantle friction) accelerated it to the 1:1 resonance.~\footnote{~We would add that, as an alternative possibility, Mercury might had started in a prograde mode and got decelerated by tides all the way down to synchronism, crossing the higher resonances. This possibility, however, is much less probable, as it requires a very small eccentricity at the start of Mercury's history.} While in the synchronous state, Mercury acquired much of its craters, whereafter its synchronism was destabilised by a huge impact. The fortuitously directed impact drove Mercury to faster spin rates. Assisted by a high eccentricity, the planet reached the state of pseudosynchronous rotation which is stable $\,${\it{in the framework of the CTL model}}. The chaotic evolution of Mercury's eccentricity afterwards got it finally trapped as in \citet{cl2004}.

 As demonstrated in Section \ref{sec:scen3}, an impact causing the disruption of the synchronous state should have left a crater of at least 250 to 400 km in diameter, depending on the velocity and the density of the impactor. For a synchronous Mercury to reach directly the 3:2 resonance without help from tides or planetary perturbations, the crater's diameter should have been at least 600 km.

 The hypothesis by \citet{wcllr2012} is based on three assumptions: that the early Mercury was impacted heavily, that the crater counts are evidence of a past synchronous rotation, and that there is a stable state of pseudosynchronous rotation. Above that, this scenario implies that Mercury was initially retrograde.

 Several studies confirm that the early Mercury was impacted intensively, in particular during the Late Heavy Bombardment \citep{scmsh2008,sbcffhmps2011}.
 The associated planetesimals probably originated from the Main Asteroid Belt that was disturbed by the inward migration of Jupiter some 3.9 -- 3.8 Ga ago, when the age of Mercury was about 0.7 Ga \citep{gltm2005}; the resulting craters are known as {\it{Population 1}}. There is a {\it{Population 2}} of younger craters that could be dominated by Yarkovsky-effect-driven
 near-Earth asteroids \citep{mv2003,smiyk2005}. These impacts probably triggered an episode of volcanism \citep{eh2005} that was global during the LHB and occured
 $\,${\it{only over small patches or within large impact basins}}$\,$ \citep{mcfhbs2013}. It is unclear as to what extent this volcanism played a role in internal melt
 generation. From the global distribution of large impact basins \citep{fhbzsnskscppop2012}, 15 craters bigger than 600 km are $\,${\it{certain or probable}}, while 21 are
 $\,${\it{suggested but unverified}}. If we set the limit at 300 km (collisions strong enough to leave the synchronous state but not to reach the 3:2 resonance without external help), then we have 46 $\,${\it{certain or probable}}$\,$ impact basins, and 61 $\,${\it{suggested but unverified}}.

 The counts of craters in \citet{wcllr2012} are nearly pre-MESSENGER, in that they combine the Mariner-10 data with those from MESSENGER's first two flybys, not with those obtained after the orbital insertion. In \citet{fhbzsnskscppop2012}, the inhomogeneous distribution of craters is emphasised, and the authors admit that it is consistent with a past synchronous spin. They also mention another possibility: differential resurfacing of Mercury would explain the non-uniform distribution of smooth plains \citep{drsmbdmehwc2009} that might have buried degraded basins. Similarly, \citet{sbcffhmps2011} say: $\,$``{\it{the crater size-frequency distribution on Mercury shows pronounced regional variations that we argue are primarily the consequence of differences
 in the extent and age of volcanic plains emplacement}}".

 The initially retrograde spin of Mercury is an interesting option.  \citet{dt1993} showed that collisions affect planets' obliquities.  \citet{ki2007} got a distribution for their values.
 \citet{cl2010} pointed that an oblique Mercury is less likely to be trapped in a spin-orbit resonance than a non-oblique one.
 So most studies, including ours, still start with the obliquity damped to $0^{\circ}$
 (prograde spin) or $180^{\circ}$ (retrograde spin).

 Finally, we would reiterate that the stability of a pseudosynchronous equilibrium depends on the tidal model chosen. For physics-based rheologies, pseudosynchronous spin is
 transient and cannot be used to help the angular velocity grow from the 1:1 to 3:2 state. This gap has to be covered in one jump, which requires a significantly larger  impactor.


 \section{The triaxiality-caused torque}\label{triaxial}

 The polar torque will be approximated with its quadrupole part \citep[see, e.g.,][]{danb}:
 \ba
 {\cal{T}}^{\rm{^{\,(TRI)}}}_z&=&\frac{3}{2}~(B-A)~\frac{{G}\,M_{star}}{r^3}~\sin2\psi~
 \approx~-~\frac{3}{2}~(B-A)~n^2~\frac{a^3}{r^3}~\sin2(\theta-f)\,~.
 \label{tri.eq}
 \ea
 As usual, $\,G\,$ is the Newton gravitational constant,  $\,M_{star}\,$ is the mass of the star, $\,r\,$ is the distance between the centres of mass, while $\,\psi\,$ is the angle between the planetocentric direction towards the star and the minimal-inertia axis $\,x\,$, as in Figure \ref{Figure}. This angle is equal to the difference between the planet's true anomaly $\,f\,$, as seen from the star, and the planet's sidereal angle $\,\theta\,$, as measured from the line of apsides. These and other notations are listed in Table 1, the numerical values being presented in Table 2. The mean motion is $\,${\it{defined}}$\,$ as $\,n\equiv\,\stackrel{\bf\centerdot}{\cal{\,M}}\,$, though $\,\sqrt{G(M_{planet}+M_{star})/a^{3}\,}\,$ often remains a good approximation.

 To spare ourselves of the necessity to compute the true anomaly $\,f\,$, we shall express the torque through the mean anomaly $\,\mathcal{M}\,$. To do so, we  shall rewrite the expression (\ref{tri.eq}) as
 \begin{equation} \mathcal{T}_z^{^{\,(TRI)}}\,=~-~\frac{3}{2}~(B-A)~n^2~\frac{a^3}{r^3}~\left(\,\sin2\theta~\sin2f~-~\cos2\theta~\cos2f\,\right)
 \label{eq:tri3}
 \end{equation}
 and shall insert therein the following formulae \citep[see, e.g.,][]{d2002}:
 \begin{eqnarray}
 \left(\frac{a}{r}\right)^3\;\cos(2f) & = & \sum_{k=-\infty}^{+\infty}\,X_k^{-3,\,2}(e)\;\cos(k\cal{M})\,~,
 \label{eq:duriezcos}
 ~\\
 \left(\frac{a}{r}\right)^3\;\sin(2f) & = & \sum_{k=-\infty}^{+\infty}\,X_k^{-3,\,2}(e)\;\sin(k\cal{M})\,~,
 \label{eq:duriezsin}
 \end{eqnarray}
 $X_k^{n,m}(e)\,$ being the classical Hansen coefficients. These are computed through the formula
 \begin{equation}
	\label{eq:hansen} X_k^{n,m}(e)~=~\left(1+{{z}}^2\right)^{-n-1}\sum_{p=0}^{\infty}\left(-{{z}}\right)^p\sum_{h=0}^p~C_{n+m+1}^{p-h}~C_{n-m+1}^h~
J_{k-m+p-2h}(ke)\,~,
\end{equation}
where $\,{z}=(1-\sqrt{1-e^2})/e\,$, while $\,C^a_b\,$ are the binomial coefficients, and $\,J_k(x)\,$ are the Bessel function of the first kind. The Hansen coefficients are
related to the eccentricity functions via $\,G_{lpq}\,=\,X_{l-p+q}^{\,-(l+1),~l-2p}\,$. Altogether, formulae (\ref{eq:tri3} - \ref{eq:duriezsin}) entail:
 \begin{equation} \mathcal{T}_z^{^{(TRI)}}~=~-~\frac{3}{2}~(B-A)~n^2\sum_{q=-\infty}^{+\infty}~G_{20q}(e)~\sin\left(\,2\,\left[\theta\,-\,
 \left(1\,+\,\frac{q}{2}\right)~\cal{M}\right]\,\right).
 \label{eq:tri4}
 \end{equation}
In practice, the infinite series may be truncated to a sum over $\,q\,=\,-\,4\,,\,.\,.\,.~\,,\,6\,$. To our study, most relevant resonances are the 1:1, 3:2, 2:1, 5:2, 3:1,
7:2 and 4:1 ones, i.e.,
 those corresponding to $\,q\,=\,0\,,\,\ldots\,,\,6\,$ for the prograde rotation, and the resonances -2:1, -3:2, -1:1, -1:2, 1:2 and 1:1 for the retrograde rotation,
 i.e. $\,q\,=\,-6\,,\,\ldots\,,\,0\,$. In the latter case, the terms corresponding to $\,q\,=-6,-5\,$ have been added and proved to be unsignificant (Sec.\ref{sec:scen3})
 For $\,e\,>\,0.3\,$, computation of the eccentricity function $\,G_{lpq}(e)\,$ requires a lot of computer time, as the convergence of the series (\ref{eq:hansen}) slows down. So we computed these functions once for the whole range of eccentricities, and interpolated them during the computation, using cubic splines. To that end,
 the GNU Scientific Library \citep{gdtgjabr2009} was employed.

  \begin{deluxetable}{lr}
 \tablecaption{Symbol key \label{nota.tab}}
 \tablewidth{0pt}
 \tablehead{
 \multicolumn{1}{c}{Notation}  &
 \multicolumn{1}{c}{Description}\\
 }
 \startdata
 $\xi$ & \dotfill the moment of inertia coefficient of Mercury\\
 $R$ & \dotfill the radius of Mercury \\
 ${\cal{T}}^{\rm{^{\,(TRI)}}}_z$ & \dotfill $\,.\,.\,.\,$ the polar component of the triaxiality-caused torque acting on Mercury\\
  ${\cal{T}}^{\rm{^{\,(TIDE)}}}_z$ & \dotfill the polar component of the tidal torque acting on Mercury\\
 $M_{planet}$ & \dotfill the mass of Mercury \\
 $M_{star}$ & \dotfill the mass of the Sun \\
 $a$ & \dotfill the semimajor axis of Mercury \\
 $r$ & \dotfill the instantaneous distance between Mercury and the Sun \\
 $f$ & \dotfill the true anomaly of Mercury \\
 $e$ & \dotfill the orbital eccentricity \\
 ${\cal{M}}$ & \dotfill the mean anomaly of Mercury \\
 $C$ & \dotfill the maximal moment of inertia of Mercury \\
 $B$ & \dotfill the middle moment of inertia of Mercury \\
 $A$ & \dotfill the minimal moment of inertia of Mercury \\
 $n$ & \dotfill the mean motion ~\\
 $G$ & \dotfill Newton's gravitational constant \\
 $\tau_{_M}$ & \dotfill the viscoelastic characteristic time (Maxwell time) \\
 $\tau_{_A}$ & \dotfill the inelastic characteristic time (Andrade time)\\
 $\mu$ & \dotfill the unrelaxed rigidity modulus \\
 $J$ & \dotfill the unrelaxed compliance \\
 $\alpha$ & \dotfill the Andrade parameter \\
 \enddata
 \end{deluxetable}


\begin{table}[htbp]
 \centering
 \caption{\small{~The numerical values of the parameters entering the model. The value of the dimensionless moment of inertia $\,\xi\,$ is taken from \citep{mpshgjygpc2012}, and the
 triaxiality ${\textstyle{(B\,-\,A)}}/{\textstyle{C}}$ is derived from the gravity field coefficient $C_{22} = 8.088\times10^{-6}$ \citep{szpshlmnpmjtprght2012}.
 The value of Mercury's radius $\,R\,$ was borrowed from \citep{archinal}. The values of both the  semimajor axis $\,a\,$ and the mean motion $\,n\,$ are secular,
 obtained from the secular planetary theory of \cite{b1982}.
 }}
 \label{table2}
\begin{tabular}{lcc}
\hline
\hline
~\\
 Parameter                                   &                Notation                  &                     Numerical value               \vspace{3mm}~\\
 \hline
 \hline
                                                                                                                                                        ~\\
 Semimajor axis                              &                  $a$                     &                     $5.791\,\times\,10^7\,$ km    \vspace{3mm}~\\
 \hline
                                                                                                                                                        ~\\
 Mean motion                                 &   $n
 $  &       $26.0879\,$ rad/yr   \vspace{3mm}~\\
 \hline
                                                                                                                                                                ~\\
 Radius of Mercury                        &                  $R$                     &                             $2440\,$ km                   \vspace{3mm}~\\
 \hline
                                                                                                                                                                ~\\
 Dimensionless moment of inertia $~\qquad~$  &     $\xi\,\equiv\,{\textstyle{C}}/{\textstyle{(M_{planet}~R^2)}}$      &     $0.346$      \vspace{3mm}~\\
 \hline
                                                                                                                                                                ~\\
  Triaxiality of Mercury                   &  ${\textstyle{(B\,-\,A)}}/{\textstyle{C}}\,\equiv\,{\textstyle{4C_{22}}}/\xi$    &     $9.350\,\times\,10^{-5}$  \vspace{3mm}~\\
 \hline
 ~\\

 Ratio of masses of the Sun and Mercury  &     ${\textstyle{M_{star}}}/{\textstyle{M_{planet}}}$     &   $6.0276\,\times\,10^6$               \vspace{3mm}~\\
  \hline
  ~\\
 unrelaxed rigidity &     $\mu$     &   $8\times10^{10}\,$ Pa               \vspace{3mm}~\\
 \hline
 ~\\
 Maxwell time  &     $\tau_{_M}$     &   $500\,$ yr               \vspace{3mm}~\\
 \hline
 ~\\
 Andrade time  &     $\tau_{_A}$     &   $500\,$ yr               \vspace{3mm}~\\
 \hline
 ~\\
 Andrade parameter  &     $\alpha$     &   $0.2$               \vspace{3mm}~\\
 \hline
 ~\\
 Newton's gravitational constant  &     $G$     &     $66468$ m$^3$ kg$^{-1}$ yr$^{-2}$              \vspace{3mm}~\\
 \hline
 \hline
 \end{tabular}
 \end{table}

 \section{The tidal torque}\label{tidal}\label{tid.sec}

 Modeling of a tidal torque acting on a telluric body has long been subject to misconceptions. One of them has it that certain rheologies should be regarded ``unphysical". The
 grounds for this belief were that $\,{\cal{T}}^{\rm{^{\;(TIDE)}}}\propto\,k_2/Q\,$, wherefore a quality factor $\,Q\,$ scaling as a positive power of the tidal frequency would
 render an infinite torque in the zero-frequency limit. The misunderstanding stemmed from confusing the $\,${\it{seismic}}$\,$ quality factor $\,Q\,$ with the $\,${\it{tidal}}$\,$
 quality factor{\underline{s}} defined as $~1/Q_l\,=\,\sin\epsilon_l\,$, where $\,\epsilon_l\,$ are phase lags. The matter is resolved by taking into account that the seismic $\,Q\,$
 factor reflects the rheology, while the tidal $\,Q\,$ factors are defined by the interplay of rheology and self-gravitation, a circumstance that excludes unphysical divergencies in
 the quality functions $~k_l/Q_l\,=\,k_l(\omega_{\textstyle{_{lmpq}}})\,\sin\epsilon_l(\omega_{\textstyle{_{lmpq}}})~$.

 As a result of another serious oversight, an expression for the torque, valid solely for the CTL model, was illegitimately employed by some authors in the constant angular
 lag context. A study of this, once-popular error can be found in \citet{em2013}.

 Another common misconception is the belief in ``pseudosynchronous" spin of terrestrial planets and moons. While the stability of such states for bodies with oceans remains an open question, for telluric objects such states are unstable and therefore transient, as explained in \citet{me2013}. The explanation can be extended also to molten planets, except in the case of extremely low viscosity.

 \subsection{General facts}\label{tt}

 A consistent way of treating linear tides comprises two steps:
 \vspace{1mm}

 (1) ~Both the tide-raising potential of the secondary body (in our case, of the star) and the incremental tidal potential of the distorted primary (in our
 case, of the planet) must be expanded into Fourier series. The terms in both series are numbered by four integers $\,lmpq\,$.

 (2) ~Each $\,lmpq\,$ term in the latter series should be related to its $\,lmpq\,$ counterpart in the former series. The interrelation implies introduction of the dynamical
 Love numbers $\,k_l\,$ and phase lags $\,\epsilon_l\,$ as functions of a Fourier tidal mode $\,\omega_{\textstyle{_{lmpq}}}\,$.
 \vspace{1mm}

 An $\,lmpq\,$ term of the induced tidal potential is proportional to the $\,lmpq\,$ term of the tide-raising potential, multiplied by some geometric
 factor and by the function $~k_l\,\cos\epsilon_l\,=\,k_l(\omega_{\textstyle{_{lmpq}}})\,\cos \epsilon_l(\omega_{\textstyle{_{lmpq}}})\,$.
 Similarly, an
 $\,lmpq\,$ term $~${\it{of the secular part}} of the tidal torque will be proportional to the $\,lmpq\,$ term of the tide-raising potential, multiplied by a geometric
 factor and by the so-called quality function $~k_l\,\sin\epsilon_l\,=\,k_l(\omega_{\textstyle{_{lmpq}}})\,\sin\epsilon_l(\omega_{\textstyle{_{lmpq}}})\,$. These
 results, now classical, were obtained by \citet{k1964}. For a significant precursor, see \citet{darwin}.
 As was pointed out later by \citet{e2012}, the torque also contains an oscillating part.

 The mode-dependencies of the Love numbers and phase lags must be derived from the constitutive equation of the material of the perturbed body, with  self-gravitation
 taken into account. Instead, it has been common in the literature to use simplistic $\,${\it{ad hoc}}$\,$ models which render mathematically convenient
 mode-dependencies incompatible with physics. This often resulted not only in distortion of time scales, but also in serious qualitative misjudgments.
In particular, \citet{k1964} and some other early authors believed in a frequency-independent $\,k_l\,\sin\epsilon_l\,$, an assumption at odds with
 the results of seismic and geodetic measurements. The erstwhile popular CTL (constant time lag)
 model allows the emergence of the afore-mentioned ``pseudosynchronous" spin states and also leads to considerable
 alteration of the probabilities of entrapment into spin-orbit resonances. Hence the necessity to reexamine the tidal entrapment process,
 employing a realistic rheology.

 \subsection{An expression for the tidal torque}

 We shall derive a physics-based expression for the tidal torque acting on a terrestrial planet which is not too close to the star (${\textstyle R}/{\textstyle a}\ll1$), has a
 small obliquity ($i\simeq 0$), and not very large an eccentricity ($e < 0.4$). ~All three conditions are obeyed by Mercury now, and we have no reason to suspect that they
 were ever violated in the past.$\,$\footnote{~In our modeling of Mercury's orbital evolution, the eccentricity reached the values between $\,e=0.4\,$ and $\,e=0.45\,$ in
 only four histories out of a thousand, and over very limited time spans. This happened for three simulations covering $\,0.5\,$ Gyr and for one simulation covering $\,0.1\,$ Gyr. The
 obliquity of Mercury never exceeded 3.3 arcseconds \citep{peale2005}.}\label{eccentricity}
 As explained in detail in the Appendix \ref{A}, under these assumptions the polar component of the tidal torque,
 \ba
 \nonumber
 {\cal T}_z^{^{\,(TIDE)}}=~\sum_{l=2}^{\infty}\,\left(\frac{R}{a}\right)^{\textstyle{^{2l+1}}}\frac{G\,M_{star}^{\textstyle{^{\,2}}}}{a}
 ~\sum_{m=1}^{\it l}\frac{({\it l} - m)!}{({\it l} + m)!} \;2\;m\;\sum_{p=0}^{\it l}F_{{\it l}mp}(\inc)
 \sum_{q=-\infty}^{\infty}G_{{\it l}pq}(e)~~~~~~~~~~~~~~~~~~~~~~~~\\
                                   \nonumber\\
 \sum_{h=0}^{\it l}F_{lmh}(\inc)\sum_{j=-\infty}^{\infty} G_{{\it l}hj}(e)
 ~k_{\textstyle{_{l}}}~\sin\epsilon_{l}~\cos\left[\,
 2~(h\,-\,p)~(\omega\,+\,{\cal M})~+~(q\,-\,j)~{\cal M}
 \,\right]
   ~~_{\textstyle{_{\textstyle ,}}}~~~~\qquad
 \label{6}
 \ea
 can be approximated with
 \bs
 \ba
 \nonumber
 {\cal T}_z^{^{\,(TIDE)}}=~\frac{\textstyle 3}{\textstyle2}\,\frac{GM_{star}^{\textstyle{^{\,2}}}}{a}\left(\frac{R}{a}\right)^5 \sum_{q,\,j=-1}^{7}G_{\textstyle{_{\textstyle{_{20\mbox{\it{q}}}}}}}(e)
 \,G_{\textstyle{_{\textstyle{_{20\mbox{\it{j}}}}}}}(e)\,
 k_2(\omega_{\textstyle{_{\textstyle{_{220\mbox{\it{q}}}}}}})\,\sin\epsilon_{\textstyle{_{2}}}(\omega_{\textstyle{_{\textstyle{_{220\mbox{\it{q}}}}}}})
 \,\cos\left[\left(q- j\right){\cal M}\right]
 ~\\ \nonumber\\
 +~O(e^8\epsilon)~+~O(i^2\epsilon)~+~O\left(\,(R/a)^{7}\epsilon\right)~_{\textstyle{_{\textstyle ,}}}\,~\qquad~
 \label{7a}
 \ea
 where $\,R\,$ and $\,{\theta\,}$ are the radius and sidereal angle of the planet, $\,\stackrel{\bf\centerdot}{\theta\,}$ is its spin rate, $~a\,$ and $\,e\,$ are the
 semimajor axis and eccentricity of the star as seen from the planet. The notation  $\,G_{lpq}(e)\,$ stands for the eccentricity polynomials. As ever, $\,i\,$ denotes
 the inclination of the perturber on the equator of the perturbed body. Since in our case the role of the perturber is played by the star, $\,i\,$ has the meaning of obliquity.

 The moderate values of eccentricity allow us, in (\ref{6}), to approximate an infinite series over $\,q,\,j\,$ with a finite sum (\ref{7a}). The details of the truncation
 can be found in the Appendix.

 An $\,lmpqhj\,$ term of the torque contains a quality function $~k_l(\omega_{\textstyle{_{lmpq}}})\;\sin
 \epsilon_l(\omega_{\textstyle{_{lmpq}}})~$. The dynamical Love numbers $\,k_l(\omega_{\textstyle{_{lmpq}}})\,$ are positive definite, even functions of the
 mode.~\footnote{~So it would be equally legitimate to write $\,k_l(\omega_{\textstyle{_{lmpq}}})\,$ as $\,k_l(\chi_{\textstyle{_{lmpq}}})\,$, where $\,\chi_{\textstyle{_{lmpq}}}\equiv\,|\,\omega_{\textstyle{_{lmpq}}}\,|\,$
 are the positive definite physical forcing frequencies corresponding to the Fourier modes $\,\omega_{\textstyle{_{lmpq}}}\,$.}
 The phase lags $\,\epsilon_l(\omega_{\textstyle{_{lmpq}}})\,$ are odd functions whose sign always coincides with that of the tidal mode $\,\omega_{\textstyle{_{lmpq}}}\,$, as explained
 in Appendix A. Thus each factor $\,k_l(\omega_{\textstyle{_{lmpq}}})\,\sin\epsilon_l(\omega_{\textstyle{_{lmpq}}})\,$ can be written as
 $\,k_l(\omega_{\textstyle{_{lmpq}}})\,\sin|\,\epsilon_l(\omega_{\textstyle{_{lmpq}}})\,|~\,\mbox{Sgn}\,(\omega_{\textstyle{_{lmpq}}})\,$. Hence the expression (\ref{7a}) can be cast as
 \ba
 \nonumber
 {\cal T}_z^{^{\,(TIDE)}}=~\frac{\textstyle 3}{\textstyle2}\,\frac{GM_{star}^{\textstyle{^{\,2}}}}{a}\left(\frac{R}{a}\right)^5 \sum_{q,\,j=-1}^{7}G_{\textstyle{_{\textstyle{_{20\mbox{\it{q}}}}}}}(e)
 \,G_{\textstyle{_{\textstyle{_{20\mbox{\it{j}}}}}}}(e)\,
 k_2(\omega_{\textstyle{_{\textstyle{_{220\mbox{\it{q}}}}}}})\,\sin|\,\epsilon_2(\omega_{\textstyle{_{\textstyle{_{220\mbox{\it{q}}}}}}})\,|
 \,~\mbox{Sgn}\,(\,\omega_{\textstyle{_{\textstyle{_{220\mbox{\it{q}}}}}}}\,)~\qquad\qquad
 ~\\ \nonumber\\
 \cos\left[\left(q- j\right){\cal M}\right]
  +~O(e^8\epsilon)~+~O(i^2\epsilon)~+~O\left(\,(R/a)^{7}\epsilon\right)~_{\textstyle{_{\textstyle .}}}\,~\qquad~
 \label{7b}
 \ea
The terms with $\,q\neq j\,$ are oscillating. While they influence the fate of each trajectory, the overall statistics remains virtually insensitive to them and is defined overwhelmingly by the secular part \citep{makarovetal2012}. The terms with $\,q=j\,$ constitute the secular part:
 \ba
 \nonumber
 \langle\,{\cal{T}}_z^{\rm{_{\,(TIDE)}}}\rangle_{\textstyle{_{\textstyle_{\textstyle{_{l=2}}}}}}~=~~\quad~\quad~~\quad~\quad~\quad~\quad
 ~\quad~\quad~\quad~\quad~\quad~\quad~\quad~\quad~\quad~\quad~\quad~\quad~\quad~\quad~\quad~\quad~\quad~\quad~\quad~\quad\\
 \nonumber\\
 \frac{3}{2}~G\,M_{star}^{\,2}\,R^5\,a^{-6}\sum_{q=-1}^{7}\,G^{\,2}_{\textstyle{_{\textstyle{_{20\mbox{\it{q}}}}}}}(e)~k_2(
 \omega_{\textstyle{_{\textstyle{_{220\mbox{\it{q}}}}}}})~\sin|\,\epsilon_2(\omega_{\textstyle{_{\textstyle{_{220\mbox{\it{q}}}}}}})\,|
 \,~\mbox{Sgn}\,\left(\,\omega_{220q}\,\right)+O(e^8\,\epsilon)+O(\inc^2\,\epsilon)~~.~\quad~\quad~
 \label{overall}
 \label{7c}
 \ea
 \label{7}
 \label{tid.eq}
 \es
 In neglect of the apsidal and nodal precessions, the Fourier tidal modes $\omega_{lmpq}$ look as
 \ba
   \label{eq:LL9}
   \label{2}
   \omega_{lmpq} = (l-2p+q)n-m\dot{\theta},
 \ea
where $l>2$ is an integer termed {\it{the degree}}, $m\,=\,0,\ldots,l$ is an integer known as {\it{the order}}, while $p$ and $q$ are two other integers, p varying from $0$
through $l$, and $q$ taking values from $-\infty$ to $\infty$. Each quality function can be expressed as a function of the planetary spin $\,\stackrel{\bf\centerdot}{\theta\,}\,$:
 \bs
 \ba
 \nonumber
 k_l(\omega_{\textstyle{_{lmpq}}})\;\sin|\,\epsilon_l(\omega_{\textstyle{_{lmpq}}})\,|~\,\mbox{Sgn}\,(\omega_{
 \textstyle{_{lmpq}}})~\quad\quad\quad\quad\quad\quad\quad\quad~\quad\quad\quad\quad\quad\quad\quad\quad\quad\quad\quad\quad\quad\quad\quad\quad
 ~\\  \nonumber\\
 =~k_l(\,(l-2p+q)\,n\,-\,m\,\dot{\theta}\,)~~\sin|\,\epsilon_l(\,(l-2p+q)\,n\,-\,m\,\dot{\theta}\,)\,|~~\mbox{Sgn}\,(\,(l-2p+q)\,n\,-\,m\,\dot{\theta}\,)\,~,~\quad
 \label{8a}
 \label{formulaa}
 \ea
 where $\,(\,(l-2p+q)\,n\,-\,m\,\dot{\theta}\,)\,$ denotes a functional dependence on the argument $\,(l-2p+q)\,n\,-\,m\,\dot{\theta}\,$, not
 multiplication by a factor of $\,(l-2p+q)\,n\,-\,m\,\dot{\theta}\,$. In our case, (\ref{formulaa}) becomes:
 \ba
 \nonumber
 k_2(\omega_{\textstyle{_{220\mbox{\it{q}}}}})\;\sin|\,\epsilon_2(\omega_{\textstyle{_{220\mbox{\it{q}}}}})\,|~\,\mbox{Sgn}\,(\,\omega_{
 \textstyle{_{220\mbox{\it{q}}}}}\,)~\quad\quad\quad\quad\quad\quad\quad\quad~\quad\quad\quad\quad\quad\quad\quad\quad\quad\quad\quad\quad
 ~\\  \nonumber\\
 =~k_2(\,2(n-\dot{\theta})\,+\,q\,n\,)~~\sin|\,\epsilon_2(\,2(n-\dot{\theta})\,+\,q\,n\,)\,|~~\mbox{Sgn}\,(\,2(n-\dot{\theta})
 \,+\,q\,n\,)\,~.~\quad
 \label{formulab}
 \label{8b}
 \ea
 \label{formula}
 \label{8}
 \es
  Then the entire sum (\ref{7a}), or its equivalent (\ref{7b}), or its secular part (\ref{7c}), can be interpreted as a function of $\,\stackrel{\bf\centerdot}{\theta\,}$, the mean motion and eccentricity treated as slow parameters.

\begin{figure}[htbp]
 \vspace{3.5mm}
 \centering
 \includegraphics[angle=0,width=0.70\textwidth]{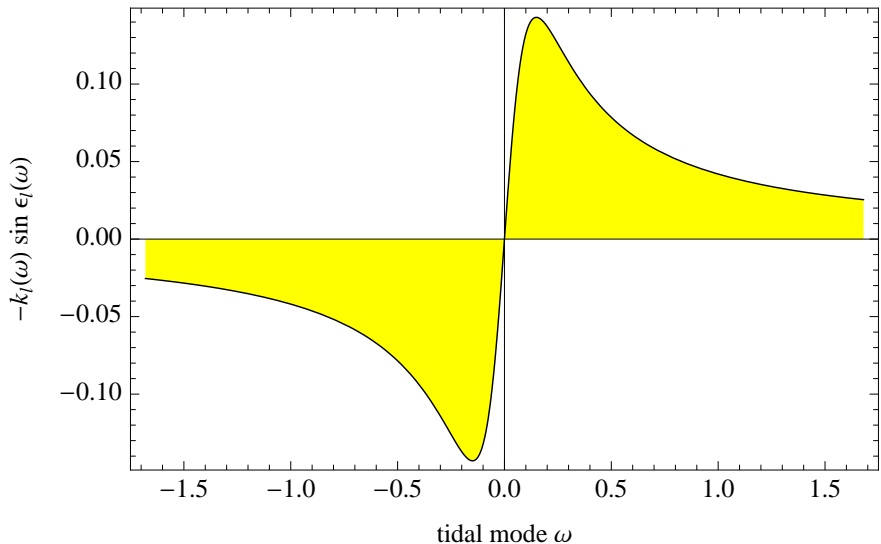}
 \caption{\small{~A typical shape of the quality function $\,k_l(\omega)\,\sin\epsilon_l(\omega)\,$, ~where $\,\omega\,$ is a shortened notation for the tidal Fourier mode
 $\,\omega_{\textstyle{_{lmpq}}}\,$.
 \label{Fig1}}}
 \end{figure}

 In each term of the sums (\ref{7}), the factor $\,k_2\,\sin\epsilon_2\,=\,k_2\,\sin|\epsilon_2|\,\mbox{Sgn}\,(\omega_{\textstyle{_{220\mbox{\it{q}}}}})\,$, expressed
 as a function of $\,\stackrel{\bf\centerdot}{\theta\,}$, has the shape of a kink, as shown in Figure \ref{Fig1}.
 The shape ensues from interplay of rheology (which dominates at higher frequencies) and self-gravitation (which takes over in the zero-frequency limit), see the subsections \ref{rheo1} and \ref{rheo2}. The emergence of this shape is natural, for each term should transcend zero and change its sign $\,${\it{continuously}}$\,$ when the rotation rate goes through the appropriate spin-orbit resonance.

\begin{figure}[htbp]
 \vspace{3.5mm}
 \centering
 \includegraphics[angle=0,width=0.98\textwidth]{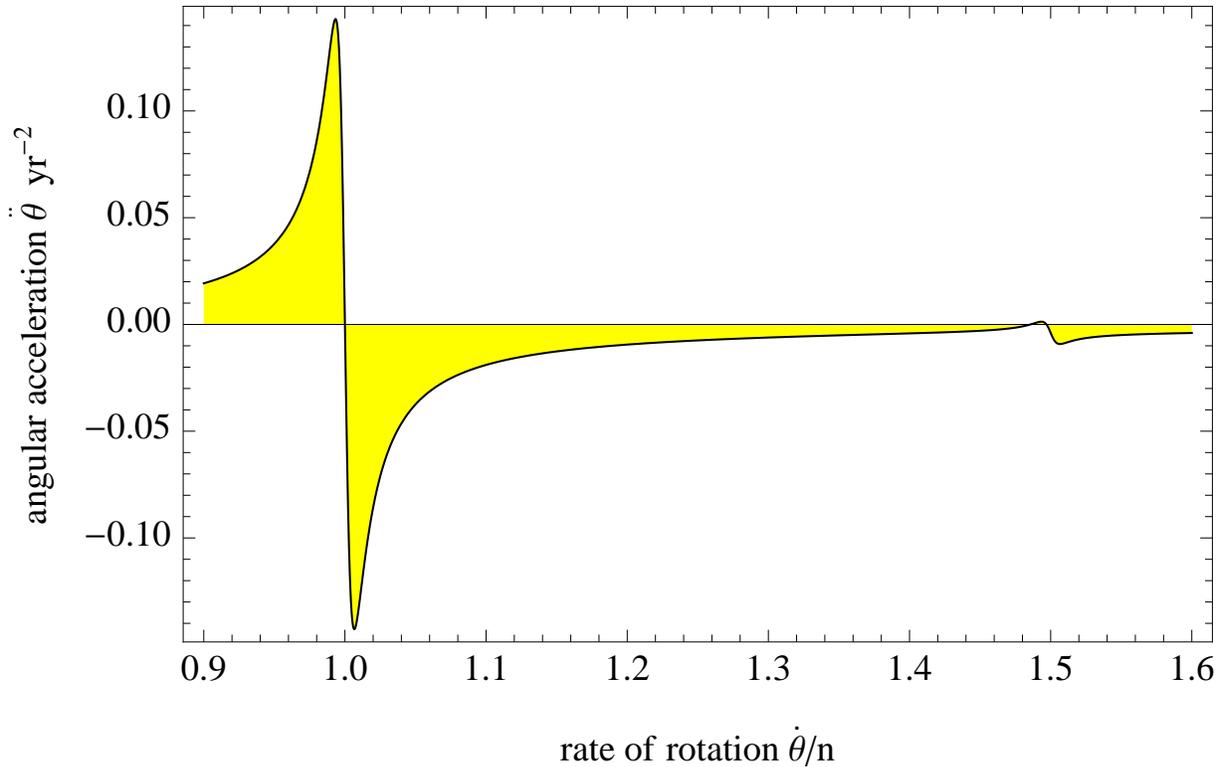}
 \caption{\small{~The angular acceleration provided by the secular part of the tidal torque. The torque is expanded into the Darwin-Kaula series and expressed as a function of the dimensionless spin rate, $\,\dot{\theta}/n\,$. The left kink comes from the leading, semidiurnal term $\,\left\{lmpq\right\}\,=\,\left\{2200\right\}\,$. It is centered in the point $\,\dot{\theta}/n=1\,$ corresponding to the 1:1 spin-orbit resonance. The right kink comes from the term with $\,\left\{lmpq\right\}\,=\,\left\{2201\right\}\,$, which is vanishing in the $\,3:2\,$ resonance. The right kink is considerably displaced vertically, owing to the bias provided by the slopes of the other kinks, mainly by the leading, semidiurnal kink.
 \label{Fig2}}}
 \end{figure}

 In the sums (\ref{7a}) and (\ref{7b}), the kink-shaped factors $\,k_2\,\sin\epsilon_2\,$ are multiplied by $\,G^{\,2}_{\textstyle{_{\textstyle{_{20q}}}}}(e)\,$. So the sum (\ref{7b}),
 treated as a function of $\,\stackrel{\bf\centerdot}{\theta\,}$, is a superposition of nine kinks of different magnitudes.
 The kinks are centered at different resonant values of $\,\stackrel{\bf\centerdot}{\theta\,}\,$. The resulting curve depicting the sum (\ref{7}) crosses zero in points {\it{very close}}
 to the resonances $\,\dot{\theta}\,=n\,\left(1+\,{\textstyle q}/{\textstyle 2}\right)\,$, but not exactly in the resonances --- like the little kink near
 $\,\dot{\theta}/n\,=\,3/2\,$ in Figure \ref{Fig2}.$\,$\footnote{~As ensues from (\ref{6}) and (\ref{8a}), the $\,lmpq\,$ term of the secular ($j=q$) part of the torque is
 decelerating for $\,\omega_{\textstyle{_{lmpq}}}<0\,$ and accelerating for $\,\omega_{\textstyle{_{lmpq}}}>0\,$. Equivalently, it is negative on the right of the $\,lmpq\,$ resonance
 and positive on its left -- see the two kinks depicted in Figure \ref{Fig2}.}

 Sitting on the right slope of a more powerful kink (the one with $\,q\,=\,0\,$), the little kink (corresponding to $\,q\,=\,1\,$) experiences a vertical displacement
 called {\it{bias}}. Although the bias of the little kink is caused by all the other kinks, the influence of its left neighbour is clearly dominant.
 The vertical bias of the little kink in Figure \ref{Fig2} is a feature immanent to any spin-orbit resonance, if the eccentricity is non-zero. The zero-eccentricity case is
 exceptional because, in the quadrupole approximation,$\,$\footnote{~In higher-degree approximations, we would be taking into account the $\,lmpq\,$ terms with $\,l>2\,$. However,
 from the tables of the $\,F(i)\,$ and $\,G(e)\,$ functions, it can be observed that only the terms with $\,q = p = l - m = 0\,$ come out non-vanishing for $\,e = i = 0\,$. In
 the quadrupole approximation, the sole surviving term will be $\,\{lmpq\}\,=\,\{2200\}\,$, while in a higher-degree approximation we would get the extra modes
 with $\,\{lmpq\}\,=\,\{ll00\}\,$, for each $\,l\,$ included.} it comprises only the semidiurnal component of the tidal torque (the one corresponding to the principal tidal
 frequency $\,\omega_{2200}\,$).

 A constant bias of the secular torque was also considered by \citep{gp1968}, who pointed out its crucial role in the capture probability calculation. However,
 in our model the bias has a physical meaning very different from that in the CTL model. In our case, the bias is generated by the inputs from the non-resonant modes
 called into being by non-uniform orbital motion. Without the non-vanishing bias terms, capture into a spin-orbit resonance would be inevitable, and the tidal dissipation
 would completely cease once the rotator is locked in the resonance. Near a resonance $\,q\,'\,$, the spin rate $\,\stackrel{\bf\centerdot}{\theta\,}$ is close to
 $\,n\,\left(1\,+\,{\textstyle q\,'}/{\textstyle 2}\right)\,$, and the right-hand side of the sum (\ref{7b}) can be conveniently decomposed into two parts. The first one
 is the $\,q=q\,'\,$ term, a kink-shaped function transcending zero at $\,\stackrel{\bf\centerdot}{\theta\,}\,=\,n\,\left(1\,+\,{\textstyle q\,'}/{\textstyle 2}\right)
\,$. The second part, {\it{bias}}, comprises all the $\,q\neq q\,'\,$ terms of the sum. This way, the bias is the input of all the $\,q\neq q\,'\,$ terms into the values
assumed by the overall torque in the vicinity of the $\,q=q\,'\,$ resonance. In the said vicinity, the bias is a slowly changing function, which can, to a good approximation,
be treated as constant. For not too large eccentricities, the bias is usually negative -- this is why the little kink in Figure \ref{Fig2} is shifted downwards.

  While the $\,q=q\,'\,$ term assumes a zero value at $\,\dot{\theta}/n\,=\,\left(1\,+\,{\textstyle q\,'}/{\textstyle 2}\right)\,$, the bias displaces the location of
  the zero. The 3:2 kink in Figure \ref{Fig2} goes through nil in a point located a bit to the left of $\,\dot{\theta}/n\,=\,3/2\,$. Being extremely small, the shift of
  the equilibrium away from the resonance has no practical consequences, since the net nonzero tidal torque is compensated by an opposing secular triaxiality-caused torque,
  as explained in \citet{me2013}.

 \subsection{The tidal torque and the role of rheology}\label{rheo1}

  The torque (\ref{7}), taken as a function of the spin $\,\stackrel{\bf\centerdot}{\theta\,}$, is a superposition of kinks. Depicting a typical kink and the way kinks
  superimpose, Figures \ref{Fig1} and \ref{Fig2} serve illustrative purposes and do not correspond to realistic
  parameters. Each kink is
  stretched horizontally, its extrema being spread apart, to emphasise that the transition through nil is continuous.

\begin{figure}[htbp]
 \centering
 \includegraphics[angle=0,width=0.90\textwidth]{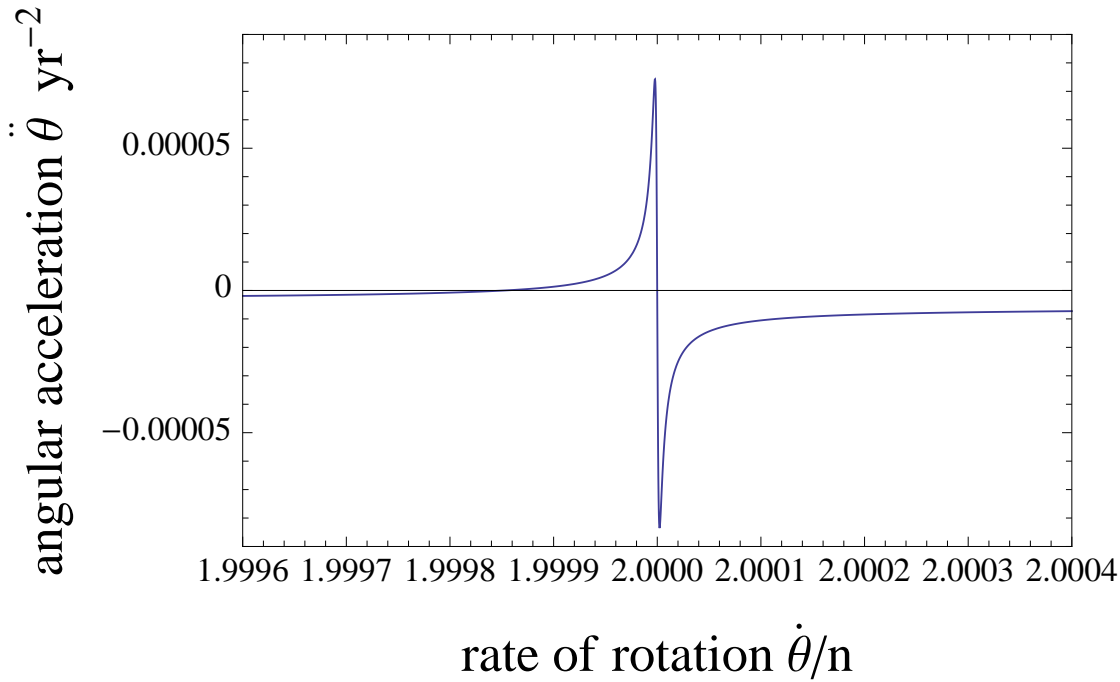}
 \caption{\small{Angular acceleration of Mercury caused by the secular tidal torque (\ref{7}), in the vicinity of the 2:1 spin-orbit resonance.
 \label{2-1kink.fig}}}
 \end{figure}

 Analogous figures for realistic parameters of Mercury would have sharper features, like in Figure \ref{2-1kink.fig}. Despite the sharp-looking peaks, the dependency is
 continuous, and the middle part has a finite slope, even though in the picture it looks almost vertical.
 The form of the plot is dictated by the shape and magnitudes of the quality functions  $~k_l(\omega_{\textstyle{_{lmpq}}})\,\sin\epsilon_l(\omega_{\textstyle{_{lmpq}}})~$
 expressed as functions of the spin rate via formulae (\ref{8}). The quality functions show up in each term of the tidal torque, and the shapes of these functions are
 crucial for determining the probabilities of entrapment. The shapes are defined by the size and mass of the body and by its rheological properties. The rheological
 properties are accounted for by a constitutive equation, i.e., by a law interconnecting the strain and stress. Within linear rheologies, a constitutive equation can
 be rewritten in the frequency domain where each Fourier harmonic $\,\bar{u}_{\gamma\nu}(\chi)\,$ of the strain gets expressed, algebraically, as the appropriate
 harmonic of the stress $\,\bar{\sigma}_{\gamma\nu}(\chi)\,$ multiplied by the complex compliance $\,\bar{J}(\chi)\,$ appropriate to this harmonic:
 \ba
 2\;\bar{u}_{\gamma\nu}(\chi)\,=\;\bar{J}(\chi)\;\bar{\sigma}_{\gamma\nu}(\chi)\,~,
 \label{LLJJKK}
 \ea
 $\chi\,=\,\chi_{\textstyle{_{lmpq}}}\,$ being the physical forcing frequency, and $\,\gamma\nu\,$ being the tensorial indices. The relation (\ref{LLJJKK}) is an exact
 analogue to a complex harmonic of current written as the appropriate harmonic of voltage, multiplied by the inverse complex impedance. The roles of the current, voltage
 and inverse impedance are played by $\,2\,\bar{u}_{\gamma\nu}(\chi)\,$, $\,\bar{\sigma}_{\gamma\nu}(\chi)\,$ and $\,\bar{J}(\chi)\,$, correspondingly.

 The complex compliance $\,\bar{J}(\chi)\,$ carries all information about the response of the material, under the assumption that the material is isotropic.
 Modeling the mantle rheology, i.e., deriving $\,\bar{J}(\chi)\,$, we rely on a combined model which is Andrade at higher frequencies and Maxwell in the low-frequency limit.
 As was demonstrated in \citet{e2012,e2012b}, the mathematical formalism of the Andrade model can be modified in a manner permitting a switch to the Maxwell model at
 low frequencies. The necessity for this switch originates from the fact that at different frequencies different physical mechanisms dominate the internal friction.

  The function $\,\bar{J}(\chi)\,$ can be used to determine the tidal response of the planet, i.e., the quality
  functions $\,k_l(\omega_{\textstyle{_{lmpq}}})~\sin\epsilon_l(\omega_{\textstyle{_{lmpq}}})\,$. Carrying out this part of the problem, we assume that the
  larger part of tidal dissipation of energy is taking place in the mantle, and we approximate the planet with a homogeneous ball endowed with the rheological
  properties representative of the mantle. Despite the evident simplification, the model is likely to reflect the qualitative features of the spin-orbit
 interactions, and to entail a reasonable comparison of the entrapment probabilities for different resonances. The main qualitative feature faithfully represented by
 the model is the kink shape of the terms in the expansion of torque.

 Further improvement of the model and its enrichment with details (friction between layers, etc) will certainly alter the magnitude and detailed shapes of the kinks.
 It is however highly improbable that the basic kink shape varies considerably (though it may, in principle, acquire secondary ripples due to interaction between the
 layers of the planet).

 With all these assumptions taken into account, the complex compliance can be used to write down the functions $\,k_l(\omega_{\textstyle{_{lmpq}}})~\sin\epsilon_l(\omega_{\textstyle{_{lmpq}}})\,$.

 \subsection{$k_l/Q_l\,$ as functionals of the complex compliance of the mantle}\label{rheo2}

  The expression (\ref{7c}) contains the so-called quality function $~k_l(\omega_{\textstyle{_{lmpq}}})~\sin\epsilon_l(\omega_{\textstyle{_{lmpq}}})
 ~=~$ $k_l(\omega_{\textstyle{_{lmpq}}})~\sin|\,\epsilon_l(\omega_{\textstyle{_{lmpq}}})\,|~\mbox{Sgn}\,(\omega_{\textstyle{_{lmpq}}})~$ defined by the
 rheology and self-gravitation of the planet. As shown in Efroimsky (2012a,b), the absolute value of this function is calculated as
 \ba
 k_l(\,\omega_{\textstyle{_{lmpq}}}\,)\;\sin|\,\epsilon_l(\,\omega_{\textstyle{_{lmpq}}}\,)\,|\;=\;-\;\frac{3}{2\,({\it l}\,-\,1)}\;\,\frac{A_l\;{\cal{I}}}{\left(\;{\cal{R}}
 \;+\;A_l\;\right)^2\;+\;{\cal{I}}^{\textstyle{^{\,2}}}}~~~,~~~~~
 \label{pppp}
 \ea
 where the dimensionless real and imaginary parts of the compliance $\,\bar{J}(\chi)~$ of the mantle are
 \ba
 {\cal R}\;=\;1\;+\;(\chi\tau_{_A})^{-\alpha}\;\cos\left(\,\frac{\alpha\,\pi}{2}\,\right)
 \;\Gamma(\alpha\,+\,1)~~~,\quad\quad\quad\quad\quad~\quad\quad\quad\quad\quad\quad\quad
 \label{}
 \ea
 \ba
 {\cal I}\;=\;-\;(\chi\tau_{_M})^{-1}\;-\;(\chi\tau_{_A})^{-\alpha}\;\sin\left(
 \,\frac{\alpha\,\pi}{2}\,\right)\;\Gamma(\alpha\,+\,1)~~~;~~~~~~~~~~~~~~\quad\quad\quad\quad\quad
 \label{}
 \ea
 $\Gamma\,$ is the Gamma function, while $\,\chi\,$ is a shortened notation for the physical forcing frequency
 \ba
 \chi_{\textstyle{_{lmpq}}}\,\equiv~|\,\omega_{\textstyle{_{lmpq}}}\,|~=~|\,(l-2p+q)\,n\,-\,m\,\dot{\theta}\,|\,~,
 \label{}
 \ea
 $\alpha\,$ is the Andrade parameter (about $\,0.2\,$ for silicates with partial melt), while $\,\tau_{_M}\,$ and $\,\tau_{_A}\,$ are the Maxwell time and the
 Andrade time of the mantle. Recall that the Maxwell time is the ratio of viscosity to rigidity: $\,\tau_{_M}=\eta/\mu\,=\,\eta\,J\,$.
 On the interrelation of the Maxwell and Andrade times, see Appendix \ref{B} below.

 The quantities $\,A_l\,$ entering (\ref{pppp}) are furnished by
 \ba
 A_l~=~\frac{3\,\mu\,(2\,l^2\,+\,4\,l\,+\,3)}{4~l~\pi~G~\rho^2~R^2}~=~\frac{3\,(2\,l^2\,+\,4\,l\,+\,3)}{4~l~\pi~G~\rho^2~R^2~J}\,~,
 \label{fdc}
 \ea
  where $\,\mu\,$, $\,J\equiv1/\mu\,$, $\,\rho\,$ and $\,R\,$ are the unrelaxed rigidity, compliance, mean density and radius of the planet,  while
 $\,G\,$ is the gravitational constant. The presence of the factors (\ref{fdc}) in the denominator of (\ref{pppp}) is owed to the competition between rheology and self-gravitation (the latter playing effectively the role of extra rigidity).

 To avoid numerical complications in the zero-frequency limit (i.e., on crossing resonances), it is convenient to multiply both the numerator and denominator of (\ref{pppp})
 by $\,\chi\,$:
 \ba
 k_l(\omega_{\textstyle{_{lmpq}}})~\sin|\,\epsilon_l(\omega_{\textstyle{_{lmpq}}})\,|~=~ -~\frac{3}{2\,(l\,-\,1)}~\frac{A_l~\mathcal{I\,}'~\chi}{\left(\mathcal{R}'\,+~A_l~\chi\right)^2~+~\mathcal{I\,}'^{\,2}}\,~,
 \label{rheology}
 \ea
 where
 \ba
 \mathcal{R}' & = & \chi~+~\chi^{1-\alpha}~\tau_{_A}^{-\alpha}~\cos\left(\alpha\frac{\pi}{2}\right)~\Gamma(\alpha+1)\,~,\\
 \label{real}
 \nonumber\\
 \mathcal{I\,}' & = & -~\tau_{_M}^{-1}-\chi^{1-\alpha}~\tau_{_A}^{-\alpha}~\sin\left(\alpha\frac{\pi}{2}\right)~\Gamma(\alpha+1)\,~.
 \label{imaginary}
 \ea
 Numerical implementation of the tidal model (\ref{7}) takes a much longer computation time than the model (\ref{eq:tidecl}). So we run a set of 1,000 simulations,
 with the tidal torque containing only the secular terms (with $q=j$). With various Maxwell times, we do this for a rigid Mercury initially prograde (Section \ref{sec:scen1}) and retrograde (Section \ref{sec:scen3}). For a differentiated Mercury with core-mantle friction, we present a semi-analytical model combined with limited numerical tests -- for the planet initially prograde (Section \ref{sec:scen2}) and retrograde (Section \ref{sec:scen3}).

 \section{A characteristic time of despinning}\label{despin.sec}

 Before integrating the law of motion (\ref{eq.eq}), it is of both theoretical and practical  interest to estimate the characteristic time scale of secular deceleration. This time scale is greatly influenced by the tidal dissipation rate which, in its turn, is dependent on the instantaneous rate of rotation, on the ratio of planet's radius $\,R\,$ to the orbital separation $\,a\,$, on the eccentricity $\,e\,$, and on other physical parameters. Theory tells us that some bodies spin down quickly, on the time scale of $\,10^5\,$--$\,10^6\,$ yr (e.g., the Moon), while others need time intervals comparable to their lifetime, to slow down appreciably \citep{cel2011}. So the required computing time may be prohibitively long for a definite result to be obtained.

 Traditionally \citep[e.g.,][]{ras}, a characteristic time of despinning is defined as
\be
\tau_{\textstyle{_{\rm{despin}}}}~=~\frac{\,|\stackrel{\bf\centerdot}{\theta\,}|\,}{|\stackrel{\bf\centerdot\,\centerdot}{\theta\,}|}~\,~.
\label{eqdespin}
\ee
 Representing a time span over which the spin slows considerably, this quantity should not be taken for the actual time a planet needs to slow down from a given $\,\stackrel{\bf\centerdot}{\theta\,}$ to, say, the synchronous state. The actual time of despinning can be obtained only by numerical integration of (\ref{eq.eq}).

\begin{figure}[h]
\centering
\resizebox{15.2cm}{!}{\includegraphics [angle=0,width=0.99\textwidth]{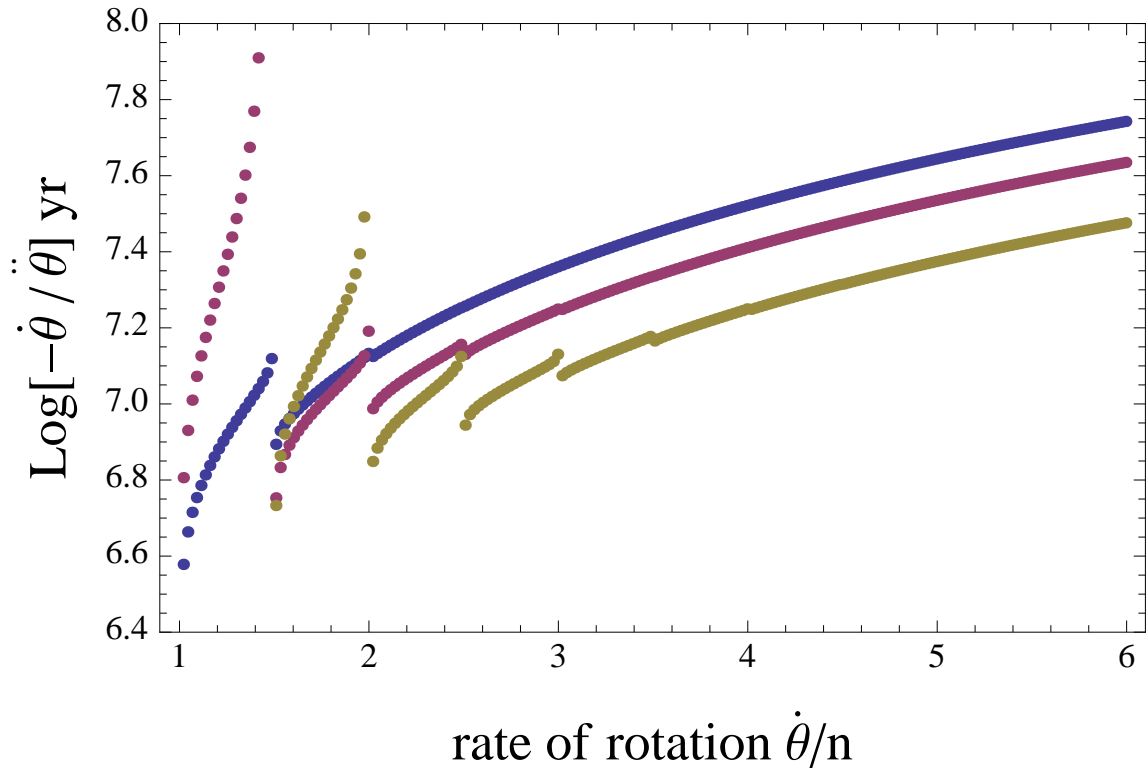}}
\caption{\small{~Decimal logarithm of the characteristic spin-down time of Mercury as a function of the spin rate, for three values of the eccentricity: $\,e=0.1\,$ (upper curve),
 $\,e=0.2056\,$ (middle curve), $\,e=0.3\,$ (lower curve). The curves are computed for discrete sets of points, to avoid the resonances. For all values of $\,e\,$,
 Mercury's spin is accelerating when $\,\dot\theta/n\,<\,1\,$ (not shown in this figure). For a sufficiently large $\,e\,$, the spin may be accelerating also over some
 spans of values of $\,\dot\theta/n\,$ larger than unity. For example, for the value $\,e=0.3\,$, the spin is accelerating when $\,\dot\theta/n\in\left[\right.1\,,\;1.5\left.\right)\,$. It is for this reason that the segment of ~Log$\,[-\dot{\theta}/\ddot{\theta}\,]\,$ between $\,\dot\theta/n=1\,$ and $\,\dot\theta/n=1.5\,$ is not plotted here. At this eccentricity, the secular acceleration, $\,\ddot\theta\,$ is positive for $\,\dot\theta/n<1.5\,$.}}
\label{despin.fig}
\end{figure}

 The decimal logarithms of characteristic spin-down times are shown in Figure \ref{despin.fig} for three values of eccentricity: $\,e=0.1\,$ (upper curve), $\,e=0.2056\,$ (middle curve), and $\,e=0.3\,$ (lower curve). Generally, these times are of the order of $\,10^7\,$ yr or longer, while the integration step size has limitations
 imposed by the kink shape of the tidal torque components.$\,$\footnote{~When the planet traverses a low-order resonance, a considerable slow-down occurs due to the left shoulder of the appropriate tidal kink. In Figures \ref{Fig2} and \ref{2-1kink.fig}, the right shoulder of each kink is, typically, negative (decelerating), while the left shoulder of each kink is, typically, positive (accelerating). We prefer to say `{\it{typically}}', because there may exist situations where
 a kink corresponding to a higher resonance is located, fully or partially, below the horizontal axis -- see the small kink
 in Figure \ref{Fig2}. Even though, despinning leftwards through this kink will result in a decrease of the overall despinning action of the tidal torque.}

 \section{Numerical modeling of the history of Mercury's eccentricity \label{sec:eccentricity}}

\subsection{Background}

 Classical methods to investigate the past orbital history of Mercury$\,$\footnote{~We also note an effective secular method
which was developed by \citet{blf2012} and was published posteriorly to our numerical integrations.} rely either on numerical integrations of the planetary complete
equations of motion \citep{qtd1991,lrjgcl2004a,lcgjlr2004b,bl2008,lg2009} or on numerical integration of
their averaged counterparts \citep{l1988}, or on fully analytical approximations \citep{b1974}. In this section, we present a method employed by us to obtain the
eccentricity history of Mercury over Gyr timescales. The method is based on the statistical formulae developed in \citet{l2008}.

 The earliest example of the effect of chaos in the solar system was the demonstration by Jack Wisdom that the Kirkwood gaps in the asteroid belt coincided with
 chaotic zones and that motion in these chaotic zones was unstable to planet crossing -- see \citet{wisdom} and references therein. A hint on chaotic behaviour
 could be found also in \citet{nmc1989}.$\,$\footnote{~Although the behaviour observed by \citet{nmc1989} was in all likelihood chaotic, their LONGSTOP tool never
 calculated orbits over intervals sufficiently long to reliably prove the chaotic character of the observed motion.} The chaotic behavior of the entire Solar System
 was demonstrated, with averaging, by \citet{l1989} and, with no averaging involved, by \citet{sw1992}. Numerical simulations show that no precise solution can be
 obtained by integration back in time over $\sim$ 50 Myr or longer \citep{lfgm2011}.$\,$\footnote{~A more precise statement is that the behaviour of the solar system
 can display both chaos and near-integrability, with tightly-packed regions of both chaotic and regular orbits \citep{guzzo2005,guzzo2006,hayes}.} Statistical
 considerations over a large number of possible orbits are thus needed.

 Within the statistical approach, \citet{l2008} studied the possible orbital history of Mercury, using the averaged planetary equations. His dynamical model
 contains the eight planets, with their mutual secular gravitational perturbations obtained through a second-order perturbation method, along with
 perturbations from the General Relativity and the
 Moon \citep{l1986}. Note that the averaged equations imply constant secular semi-major axes of the planets.
 After numerical integration of a large number of histories of the Solar System over 5 Gyr,  starting with
very close initial conditions, some statistics can be obtained about the values of the eccentricity and inclination of each planet
over this timespan. In particular, the eccentricity of Mercury $\,e\,$ statistically follows a Rice density function (see Laskar 2008):
\begin{equation}
f_{T,s}(e) ~=~ \frac{e}{\sigma^2} ~\exp \bigg( -~ \frac{\,e^2\,+\,s^2\,}{2~ \sigma^2}  \,\bigg)~ I_0\bigg(\frac{e s }{\sigma^2}\bigg)\,~,
\label{jueq1}
\end{equation}
where the parameter $\sigma$ obeys the approximation
\begin{equation}
\sigma^2(T) ~=~ b_0 ~+~ b_1 ~T
\label{jueq2}
\end{equation}
corresponding to a diffusion process, with $\,b_0 = 2.07 \times 10^{-3}\,$, $\,b_1 = 1.043 \times 10^{-3}\,$, $\,s=0.1875\,$,
and with $\,T\,$ being the time in Gyr. The notation $\,I_o\,$ stands for the modified Bessel function of the first kind.
 This function is
 often approximated by a series expansion, but in the case of large values of the argument, it is adequately approximated with an exponential function \citep{ptvf1992}. It
 is important to note that $\,s\,$ and $\,\sigma\,$ are not the mean and standard deviation of the Rice distribution.

\subsection{Method}

Our goal here is to produce a synthetic time series of Mercury's eccentricity, based on the density function (\ref{jueq1}). To this end, we use a simple Wiener
process (or a Brownian motion) of the type
\begin{equation}
e(T+\delta_t) ~=~ e(T) ~+~ \sigma(\delta_t)~\delta_e \,~,
\label{jueq3}
\end{equation}
where $\,e(0)\,$ corresponds to the initial eccentricity and $\,\sigma\,$ is the standard deviation. Each of the eccentricity values are separated in time by a
constant time step $\,\delta_t\,$, and the process has independent and stationary increments $\,\delta_e\,$. For a Wiener process, the increments $\,\delta_e\,$ obey a
Gaussian distribution $\,\mathcal{N}(0,1)\,$.

 To obtain the standard deviation needed in (\ref{jueq3}), we fit a Gaussian distribution
\begin{equation}
f(e;\bar{e},\sigma^2) ~=~ \frac{1}{\sqrt{2 \pi \sigma^2}} ~\exp\bigg[-\,\frac{(e-\bar{e})^2}{2~ \sigma^2}\,\bigg]
\end{equation}
to the Rice density function (\ref{jueq1}). $\bar{e}$ is the mean value of the Gaussian distribution, and its standard deviation corresponds to $\,\sigma^2(t) = 0.0009\, t\,$. We
then applied a few corrections to the time series.

First, the Rice density has a characteristic drift of its mean value and most probable value over time. In order for the Gaussian distribution to still maintain a
good fit in the complete time interval, this can be compensated by adding a small drift $\,D = 0.0023\,T\,$ to the value of the eccentricity at a time $\,T\,$. By fitting
the drift, we obtain an initial mean $\,\bar{e}=0.1876\,$, which will be used also for the initial value $\,e(0)\,$.
\medskip

Secondly, we make a change of the timescale: $\,T^*=\,T\,+\,2.33\,$ Gyr. This step is necessitated by the fact that the first 250 Myr of integration computed by
\citet{l2008} have a very different slope of $\,\sigma^2(t)\,$, and have not been taken into account in his fit of (\ref{jueq1}) -- see Laskar (2008,  Figure 15b). To
be able to fit the Rice distribution with its current parameters for all values of $\,T\,$, a Gaussian distribution with a standard deviation linear in time has to
start at an initial time of $\,-2.33\,$ Gyr. Also, due to the lack of information about the first $\,0.5\,$ Gyr of evolution, we keep only the part corresponding to
the interval $\,T^*\in\,[\,0.5\,,~4\,]\,$ Gyr.
\medskip

\begin{figure}[h]
\centering
\resizebox{8.cm}{!}{\includegraphics [angle=270,width=\textwidth] {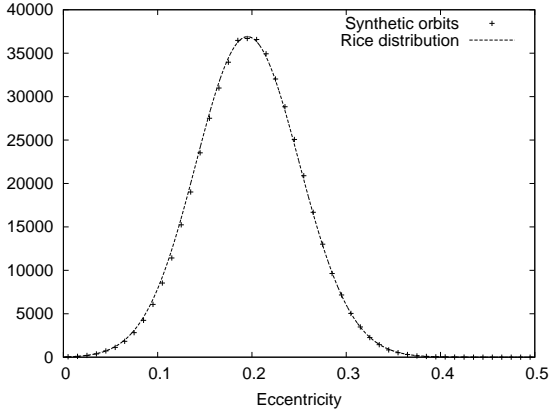}}
\caption{The Rice distribution function for $T=1$ Gyr from Eq.\ref{jueq1} (line) and its approximation obtained from the distribution of 500,000 synthetic orbits
evolved for 1 Gyr with the process of Eq.\ref{jueq3} (points).}
\label{jufig4}
\end{figure}

 One of the discrepancies between the Rice distribution and the Gaussian distribution lies in the distribution of very low eccentricity values. The Rice density is
 defined in the range $[0\,,\,+\infty]$, and its distribution for low $e$ tends to be more linear with increasing $\,T\,$. On the other
hand, the Gaussian distribution is defined on $[-\infty\,,\,+\infty]$, allowing the possibility of negative eccentricities, and is completely symmetric with respect
to its mean value. We can see from \citet{l2008} that the eccentricity distribution of Mercury does not show a strong linear behaviour for low $\,e\,$
(which is not the case for the other planets of the Solar System). The orbits which have $\,e\,<\,0\,$ at any time during the 3.5 Gyr time span have been discarded
from the final results, with the side effect of slightly depleting the distribution of low $\,e\,$, compared to a non-modified Gaussian distribution.$\,$\footnote{~Synthetic
orbits obtained via the Brownian-motion method can have negative eccentricities. While the Rice distribution prohibits this, the Gaussian distribution does not.} We
have checked that it does not introduce any bias, since this reduced distribution matches the Rice distribution for low $\,e\,$ even more closely. Figure \ref{jufig4}
shows the Rice distribution
function for $\,T=1\,$ Gyr computed with the equation (\ref{jueq1}), compared with the distribution of 500,000 synthetic orbits obtained through the process (\ref{jueq3}).

\begin{figure}[h]
 \centering
 \resizebox{7.9cm}{!}{\includegraphics [angle=270,width=\textwidth] {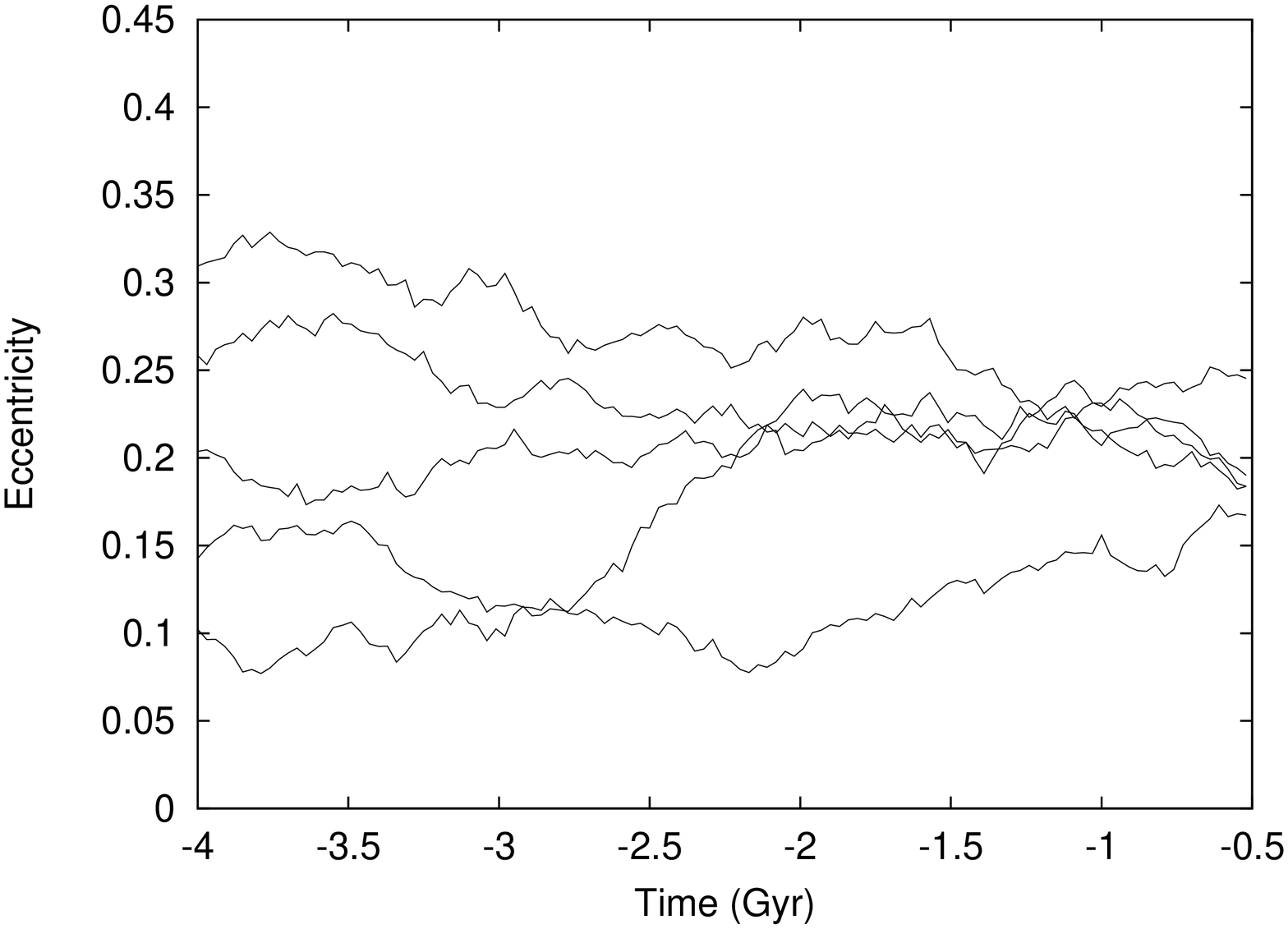}}
 \resizebox{7.9cm}{!}{\includegraphics [angle=270,width=\textwidth] {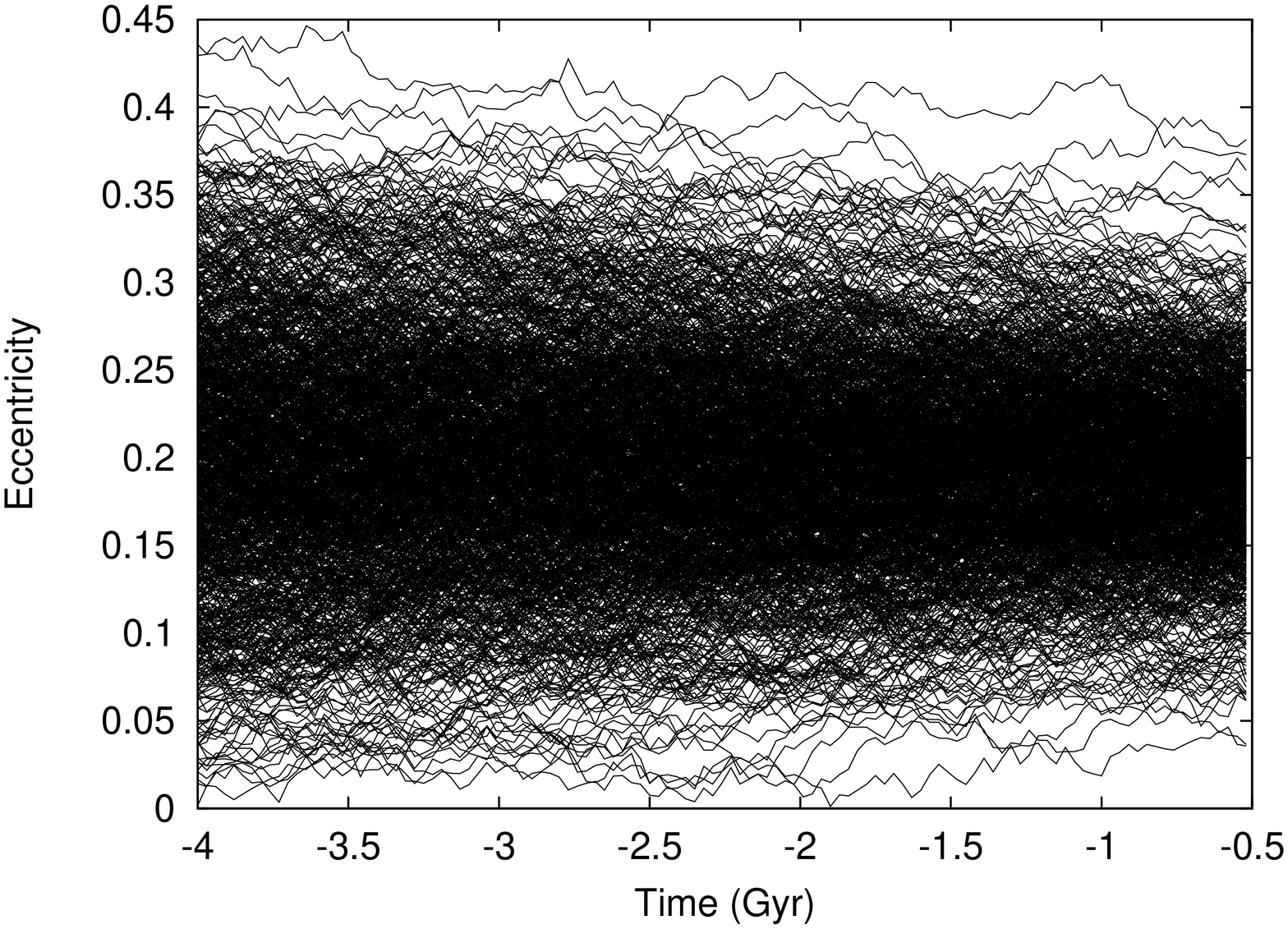}}
 \caption{Evolution over time of five and one thousand synthetic orbits obtained with the Wiener process (Eq.\ref{jueq3}).}
 \label{jufig2}
 \end{figure}

 Our third correction concerns the time direction. The secular planetary equations used by \citet{l2008} are perfectly time-reversible and have been used to explore the
 past orbital history of Mercury by \citet{cl2004} over 4 Gyr. The process described above can then be applied by only reversing its time evolution
 (see some examples in Figure $\ref{jufig2}$).

 \section{Numerical simulation of Mercury's spin rate \label{sec:correia}}

 \subsection{Methodology}

 \par
 Our next goal is to study the influence of the realistic tidal model on the probability of Mercury's capture into spin-orbit resonances. \citet{m2012} has recently accomplished this work for a constant eccentricity. Here we use our synthetic trajectories, to include the statistical variations of Mercury's eccentricity in the study. We will build statistically relevant samples of simulations of the despinning of Mercury, with two kinds of tidal model: the one from \citet{cl2004} and our model (\ref{7b}).

 \par
  We integrate numerically the equation (\ref{eq.eq}) for the spin rate of Mercury. The torques entering this equation depend on the eccentricity $\,e\,$ which is calculated, at each step, from the synthetic distribution. The value of $\,e\,$ at a given time is estimated, by cubic spline interpolation,
 from the synthetic evolution trajectory sampled at a regular grid of time points \citep{db2008}. The interpolation error is of the $\,4^{th}\,$ order \citep{s1969}, so a decrease of the sampling step of the trajectory by a factor of $\,\alpha\,$ reduces the
 interpolation error by a factor of $\,\approx\alpha^4\,$. For generating eccentricities, we used the interpolation routines from the GNU Scientific Library \citep{gdtgjabr2009}.
 The numerical integrator was the Adams-Bashforth-Moulton $\,10^{th}$-order predictor-corrector scheme \citep{hnw1987}. Through the simulations, the value of the mean
 motion was kept constant: $\,n\,=\,26.0878\,$ rad/yr.

 \subsection{Test: reproducing the results from \citet{cl2004}}

 To check our code, in general, and our generator of eccentricity histories, in particular, we reproduced the results from \citet{cl2004}.
 To that end, , we ran 10 sets of 1,000 runs, all starting with a spin period of 20 days. As stated in {\it{Ibid.}}, the choice of the initial spin rate is not critical, and high-order resonances are unimportant. To save the computer time, we integrated over only 3 Myr, which is long enough to slow down Mercury and to reach the Type I captures. Each set of 1,000 simulations uses the same set of 1,000 synthetic eccentricity trajectories, the differences between our 10 sets being only in the initial value of the sidereal angle $\,\theta\,$. We use these 10 sets to check how the number of trapped trajectories can change from one set to another, with the same eccentricities. We used the same values of the physical parameters as in {\it{Ibid.}}. In particular, in this subsection we set $\xi = 1/3$ and $(B-A)/C = 1.2\times10^{-4}$.

\begin{figure}[ht]
\centering
\includegraphics[width=7.5cm]{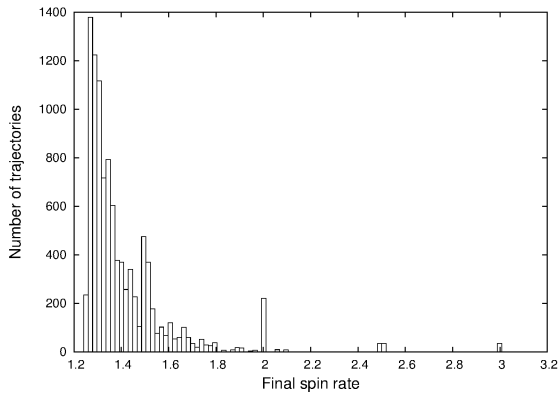}
\caption{\small{~Distribution of the final spin rates $\,\dot{\theta}/n\,$ of our 10,000 trajectories integrated using the Constant Time Lag (CTL) tidal model. The simulations cover the span of 3 Myr.\label{fig:histocl}}}
\end{figure}

 \par
 Figure \ref{fig:histocl} gives the histogram of the final spin at the end of the 10,000 runs. Small peaks correspond to low-order spin-orbit resonances. The final spin rate always exceeds $\,1.2\,n\,$. The number of trajectories trapped into the 2:1, 5:2 and 3:1 resonances can be estimated from the plot. For the 3:2 resonance, however, some of the trajectories ending at this spin rate, may not be actually trapped. The continuum distribution around 1.5 implies that some trajectories were simply caught by the end of integration while in fact traversing the resonance. So we fit a mathematical approximation to the continuum part of the histogram. Denoting with $\,f(\dot{\theta})\,$ the number of trajectories whose final spin
rate is $\,\dot{\theta}\,$, we fit the function
\begin{equation}
f(\dot{\theta})\approx\frac{a}{\dot{\theta}^b}~~,
\label{eq:thef}
\end{equation}
and get: $\,a\,=\,31,476.4\,\pm\,2,836\,$ and $\,b\,=\,12.9944\,\pm\,0.3311\,$. The numbers depend on the size of the bin, which in our case
was $\,0.0188\,n\,$. The equality $\,f(\dot{\theta})\,=\,N\,$ means that there are $\,N\,$ trajectories for which the final spin rate lies
between $\,\dot{\theta}\,-\,0.0094\,n\,$ and $\,\dot{\theta}\,+\,0.0094\,n\,$.

\par Over 10,000 trajectories, we obtain:
\begin{itemize}
\item 577 trajectories close to the 3:2 resonance ($\dot{\theta}/n$ between $1.4906$ and $1.5094$),
\item 221 trajectories close to the 2:1 resonance ($\dot{\theta}/n$ between $1.9906$ and $2.0094$),
\item 70 trajectories close to the 5:2 resonance ($\dot{\theta}/n$ between $2.4906$ and $2.5094$),
\item 35 trajectories close to the 3:1 resonance ($\dot{\theta}/n$ between $2.9906$ and $3.0094$),
\end{itemize}
 With $\,f(1.5n)\approx162.1007\,$, $\,f(2n)\approx3.8573\,$, $\,f(2.5n)\approx0.2123\,$, $\,f(3n)\approx0.0199\,$, we obtain
 415 ($\,=577\,-\,f(1.5n)\,$) captures into the 3:2 resonance, 217 into 2:1, 70 into 5:2, 35 into 3:1. We use the function $\,f\,$ rendered by
 the equation (\ref{eq:thef}) to evaluate the number of resonant trajectories in each of our 10 sets of 1,000 runs. The results are gathered in Table \ref{tab:r10sets}.
 \begin{table}[ht]
 \centering
 \caption{\small{~Evaluation of the number of captures in the 10 sets of 1,000 numerical runs. The numbers result from an estimate of the untrapped trajectories
 with the function $f$ given by the eqnuation (\ref{eq:thef}). Due to round-off errors, the \emph{Total} row does not exactly coincide with the sum of
 the 10 sets.\label{tab:r10sets}}}
 \begin{tabular}{l|cccc}
      & 3:2 & 2:1 & 5:2 & 3:1 \\
\hline
Set 0 &  37 &  20 &   8 & 5 \\
Set 1 &  47 &  15 &   5 & 1 \\
Set 2 &  51 &  22 &  10 & 4 \\
Set 3 &  34 &  26 &   8 & 8 \\
Set 4 &  56 &  21 &   5 & 1 \\
Set 5 &  35 &  28 &   6 & 4 \\
Set 6 &  40 &  23 &   6 & 1 \\
Set 7 &  29 &  17 &  11 & 3 \\
Set 8 &  39 &  26 &   5 & 7 \\
Set 9 &  49 &  23 &   6 & 1 \\
\hline
Total & 415 & 217 &  70 & 35 \\
\hline
\end{tabular}
\end{table}

 \par The statistics vary significantly among our 10 sets, which is expected due to the Poisson distribution of the number of positive outcomes. We
 got between 29 and 56 Type I captures into the 3:2 resonance, while \citet{cl2004} had 31. So our results are consistent. Our simulation
 methodology is valid and can be combined with a different tidal model.

 \section{Revisiting the Option 1: a rigid Mercury, initially prograde\label{sec:scen1}}

 Here we present the statistics of capture into spin-orbit resonances, based on the realistic tidal model, i.e., on the formula (\ref{7b}) with the tidal response (\ref{rheology} - \ref{imaginary}) plugged in. Recall that the latter formulae reflect the inputs from both rheology and self-gravitation. We integrated the law of motion (\ref{eq.eq}), with the torques (\ref{eq:tri4}) and (\ref{7c}) inserted, and with the eccentricity evolving along the synthetic trajectories described in Section \ref{sec:eccentricity} and tested in Section \ref{sec:correia}. In this section, Mercury is initially set prograde, with a rotation period of $20$ days.


   First, we set the values of the Maxwell and Andrade times of the mantle, $\,\tau_{_M}\,$ and $\,\tau_{_A}\,$. Our choice of values for the Maxwell time of Mercury is in agreement
with the range proposed by \citep{pmhms2014}.$\,$\footnote{~For the Maxwell time, \citep{pmhms2014} used a fixed reference value, to which the actual $\,\tau_{_M}\,$ was related
by some rescaling factor dependent on the temperature and on the grain size. With that factor taken into account, the actual Maxwell time of Mercury would assume values from several months to about 150 years.}
    As is explained in Appendix \ref{B} below, their values should be comparable. We set them equal, with the important caveat that $\,\tau_{_A}\,$ increases rapidly below some threshold frequency, to make the low-frequency reaction purely Maxwell.

 For the Earth, $\tau_{_M}\approx 500$ yr.
 For warmer mantles, though, the values of
 dozens of years or even years should be expected. Indeed, the viscosity $\eta$ depends on the temperature $T$ through the Arrhenius law $\eta \propto \exp(A^*/RT)$, where the gas constant is $\,R=8.3\,$ J/(mol K) and the activation energy for silicate rocks is $\,A^*\approx 6 \times 10^5\,$ J mol$^{-1}\,$. At the same time, the rigidity $\,\mu\,$ depends on $\,T\,$ slower, until $\,T\,$ approaches about three quarters of the melting temperature, the latter being increased by the confining pressure.
 $\,$\footnote{~Variation in the temperature changes the viscosity the Maxwell time: $\frac{\textstyle \Delta \tau_{_M}}{\textstyle \tau_{_M}}\,\approx\,\frac{\textstyle \Delta \eta}{\textstyle \eta}\,\approx\,-\,\frac{\textstyle \Delta T}{\textstyle T^2} ~\frac{\textstyle A^*}{\textstyle R}~$. Consider a silicate mantle of a mean $T=2300$ K and $\tau_{_M}=500$ yr. A decrease in $\eta$ and $\tau_{_M}$ by 9/10 ~(down to $\tau_{_M}=50$ yr) would correspond to an increase of the temperature by $\,\Delta T = 66$ K. Heating up the planet by another hundred degrees, we shall lower its $\,\tau_{_M}\,$ down to several years. At even higher temperatures, though, the decrease in $\,\tau_{_M}\,$ with the growth of $\,T\,$ will be less steep, as the rigidity $\,\mu\,$, too, will start decreasing.}

 \subsection{First simulations with $\tau_M=\tau_A = 500$ years\label{subsec:accel}}

 \par We present the results of 1,000 numerical runs. Only the secular tidal torque was used (the terms with  $q=j$ in the formula
 \ref{7b}). The oscillating terms were dropped to save time.

\par The capture statistics are as follows:

\begin{itemize}
	
\item resonance 4:1 : ~~~~5 captures ~($0.5\%$)$\,$,

\item resonance 7:2 : ~~~21 captures ~($2.1\%$)$\,$,

\item resonance 3:1 : ~~~40 captures ~($4\%$)$\,$,

\item resonance 5:2 : ~~103 captures ~($10.3\%$)$\,$,

\item resonance 2:1 : ~~279 captures ~($27.9\%$)$\,$,

\item resonance 3:2 : ~~444 captures ~($44.4\%$)$\,$,

\item resonance 1:1 : ~~104 captures ~($10.4\%$)$\,$.

\end{itemize}

\par
\begin{figure}[ht]
 \centering
 \includegraphics[width=0.58\textwidth]{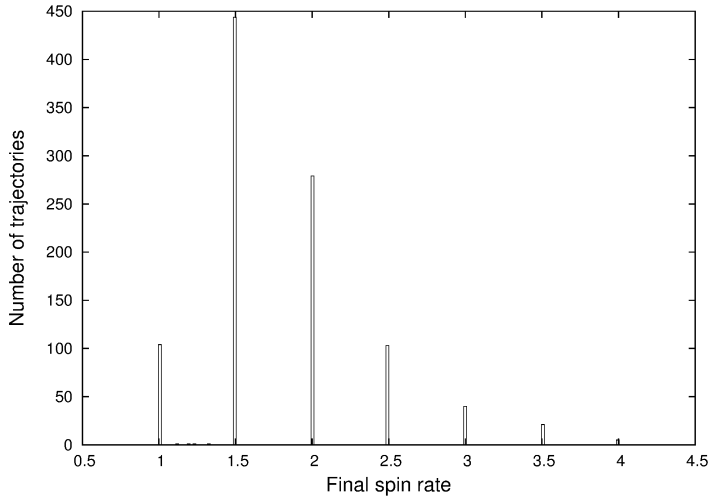}
 \caption{\small{~Distribution of the final spin rates of 1,000 trajectories integrated using the tidal model (\ref{7c}), (\ref{pppp}), with $\tau_M=\tau_A = 500$ years.
 The 3:2 resonance is the most probable destination, and is reached within the first 20 Ma.\label{fig:histot1}}}
\end{figure}

 The 4 remaining histories are between the 1:1 and 3:2 resonances and are bound to be trapped into the former,
as the secular tidal torque is negative, i.e., decelerating.
The histogram of these histories is presented in Figure \ref{fig:histot1}, to be compared with Figure \ref{fig:histocl}.

 We conclude that the 3:2 resonance is the most probable for the Type I captures. Other types of captures are much less likely, as the tidal despinning drives the spin rate to the synchronous rotation $\,\dot{\theta}=n\,$, making more than one crossing of the resonance impossible. As the evolution of the eccentricity will not change these results of resonant trapping, we can safely avoid the necessity to numerically integrate the spin evolution further in time.

 Trapping into the 3:2 resonance usually requires no more than 20 Myr, often much less.

 \subsection{Different values of the Maxwell time $\tau_M$\label{subsec:newmaxwell}}

 \par
\begin{figure}[ht]
\centering
\begin{tabular}{cc}
	\includegraphics[width=0.54\textwidth]{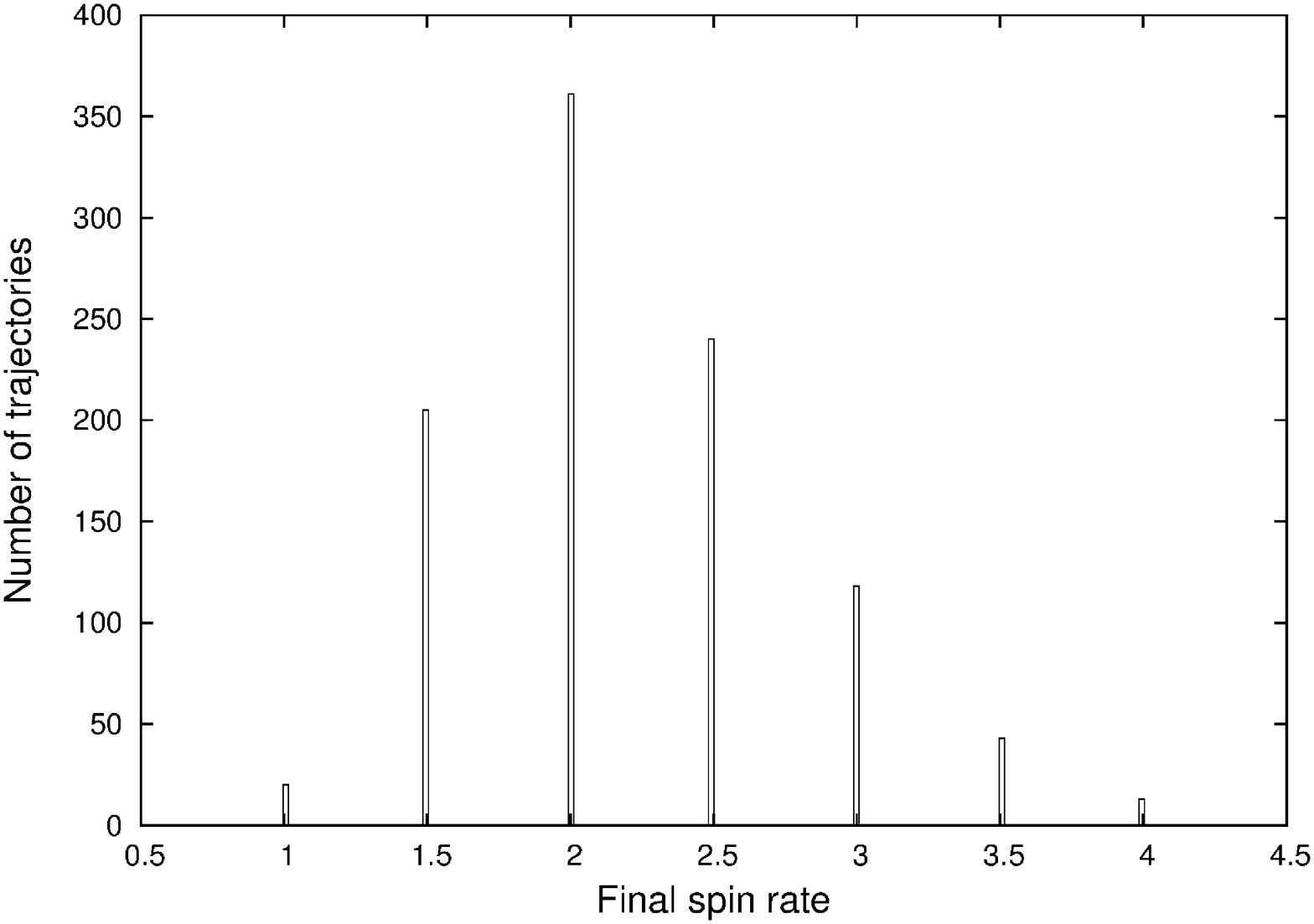} & \includegraphics[width=0.54\textwidth]{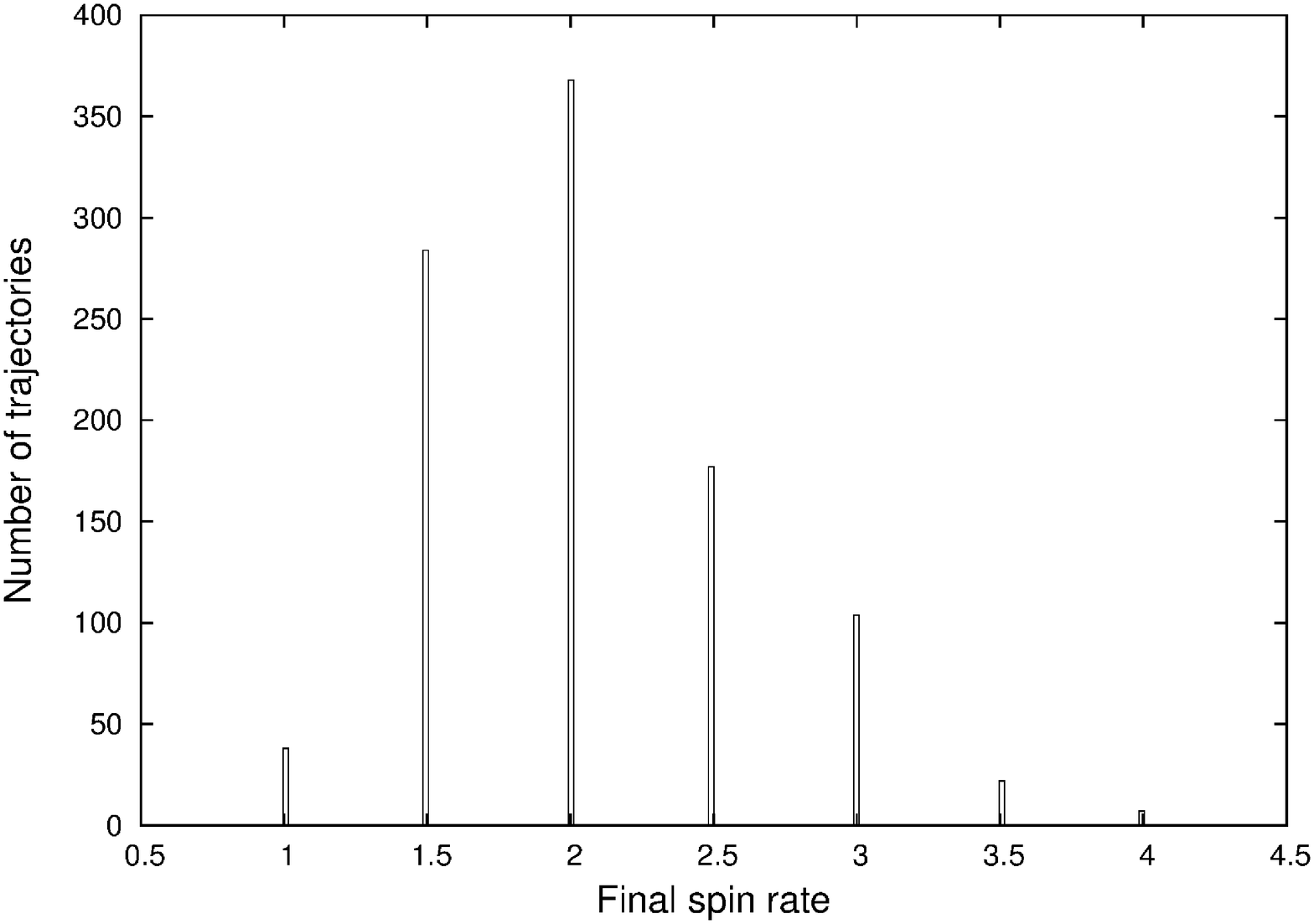} \\
	$\tau_M = 5$ yr & $\tau_M = 15$ yr
\end{tabular}
\caption{\small{~Capture statistics with the secular tidal torque, i.e., with $\,q=j\,$ in the equation (\ref{7b}), for a warmer Mercury, i.e. with a shorter Maxwell time
than before. The main difference is a higher probability for the 2:1 resonance.\label{fig:newmaxwell}}}
\end{figure}

 Mercury might have been warmer in its early life, especially if impacted. This would imply shorter values for the Maxwell time. Approximating
 the tidal torque with its secular part, we carried out the simulation for $\,\tau_{_M}=5\,$ yr$\,$ and $\,\tau_{_M}=15\,$ yr. In both cases, we set the Andrade and Maxwell times equal. The results are presented in Table \ref{tab:newmaxwell} and Figure \ref{fig:newmaxwell}.

\begin{table}[ht]
	\centering
	\caption{{~Entrapment statistics for shorter values of the Maxwell time.\label{tab:newmaxwell}}}
  \vspace{3mm}
	\begin{tabular}{|l|rr|}
\hline
		& $\,\tau_M = 5\,$ yr~ & $\,\tau_M = 15\,$ yr~ \\
		\hline
		Resonance 4:1 &  13 captures ~($\,1.3\,\%\,$)    &   7 captures ~($\,0.7\,\%\,$) \\
		Resonance 7:2 &  43 captures ~($\,4.3\,\%\,$)    &  22 captures ~($\,2.2\,\%\,$) \\
		Resonance 3:1 & 118 captures ~($\,11.8\,\%\,$)   & 104 captures ~($\,10.4\,\%\,$) \\
		Resonance 5:2 & 240 captures ~($\,24\,\%\,$)     & 177 captures ~($\,17.7\,\%\,$) \\
		Resonance 2:1 & 361 captures ~($\,36.1\,\%\,$)   & 368 captures ~($\,36.8\,\%\,$) \\
		Resonance 3:2 & 205 captures ~($\,20.5\,\%\,$)   & 284 captures ~($\,28.4\,\%\,$) \\
		Resonance 1:1 &  20 captures ~($\,2\,\%\,$)      &  38 captures ~($\,3.8\,\%\,$)\\		
\hline
	\end{tabular}
\end{table}

 \par Shorter Maxwell times favour higher-order resonances, especially the 2:1 one. As the actual final state of Mercury is the 3:2 resonance, a longer
 initial Maxwell time (hundreds of years) is a likelier option. At the same time, we should keep in mind that the final states are probabilistic. With small but not
 negligible probability, the planet can, in principle, reach the 3:2 spin-orbit resonance even with an initial Maxwell time smaller than 20 years.

\par Another consequence of a shorter initial Maxwell time is an even faster tidal despinning. While our former simulations with $\,\tau_{_M}=500\,$ years reached the
3:2 resonance in less than 20 Myr, with the smaller values we arrived at a final state in less than 10 Myr.

 \section{Revisiting the Option 2: Mercury with a liquid core, initially prograde\label{sec:scen2}\label{sec:core}}

  As indicated by the large amplitude of longitudinal forced librations \citep{mpjsh2007}, Mercury has a massive molten core decoupled from the mantle. In Appendix \ref{sec:protacalepsis}, we explain that gradual formation of a core could have taken up to billion of years. The short spin-down times obtained in this paper (10 - 20 Myr)  tell us that the core formed $\,${\it{after}}$\,$ the capture into the 3:2 resonance. There exist, however, scenarios wherein the core emerged at the earliest stages (e.g., after a collision). Hence the need to consider the case where the capture was influenced by the core. That the friction in the boundary layer can enhance capture was stated by Correia \& Laskar (2009) and, earlier, by Peale \& Boss (1977).

  In {\it{Ibid.}}, the authors argue that, in the laminar boundary approximation, entrapment of Mercury into the 2:1 resonance becomes certain even for a small degree of coupling. On examining the treatment in {\it{Ibid.}}, we found that the said study was based on a ``constant-$Q\,$" model, i.e., on an assumption that the secular tidal torque is frequency-independent.  Below we shall demonstrate that the calculation of capture probabilities in the presence of a core, developed by \citet{gp1966,gp1967,gp1968}, can be generalised to $\,${\it{any}}$\,$ realistic frequency-dependent tidal model. We shall then apply this generalisation to our model (\ref{overall}).

 \subsection{The theory of \citet{gp1966,gp1967,gp1968}}

 The first steps of our analysis follow the seminal study by \citet{gp1966,gp1967,gp1968} on the spin state of Venus. The theory describes the behaviour of the quantity
 \ba
 \gamma~\equiv~\theta~-~\left(\,1~+~\frac{\,q\,'}{2\,}\,\right)~{\cal{M}}~~,
 \label{libr}
 \ea
 which is the angle of libration about the resonance number $\,q\,'\,$. It is a resonance where $\,{\textstyle{\dot{\theta}}}/{\textstyle{n}}\,=\,1\,+\,q\,'/2\,~$, with $~{\cal M}\,$ being the mean motion, and with $\,\theta\,$ and $\,\dot{\theta}\,$ being the sidereal angle and spin rate of Mercury. Our definition (\ref{libr}) of the libration angle is consistent with that introduced by Goldreich \& Peale (1967, 1968). It differs by a factor of 1/2 from the quantity $\,\gamma\,=\,2\,\theta\,-\,(2+q\,')\,{\cal M}\,$ employed in Makarov (2012) and Makarov et al. (2012).

 For a laminar boundary between the core and the mantle, the equations of motion are
 \bs
 \begin{eqnarray}
 \ddot\gamma_{\rm m}&=&-~\omega_0^2~\sin2\gamma_{\rm m}~+~\frac{ \langle\,{\cal{T}}_z^{\rm{_{\,(TIDE)}}}\rangle }{C_{\rm m}}~-~\frac{k}{C_{\rm m}}~\left(\dot\gamma_{\rm m}~-~\dot\gamma_{\rm c}\right)~~,
 \label{first}\\
 \nonumber\\
\ddot\gamma_{\rm c}&=&\frac{k}{C_{\rm c}}~\left(\dot\gamma_{\rm m}~-~\dot\gamma_{\rm c}\right)~~,
 \label{second}
 \end{eqnarray}
 \label{moti.eq}
 \es
 where $\,\langle\,{\cal{T}}_z^{\rm{_{\,(TIDE)}}}\rangle\,$ is the secular part of the polar tidal torque acting on the mantle, $\,k\,$ is a friction constant, while the indices $\,m\,$ and $\,c\,$ designate the mantle and core, respectively (so $\,C_m\,$ and $\,C_c\,$ are the largest moments of inertia of the mantle and the core). The quantity
 \ba
 \sqrt{2}~\omega_0~=~\left[\,3~\frac{B_{\rm m}-A_{\rm m}}{C_{\rm m}}~\,\frac{M_{star}}{\,M_{star}\,+\,M_{planet}\,}~G_{20\mbox{\small{q}\,}'}(e)\,\right]^{1/2}n
 \label{freqq}
 \ea
 is the frequency which small-amplitude ($\,\sin2\gamma_{\rm m}\approx 2\gamma_{\rm m}\,$) librations would have, were there no tidal or core-mantle friction. The separatrix between rotation and libration is given by
 \be
 \dot\gamma_{\rm m}~=~
 \sqrt{2}~\omega_0~
 \cos{\gamma_{\rm m}}\,~,
 \label{separ.eq}
 \ee
 as explained in Appendix \ref{D} below.  \footnote{~Our equation (\ref{separ.eq}) coincides with a formula from \citet{gp1968}, but is slightly different from a similar equation in \citet{m2012} and \citet{makarovetal2012}. A difference in the numerical factor is due to the fact that the $\,\gamma\,$ defined in \citet{gp1968} and in the current paper differs by a factor of 1/2 from the $\,\gamma\,$ introduced in \citet{m2012} and \citet{makarovetal2012} -- see the paragraph after the formula (\ref{libr}) above. In our equations, we also keep the mass factor omitted in \citet{gp1968}.\\
 \vspace{1mm}
 $\quad$ We denote the frequency (\ref{freqq}) with $\,\sqrt{2}\,\omega_0\,$ to conform to the notation from \citet{pb1977,pb}.}

 The equations (\ref{moti.eq}) describe a non-harmonic, driven pendulum with friction. When coupling is strong (i.e., when $\,k\,$ is large), the core is participating in the librational oscillations. Opposite is a situation with weak friction, which is the case for Mercury; in that situation the core does not have time to react to the high-frequency changes in the rotation of the mantle. As the planet approaches the $\,q\,'\,$ resonance from the domain of positive $\,\dot\gamma_{\rm m}\,$,
 the mantle decelerates much faster than the core, increasing the negative difference in their spin rates. In the works by \citet{gp1966,gp1967,gp1968}, a cornerstone
 assumption was that the core is in a dynamic equilibrium with the time-average tidal torque (which was implicitly set constant). In our treatment, we shall depart from this
 assumption and shall generalise the treatment to an arbitrary form of the frequency-dependence of the torque.

 \subsection{Generalisation of the Goldreich-Peale theory}

 In the expression for the secular part (\ref{7c}) of the degree-2 torque, a $\,220q\,$ term, when expressed as a function of the Fourier mode $\,\omega_{220q}\,$, is antisymmetric. As Goldreich \& Peale (1966, 1967, 1968) were
 addressing the simplistic models with a constant time lag or a constant $\,Q\,$, an $\,lmpq\,$ term in their theory was either linear in the
 tidal mode or constant. In realistic models, each such term has the shape of a kink as in Figure \ref{Fig1}. When the formula (\ref{2}) is
 utilised to express each term as a function of $\,\dot{\theta}/n\,$, the overall sum becomes a superposition of kinks of different sizes, see Figure \ref{Fig2}.

 \subsubsection{The tidal torque near a resonance}

 In the vicinity of a resonance $\,q\,'\,$, the secular tidal torque can be written down as
 \ba
 \langle\,{\cal{T}}_z^{\rm{_{\,(TIDE)}}}\rangle_{\textstyle{_{\textstyle_{\textstyle{_{l=2}}}}}}~=~W~+~V~~.
 \label{sum}
 \ea
 Here $\,W\,$ is the $\,q\,=\,q\,'\,$ term of the sum (\ref{overall}):
 \ba
 W~\equiv~\frac{3}{2}~G\,M_{star}^{\,2}\,R^5\,a^{-6}\,G^{\,2}_{\textstyle{_{\textstyle{_{20\mbox{\it{q}}~'}}}}}(e)~k_2(
 \omega_{\textstyle{_{\textstyle{_{220\mbox{\it{q}}~'}}}}})~\sin|\,\epsilon_2(\omega_{\textstyle{_{\textstyle{_{220\mbox{\it{q}}~'}}}}})\,|
 \,~\mbox{Sgn}\,(\,\omega_{\textstyle{_{\textstyle{_{220\mbox{\it{q}}~'}}}}}\,)~~.
 \label{}
 \ea
 A kink centered around the point $\,{\textstyle{\dot{\theta}}}/{\textstyle{n}}\,=\,1\,+\,q\,'/2\,~$, this
 term is a function of the Fourier mode $\,\omega_{220q\,'}\;$ which is the negative twice of $\,\dot{\gamma}_{\rm{m}}~$:
 \ba
 \omega_{\textstyle{_{\textstyle{_{220\mbox{\it{q}}~'}}}}}~=~-~2~\dot{\gamma}_{\rm{m}}~~.
 \label{}
 \ea
 So we may treat $\,W\,$ as a function of $\,\dot{\gamma}_{\rm{m}}~$:
 \ba
 W~=~W(\dot\gamma_{\rm m})~=~-~K~G^{\,2}_{\textstyle{_{20\mbox{\it{q}}\,'}}}~k_2(\dot{\gamma}_{\rm{m}})~\sin |\,\epsilon_2(\dot{\gamma}_{\rm{m}})\,|~\,\mbox{Sgn}
 \,(\dot{\gamma}_{\rm{m}})\,~,
 \label{WW}
 \ea
 $K\,$ being a positive constant. The term $\,V\,$ in (\ref{sum}) is constituted by the rest of the sum (\ref{overall}):
 \bs
 \ba
 V~\equiv~K\,\sum_{q\neq q\,'}G^{\,2}_{20\mbox{\it{q}}}~k_2(\omega_{\textstyle{_{220\mbox{\it{q}}}}})~\sin |\,
 \epsilon_2(\omega_{\textstyle{_{220\mbox{\it{q}}}}})\,|~\,\mbox{Sgn}\,(\omega_{\textstyle{_{220\mbox{\it{q}}}}})~+~O(e^8\,\epsilon)~+~O(\inc^2\,\epsilon)~~.\quad
 \label{}
 \ea
  In the vicinity of the $\,q\,=\,q\,'\,$ resonance, the tidal modes are approximated with $\,\omega_{\textstyle{_{220\mbox{\it{q}}}}}\,=\,\omega_{\textstyle{_{220\mbox{\it{q}}\,'}}}\,+\,(q\,-\,q\,')\,n\,\approx\,(q\,-\,q\,')\,n\,$, whence $\,V\,$ is approximated as
 \ba
 V~\approx~K\,\sum_{q\neq q\,'}G^{\,2}_{20\mbox{\it{q}}}~k_2(\,(q\,-\,q\,')\,n\,)~\sin |\,
 \epsilon_2(\,(q\,-\,q\,')\,n\,)\,|~\,\mbox{Sgn}\,(q\,-\,q\,')\,~.
 \label{}
 \ea
 \label{}
 \es
 In the said vicinity, $\,V\,$ is a slowly-changing, nearly constant bias generated by the non-resonant terms. This can be understood from Figure \ref{Fig2} where a smaller kink on the right is residing on the smooth slope produced by other kinks, mainly by the larger kink on the left.

 The idea of the decomposition (\ref{sum}) belongs to \citet{gp1966,gp1967,gp1968} who employed it for calculation of entrapment probabilities, within the two
 simplest tidal models, that of a frequency-independent Q and that of a frequency-independent time lag. In the works by \citet{m2012} and \citet{makarovetal2012}, their theory was generalised to an arbitrary frequency-dependence of the tidal torque. That generalisation, however, was carried out for a homogeneous rotator. So our goal now is to embrace the case with a core.

 \subsubsection{Dynamics of the mantle near a resonance}

 The frequency-dependent part of the torque changes very quickly as the tidal frequency enters a close vicinity of a resonance. For weak coupling, the core does not react appreciably to sudden accelerations of the mantle. This means that the term $\,k/C_{\rm m}(\dot{\gamma}_{\rm m} - \dot{\gamma}_{\rm c})\,$ in the equation (\ref{first}) is small. Numerical runs of (\ref{moti.eq}) demonstrate that in this situation the resonance is traversed very quickly (within a few years).

 Near a resonance, the steady part of $\,\dot{\gamma}_{\rm m}\,$ vanishes: $\,\langle\dot{\gamma}_{\rm m}\rangle\,=\,0\,$ (while the oscillating part survives due to free librations). At the same time, the steady part of $\,\ddot{\gamma}_m\,$ does not vanish -- on the contrary, it becomes large. Time-averaging of the equations (\ref{first}) and (\ref{second}) yields:
 \bs
 \ba
 \langle\,\ddot{\gamma}_{\rm m}\rangle&=&\frac{V}{\,C_{\rm m}}~+~\frac{k}{\,C_{\rm m}}~\dot{\gamma}_{\rm c}~~,
 \label{XXX}\\
 \nonumber\\
 \ddot{\gamma}_{\rm c}&=&-~\frac{k}{\,C_{\rm c}}~\dot{\gamma}_{\rm c}~~.
 \label{YYY}
 \ea
 \label{ZZZ}
 \es
 We also assume that near resonances the mantle is in a dynamical equilibrium with the {\it constant} part $V$ of the torque.\footnote{~The core is not in equilibrium, because of the velocity offset and, thus, a nearly constant friction force.} So the friction force stays approximately constant during the short passage, and the quantities $\,\langle\,\ddot\gamma_{\rm m}\rangle\,$ and $\,\ddot\gamma_{\rm c}\,$ are approximately equal. Hence,
 \ba
 -~\frac{k}{C_{\rm c}}~\dot\gamma_{\rm c}~=~\frac{V}{C_{\rm m}}~+~\frac{k}{C_{\rm m}}~\dot\gamma_{\rm c}
 \qquad~\mbox{or, equivalently:}~\qquad
 \dot\gamma_{\rm c}~=~-~\frac{V}{k}~\frac{C_{\rm c}}{C_{\rm m}~+~C_{\rm c}}~~.~~
 \label{former}
 \ea
 Insertion of the equations (\ref{separ.eq}), (\ref{sum}) and (\ref{former}) into the equation of motion (\ref{first}) furnishes an equation equivalent to the equation (27) in \citet{pb}:
 \be
  \ddot\gamma_{\rm m}~=~-~\omega_0^2\;\sin2\gamma_{\rm m}~+~\frac{W~\dot\gamma_{\rm m}}{C_{\rm m}}~+~\frac{V}{C_{\rm m}~+~C_{\rm c}}~-~\frac{
  \sqrt{2}~k~\omega_0}{C_{\rm m}}\;\cos\gamma_{\rm m}~~.
 \label{accel.eq}
 \ee

 \subsubsection{The probability of entrapment}

 To estimate capture probabilities, it is sufficient to consider two librations bracketing the point of resonance $~\dot\gamma=0~$. As explained in \citet{gp1968}, the probability is
 \be
 P_{\rm capt}~=~\frac{\delta E}{\Delta E}~~,
 \ee
 $\delta E\,$ being the total change of kinetic energy at the end of the libration below the resonance, and $\,\Delta E\,$ being the offset from zero at the beginning of the last libration above the resonance. To obtain $\,\Delta E\,$, the torques acting on the mantle should be integrated over a cycle of libration. To obtain $\,\delta E\,$,
 the torques should be integrated over two librations symmetric around the resonance $\,\omega_{220q'}=0\,$. As a result, the odd part of the tidal torque at $\,q=q\,'\,$ doubles in the integration for $\,\delta E\,$, whereas the bias vanishes. As can be understood from the equation (\ref{accel.eq}), the core-mantle friction has two manifestations in the expression for the secular torque. The even (driving) part $\,V\,$ gets attenuated by a factor of $\,C_{\rm m}/(C_{\rm m}+C_{\rm c})\,$. The odd (restoring) part $\,W\,$ is boosted by an additional friction term proportional to $\,k\omega_0\,$. Both these effects lead to higher probabilities of entrapment. The final formula for $\,P_{\rm capt}\,$ assumes the form of
 \ba
 P_{\rm capt}~=~\frac{2}{~1~-~\frac{\textstyle \pi ~V~ C_{\rm m}}{~\textstyle (\,C_{\rm m}~+~C_{\rm c}\,)~\,(~
 \int_{-\pi/2}^{\pi/2}\,W(\dot\gamma_{\rm m})\,d\gamma_{\rm m}
 ~+~2~\sqrt{2}~k~\omega_0~)~}}~~~.
 \label{prob.eq}
 \ea
 The ``minus" sign in the denominator accounts for $V$ being negative for supersynchronous resonances. By the expression (\ref{prob.eq}), capture probabilities can be estimated semi-analytically for any resonance above 1:1 and for any combination of $\,C_{\rm m}/C_{\rm c}\,$, $~e\,$ and $\,k\,$.

 \subsection{Results for Mercury}

\begin{figure}[htbp]
 \centering
 \plottwo{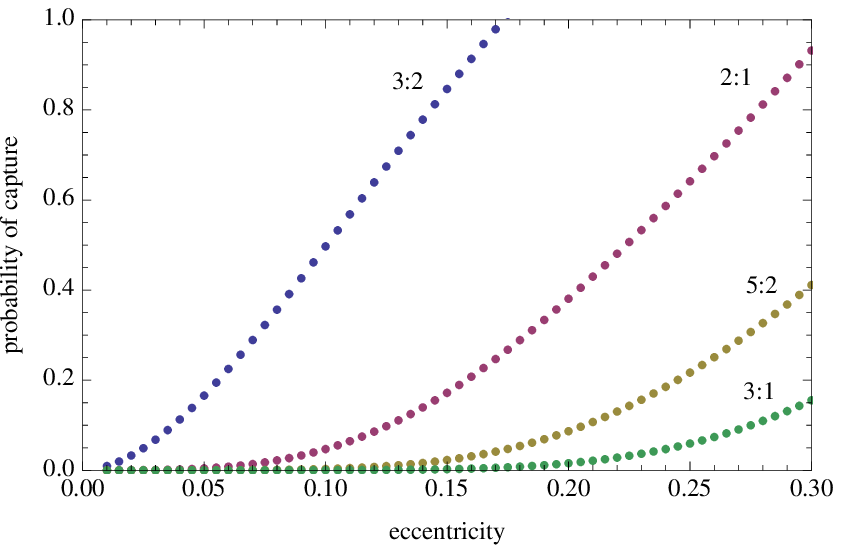}{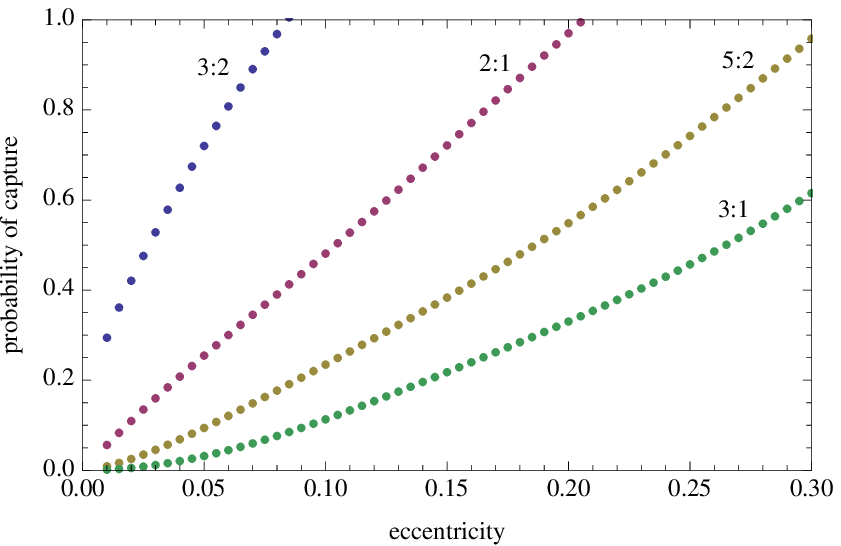}
 \caption{\small{Probabilities of capture of Mercury into the 3:2, 2:1 and 5:2 spin-orbit resonances, with the default parameters listed in Table \ref{table2}.
 Left: for a uniform, solid composition. Right: for the case with a massive molten core with $\,C_{\rm m}/({C_{\rm m}+C_{\rm c}})=0.5\,$ and $\,k_{\rm frict}=k/C_{\rm m}=10^{-5}\,$.
 \label{prob.fig}}}
 \end{figure}

 For a given quality function, the integral in the equation (\ref{prob.eq}) can be computed numerically using the formulae (\ref{WW}) and (\ref{separ.eq}). The expression (\ref{prob.eq}) gives a fast way of semi-analytical estimation of capture probabilities, and it helps us to avoid computer-heavy simulations. The results of the semi-analytical estimation of capture probability as a function of eccentricity are presented in Figure \ref{prob.fig} for the set of default parameters from Table \ref{table2}. The left panel of the graph shows the predicted probabilities for a solid uniform planet without a core. The right panel shows, for the same resonances, our calculations for a realistic model of the present-day Mercury with $\,C_{\rm m}/({C_{\rm m}+C_{\rm c}})\,=\,0.5\,$ and $\,k/C_{\rm m}\,=\,10^{-5}\,$. This value of the coefficient of friction corresponds to a kinematic viscosity of $\,0.008\,$ cm$^2$ s$^{-1}\,$, which is close to the value for water at $\,30\,$ $\degr C\,$. We conclude
that even a moderate degree of friction in the core-mantle boundary boosts the probabilities of capture into the 3:2 and higher resonances. Capture into the 2:1 resonance, for example, becomes certain at the current value of eccentricity. Furthermore, for $\,e=0.206\,$, employment of such a model for Mercury makes it more likely for the planet to be trapped in the 5:2 resonance (probability 0.58) than to traverse it.

 To verify the semi-analytical estimate, we performed 40 integrations of the equations (\ref{moti.eq}), starting with a uniformly distributed $\,\gamma_{\rm m}(0)\,$ and
 $\,\dot\gamma_{\rm m}(0)=2.009\,n\,$, for $\,e=0.1\,$ and $\,C_{\rm m}/({C_{\rm m}+C_{\rm c}})=0.5\,$. The equilibrium spin rate of the core for the 2:1 resonance is $\,\dot\gamma_{\rm c}(0)=2.02\,n\,$. The time span of each run was $\,3.3\times 10^4\,$ yr, and the maximal step was $\,2.3\times 10^{-3}\,$ yr. The 40 numerical histories rendered us 19 captures and 21 passages. The probability of capture estimated semi-analytically is 0.48 (see the right pane in Figure \ref{prob.fig}). Thus, our semi-analytical method and the spot-check simulation are in excellent agreement.

\section{Revisiting the Option 3: Mercury initially retrograde\label{sec:scen3}}

 The scenario developed by \citet{wcllr2012} comprised two steps:
 \begin{itemize} \item[Step 1.~]
  Mercury begins with a retrograde spin and gets accelerated by tides (possibly, assisted by the core-mantle friction), until trapped into the synchronous resonance.
 \item[Step 2.~]  An impact disrupts the synchronous spin, and Mercury shifts to the 3:2 resonance.
  \end{itemize}

 \par Our tidal model prohibits pseudosynchronous equilibria.
 So, if an impact is energetic enough to disrupt the synchronous spin, but not to reach the 3:2 resonance, then the tidal torque will drive Mercury into a spin-orbit resonance. If the torque is accelerating, the planet will be trapped into the 3:2 resonance. We however explain in the end of this section that the most probable outcome is a negative tidal torque which will return Mercury back into synchronism.

 To boost the rotation rate above $3n/2$, the impactor must have created a crater with a diameter of at least $600$ km. In such a case, we are back to either to the Option 1 (a prograde rigid planet, see Section \ref{sec:scen1}) or to the Option 2 (an initially prograde Mercury with a core, see Section \ref{sec:scen2}), dependent on whether the core-mantle friction is present. We have seen that in such a case entrapment into the 3:2 resonance is the most realistic outcome. So, we here just need to reinvestigate the acceleration of an initially retrograde Mercury with our tidal model, and with or without core-mantle friction.

 \subsection{Statistical hypothesis testing on crater distribution
 \label{stat.sec}}

 To support the hypothesis of a previous stay of Mercury in the 1:1 resonance, \citet{wcllr2012} looked for irregularities in the distribution of large craters. Those counts were based on limited data obtained by Mariner 10 and two flybys of MESSENGER. In the light of more accurate and complete data currently available, a re-examination of the statistical confidence of the said irregularities is warranted. Here we use the data published by Fassett et al. (2012, Table 2) for certain and verified craters larger than 300 km in diameter. We use the classical and robust Kolmogorov-Smirnov test of statistical hypotheses, to estimate the confidence level of any irregularities in the observed distribution. Therefore, ours is a completely independent verification of the result published by \citet{wcllr2012}.

 If a distribution of points on a unit sphere is absolutely random and uniform, the angular distance $\delta$ of each point from a fixed direction should follow a probability density function (PDF) that is proportional to $\sin(\delta)$ on the range $[0,\pi]$. Any significant non-randomness in the sample distribution will result in a deviation of  the sample frequency of $\delta$ from this PDF. The corresponding cumulative distribution function (CDF) is $1/2(1-\cos\delta)$. The one-sample Kolmogorov-Smirnov (KS) test determines the largest deviation of the observed sample CDF from the expected CDF and estimates the probability of such deviation to happen by chance.

 As explained in \citet{wcllr2012}, the most prominent consequence of an extended period of syncronised rotation is a bipolar excess of large craters on the caps facing and opposing the Sun, i.e., close to the directions $(0,0)$ and $(180,0)$ in Mercurian longitude and latitude. These directions are aligned with the axis of the smallest moment of inertia $A$. Mathematically, the distribution of angular distances $\delta$ from this chosen direction should be significantly flatter than the expected PDF. The tendency of a synchronised planet to be bombarded mostly from the sun's direction and the opposite direction is caused by the large fraction of asteroids on high-eccentricity orbits in the population of impactors that can collide with the planet with an impact velocity sufficiently high to leave behind a large crater.

 The null hypothesis to test is that the observed sample distribution of $\delta$ from the fixed direction $(0,0)$ follows the general CDF. This is done by a one-sample KS test using standard statistical routines. The output of the test, called p-value, quantifies the probability of the null hypothesis to be rejected while being true, given the observed sample. In astronomy it is also called false alarm probability (FAP). A small p-value indicates that it is unlikely that a given observed sample is drawn from a random, uniform distribution. In some fields, such as exoplanet detection, a high-confidence result is required to have a p-value as small as 0.01. Inversely, a large p-value indicates that there is no reason to reject the null hypothesis.
 \begin{figure}[htbp]
 \epsscale{0.9}
 \plottwo{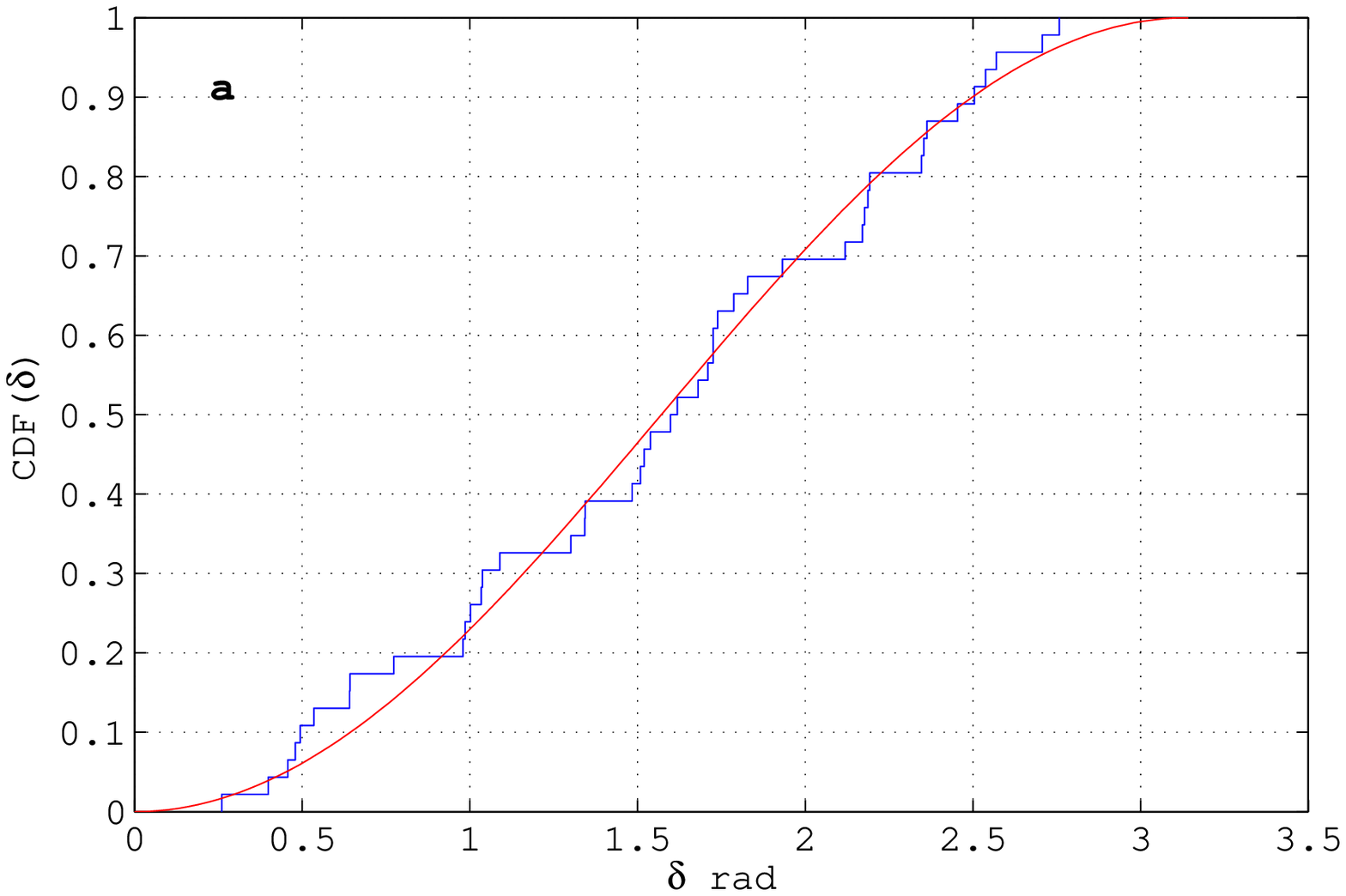}{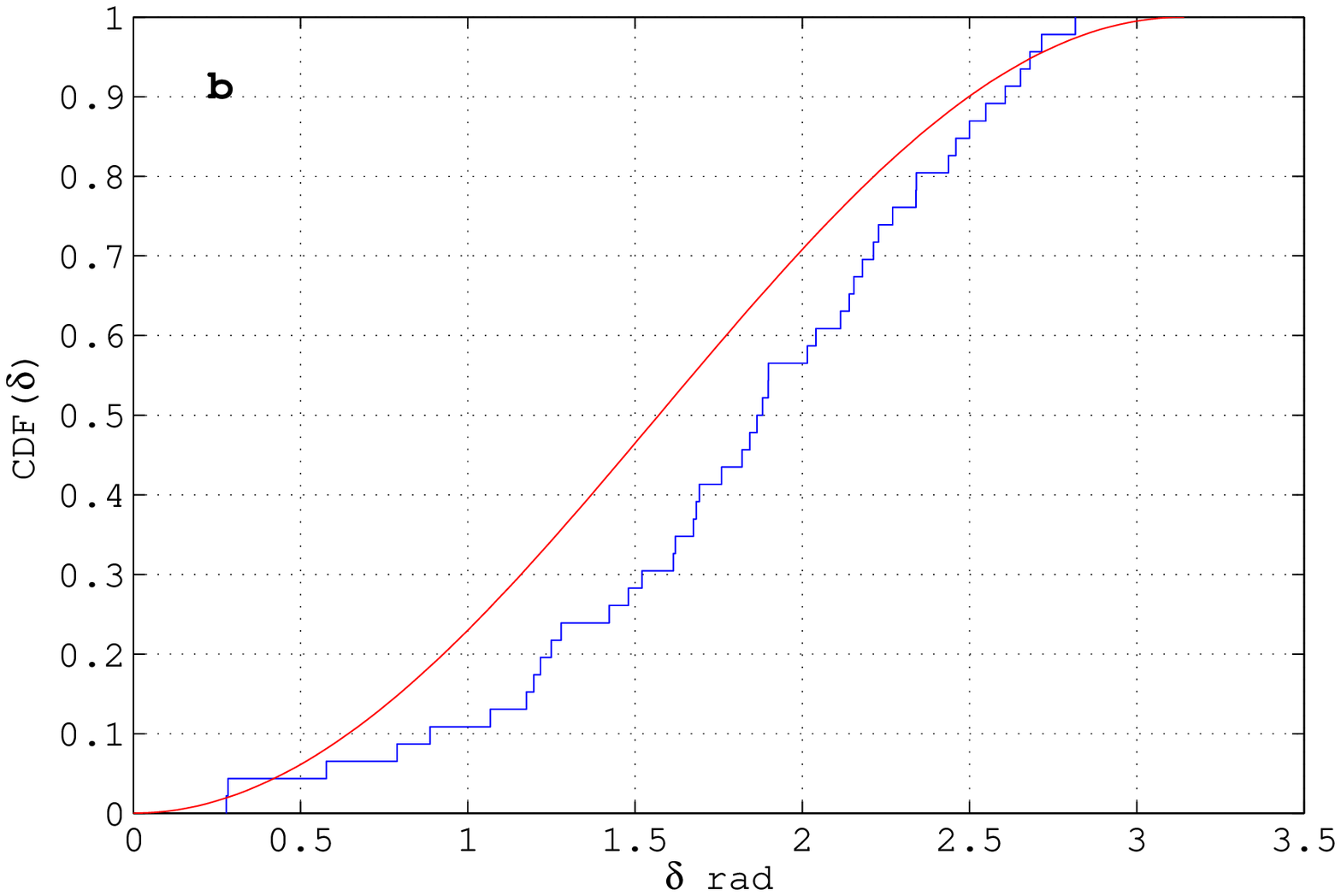}
 \caption{\small{~Sample CDF (broken line) and expected population CDF (smooth line) of angular distances
 of large and confidently detected impact craters on the surface of Mercury from (a) the presumably
 subsolar direction $(0\degr,0\degr)$ and (b) East direction $(90\degr,0\degr)$.
 \label{cdf.fig}}}
 \end{figure}

 The p-value obtained from the KS test for the presumably subsolar point $(0\degr,0\degr)$ is 0.946, so the null hypothesis is passable. There is no significant excess of crater density in the direction aligned with the minimal-inertia axis, as the hypothesis of synchronous spin predicts. Figure \ref{cdf.fig}a shows the observed sample CDF superimposed with the null-hypothesis CDF for a perfectly random-uniform distribution. Indeed, the two curves are quite close.

 The negative result for the main prediction of the synchronisation surmise does not imply that the observed distribution is random in any chosen direction. To test if there are other directions in which the observed distribution may be non-random, we performed a KS test for each node of a grid of 1146 points evenly distributed on the sphere. The smallest p-value, 0.006, was found for the coordinates $(111.0\degr, 8.2\degr)$. Such a small value is not necessarily a good reason to suspect that the sample is non-random. \footnote{~A self-consistent hypothesis test requires the direction to be chosen arbitrarily. We, however, are dealing here with the direction minimising the p-value.} What is alarming is the closeness of this direction to the due East point, $(90\degr,0\degr)$. Indeed, the secondary, and less important, prediction of the synchronisation surmise is a dipole-type asymmetry in the East-West distribution of craters. It is the ``runner in the rain" effect, when a person
  gets soaked faster on the front side than on the back. One can rationalise that the due West direction is an arbitrarily chosen point of reference, not {\it a priori} related to the sample, so the KS test is valid. The KS p-value in this case is 0.02. The corresponding pair of a sample and expected CDF are displayed in Figure \ref{cdf.fig}b. The conclusion is clear: there are considerably more of large and confidently detected craters on the West side than on the East.

 Our statistical hypothesis tests indicate that the primary prediction of the synchronous rotation theory is not supported by observation. The secondary predicted effect, i.e., the East-West asymmetry in the counts, is present at a non-negligible confidence level of 98\%.

 \subsection{A rigid and homogeneous Mercury}

 As in Section \ref{sec:scen1}, we ran 1,000 trajectories with the Maxwell times of respectively $500$, $15$ and $5$ years; but now starting from an initially retrograde spin corresponding to a period of $20$ days. In this case, the planet has to traverse spin-orbit resonances in the reverse order (i.e., -2:1, -3:2, -1:1, -1:2, 1:2, and 1:1), as can be understood from the expression (\ref{eq:tri4}) for the triaxiality-caused torque. The magnitude of the restoring component of the triaxial torque is very important for the capture process. We recall that the eccentricity-dependent coefficients of the restoring components, $G_{20q'}(e)$, scale in the leading order as $e^4/24$, $e^3/48$, $-e/2$, $1$, and $7e/2$, for $q'=(-4,-3,-1,0,1)$, respectively. Obviously, the first two resonances, -1:1 and -1:2, are intrinsically very weak for moderate eccentricities, and can in practice be ignored. It is quite unlikely that planets can ever be captured into retrograde spin-orbit resonances. The 3:2 resonance becomes stronger than the 1:1 resonance at an eccentricity as small as $\simeq 0.3$. The prograde (but subsynchronous) 1:2 resonance can, however, become significant, depending on the eccentricity. At first glance, this resonance should correspond to an unstable equilibrium, because the corresponding coefficient $G_{20(-1)}(e)$ is negative and the torque becomes diverging rather than restoring. However, our simulations have shown that, under certain circumstances, capture into this resonance is probable or even unavoidable.
As mentioned by Goldreich \& Peale (1966), this resonance is stable if the planet is captured sidewise, i.e., if its shortest axis of inertia is at $\pi/2$ to the sun direction at periastron.

\begin{figure}[ht]
 	 \centering
 	 \includegraphics[width=.55\textwidth]{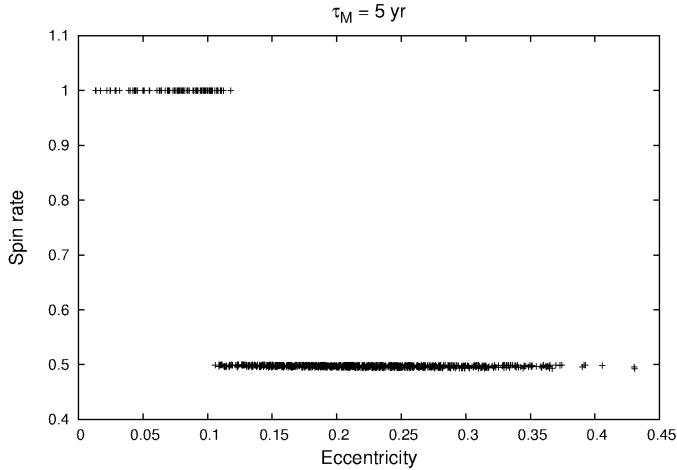}
 	 \caption{Final states for an initially retrograde Mercury and $\tau_M=\tau_A=5$ years. For a longer Maxwell time, we get only synchronous resonances.
 	 We can see that the trapping into the 1:2 is certain for an eccentricity $e$ above $\approx0.12$. \label{fig:res12back}}
 \end{figure}

 For $\tau_M=\tau_A=500$ years and $15$ years, all the simulated trajectories were eventually trapped into the synchronous resonance, successfully crossing the 1:2 resonance.
 However, the results for $\tau_M=\tau_A=5$ years were quite different, with 878 trajectories ending up in the 1:2 resonance (see Fig.\ref{fig:res12back}). Thus, if the early Mercury was warm and had a low viscosity, it could hardly avoid entrapment into this sub-synchronous resonance. The simulations also showed that, once captured, the planet would stay in this resonance indefinitely long, despite significant variations in eccentricity. Only one history spontaneously left the 1:2 resonance. (It ended up in the 1:1 one). The conclusion is that a warm Mercury with an initial retrograde spin would have an extremely slim chance of getting to synchronism. {\it{En route}} thereto, it would almost certainly have
 been trapped into the 1:2 subsynchronous state. Only external agents, such as giant impacts, could drive it through to the current 3:2 resonance. But the likely capture into the 1:2 resonance becomes another problem for the retrograde spin hypothesis, because at least two fortuitous giant impacts are now needed to account for the current state.

 \subsection{A segregated Mercury with a fluid core}

 We have not conducted large-scale numerical simulations of a differentiated-Mercury spin, because the analysis presented in Section \ref{sec:core} is fully applicable in this case too. Undoubtedly, a segregated planet with a massive, decoupled liquid core would be even more easily captured into the subsynchronous 1:2 resonance than its
 homogeneous analogue.

 Examining the hypothesis that Mercury was entrapped into synchronism for a continuous length of time, and that bombardment by heavy asteroids drove it out to the present 3:2 spin state, one has to enquire whether at that time the planet already had a liquid core. Scenarios involving a synchronous state were examined numerically by \citet{cl2012}, in the framework of the CTL tidal model. Here we offer analysis more relevant for our Maxwell-Andrade tidal model.

 Bringing a synchronous planet into a 3:2 spin-orbit resonance comprises two steps. First, the planet should be kicked out of synchronism by a mega-impact. For small and moderate eccentricities ($e<0.3$), the 1:1 resonance is the most stable equilibrium. It is the deepest well of potential energy; to leave it, a formidable external energy deposition is needed. The direction of impact should be aligned with the sidereal rotation velocity (lest Mercury finds itself in a sub-synchronous state whence it is destined to fall into either 1:2 or 1:1 resonance again). Second, Mercury's spin should evolve toward the 3:2 resonance after the impact. The state of pseudosynchronism at spin rates between the 1:1 and 3:2 resonance provides a convenient and stable transit point in the CTL, but is unstable for planets of terrestrial composition.

 There remain two possible scenarios consistent with the impact hypothesis. First, the impact could be energetic enough to drive Mercury's spin rate to values above the 3:2 resonance, $\dot\theta > 3/2\,n$. We can estimate the characteristics of the required impactor. If $u$ is the component of the relative impact velocity in the equatorial plane of Mercury (supposed to be close to the orbital plane, due to a small pre-impact obliquity), the upper-bound change of the spin rate from an approximation ignoring possible ejecta and thermal dissipation is
 \be
 \Delta\omega \leq \frac{m_i u\,R}{C}~~,
 \ee
 $m_i$ being the impactor's mass, while $C$ and $R$ being the principal moment of inertia and the radius of Mercury. In our approximation, the upper bound is achieved when the collision is grazing and on the equator. On the other hand, the change of spin rate should be greater than the gap $1/2\,n$ separating the two resonances. The upper-bound condition becomes
 \be
 m_i u \geq \frac{C\,n}{2R}~~.
 \label{mu.eq}
 \ee
 For the present-day parameters, the momentum of the impactor should have exceeded $1.13\times10^{23}$ kg m s$^{-1}$, if Mercury were a uniformly rigid body. According to \citet{lefe}, the average impact velocity is $42.5$ km s$^{-1}$. The average projection component $u$ is statistically smaller, $2/\pi \cdot 42.5=27.1$ km s$^{-1}$. The minimum mass of the impactor capable of driving a uniform Mercury analogue all the way above the 3:2 resonance is then $4.2\times 10^{18}$ kg. Using an assumed median density of asteroids of $2500$ kg m$^{-3}$, and an estimated relation between the basin and impactor diameters, we conclude that only the largest two of the certain and probable craters from \citet{fhbzsnskscppop2012} -- namely, Caloris and b30 -- could impart that much momentum under favourable circumstances. The presence of a large liquid core decoupled from the mantle lowers the threshold momentum and the mass of impactor by roughly one-half, as only the moment of the mantle's inertia,
 $C_m$, enters the equation (\ref{mu.eq}). Consequently, any impactor leaving a basin larger than $\simeq 640$ km could potentially cause the required spin-up. Still, there are only a dozen or so such craters on Mercury.

 In the alternative scenario, the planet only has to be driven out of the 1:1 resonance, the subsequent spin-up and capture into the 3:2 resonance being assured by tides. The required impactor's momentum is less stringent then. The lower-bound trajectory of departure from the 1:1 resonance is obviously the same as the zero-energy trajectory of capture, i.e., the separatrix from the equation (\ref{separ.eq}). The corresponding lower-bound change of spin rate is
 \be
 \Delta\omega_{\rm min} \geq n\left[3\frac{B-A}{C}G_{200}(e)\right]^\frac{1}{2}~~,
 \ee
 ignoring the small forced libration which is slightly dependent on the orbital eccentricity in this case. We obtain the lower-bound $\Delta\omega_{\rm min}$ of $2.0\cdot 10^{-8}$, $1.9\cdot 10^{-8}$, and $1.8\cdot 10^{-8}$ Hz, for $e=0.1$, 0.2 and 0.3, respectively. The lower-bound impactor's momentum is then
 \be
 m_i u \geq \frac{C\,n}{R}\left[3\frac{B-A}{C}G_{200}(e)\right]^\frac{1}{2}, \label{mu2.eq}
 \ee
 and the corresponding minimum mass ranges between $1.8\times 10^{17}$ and $2.0\times 10^{17}$ kg. The corresponding minimal basin diameter is approximately $510$ km. Thus, half of the impactors considered in Section \ref{stat.sec}  could do the work of driving the planet out of synchronism under favorable conditions. Again, the required impactor's size becomes even smaller for a planet with a decoupled core. The lower bound of basin diameter goes down to $\simeq 430$ km, which adds several more
 potentially capable impact event, for which we have observational evidence.

 The main obstacle for this scenario to be accepted as plausible, with or without a liquid core, is the necessity to further accelerate the spin from the post-impact value to the 3:2 commensurabilty. Within our realistic tidal model, we do not have the help from a stable intermediate equilibrium of pseudosynchronous rotation. Thus, the required spin-up should be provided by a steady positive tidal torque. This imposes a requirement on the value of eccentricity during this time span. For each non-resonant value of $\dot\theta$, there is a unique equilibrium value of eccentricity that separates the domains of negative and positive secular tidal torque. The equilibrium eccentricity
 for a uniform Mercury analogue was computed in \citet[][Figure 3]{m2012}. The area of interest here is the span of $\dot\theta$ between $1\,n$ and $1.5\,n$ where the eccentricity is descending from approximately 0.3 to approximately 0.2. The phase space trajectory of the planet depends on the initial post-impact spin rate. If the impact was just enough to drive the planet out of the 1:1 resonance, the initial spin rate would end up right above $1\,n$, in which case the eccentricity would have to be consistently above 0.3, to boost the spin further. Otherwise, Mercury quickly returns to the 1:1 resonance. A more powerful impact could set the spin rate at a higher value, which eases up the limit on eccentricity needed for a further prograde spin-up. Statistical analysis can be performed by taking into account the observed
distribution of impactors and the Monte-Carlo simulations of eccentricity evolution, but such an endeavour
is outside the scope of this paper.

\section{Discussion}

 A large variety of scenarios of Mercury's spin-orbit evolution has been considered in the literature, striving to explain its present 3:2 resonance. Differentiating the proposed theories by the degree of complexity and physical validity, the options include: (1) simplified linear tidal models versus
 physics-based rheologies resulting in realistic frequency-dependencies of $\,k_2/Q~$; ~(2) constant-in-time versus
 chaotically variable eccentricity; ~(3) initially prograde versus retrograde spin; ~(4) the possibility of powerful impacts followed by instantaneous resets of the spin rate; ~(5) the presence or absence of a massive molten core at the epoch of capture into the 3:2 resonance; ~(6) the tidal properties, such as the average viscosity, of Mercury's mantle at the time of capture. Thus a multitude of possible paths emerges, and the results happen to strongly depend upon the choice made on each of the said six items. Not all of these paths, however, render the current 3:2 resonance as the likeliest end-state.

 Reliable conclusions on the history of Mercury's spin must be derived from the actual behaviour of minerals -- a viscoelastic Maxwell response at low frequencies and the inelastic Andrade creep at higher frequencies. This complicates both numerical simulations and direct analysis of capture probabilities.  Another complication is the large, unpredictable variation of eccentricity over aeons. The onus is then on massive Monte-Carlo simulations, though we also found a limited analytical estimate to be valuable as a sanity check. In other regards, a wide scope of initial conditions and assumptions has been considered in this paper, avoiding {\it{a priori}} preferences. We explored both the prograde and retrograde initial spin; considered a planet with or without a liquid core at the early stage of its evolution; and studied both a partially molten warm mantle and a colder mantle similar to that of the present-day Earth.

 The main outcome of our study is that the 3:2 spin-orbit resonance is the likeliest end-state, provided the following set of conditions is satisfied by Mercury: (a) it did not have a massive decoupled core prior to being captured into this resonance; (b) its mantle was relatively cold and had a sufficiently high viscosity, with a Maxwell time being of the order of hundreds years; (c) the initial spin was prograde, as would be expected for most of the Solar System planets. An initially prograde undifferentiated cold Mercury gets trapped into the resonance within 10 - 20 Myr, while formation of a core requires up to a billion of years.

 Our analysis thus invalidates some previously discussed scenarios. We do not find it possible for a telluric planet locked in a spin-orbit resonance to leave it spontaneously through a change in eccentricity. In our simulations, resonant states were almost always long-term and stable. Without an external agent, such as a heavy impact, Mercury is trapped forever. This fact, combined with the short spin-down times found in this paper, implies that Mercury was captured into the 3:2 state billions of years ago, and stayed there ever since.

 Alternatives to this scenario meet substantial difficulties. If the initial spin were fast and prograde, but the core had already been segregated or/and if the mantle was warm by the moment of capture, then Mercury would very likely end up in a higher resonance, e.g., 2:1. If the initial spin were retrograde, then the hitherto unheeded sub-synchronous resonance 1:2 becomes hard to avoid, especially for a differentiated or/and warm planet.
 So two hits would be needed -- to kick the planet from the likely trap of 1:2, and then to
 kick it out from the certain trap 1:1. Hence, both for an initially prograde and an initially retrograde planet, early differentiation becomes a formidable obstacle on its way to the 3:2 spin state.

 The published studies of Mercury's differentiation are centered around the fact that, being anomalously dense, the planet should be poor in silicates. Early volatilisation is one of the attractive possibilities. The composition can be explained if the removal process in Mercury's zone of the solar nebula were only slightly more effective for silicates than for iron. The fractionation required to produce an iron-rich planet is achieved through a combination of gravitational and drag forces acting in the nebula during the formation of proto-Mercury \citep{w1978}. As an alternative, a slow-volatilisation model was suggested, in which the solar wind, after the formation of Mercury, could partially volatilise the mantle \citep{c1985,fc1987}. Both these models imply gradual differentiation of the planet. As explained in Appendix \ref{sec:protacalepsis}, the process of core formation required hundreds of millions to a billion of years. Simulations without core-mantle friction furnishes entrapment within less than 20 Myr, an age indeed much younger than that of differentiation.

 A very different model of rapid differentiation relies on an impact
 energetic enough to melt down the originally chondritic proto-Mercury
 (Benz et al. 1988, 2007). The event would have resulted in a very warm tumbling Mercury wherein the core-mantle boundary would have formed very quickly. An impact, which could have taken place when Mercury's age was about 20 Myr or larger, is consistent with our core-less computation: the capture happens first, the impact and the subsequent differentiation comes later. An earlier impact poses a nontrivial problem for our main scenario. The hypothesis of an early impact becomes questionable also in the light of the recent findings by MESSENGER. As was pointed out by \citet{pehmbgeghlmnsrsss2011}, the abundance of potassium {\textit{is inconsistent with those physical models of Mercury formation, which require extreme heating of the planet}}. This observation prompted \citet{wurm} to explore the action of photophoretic forces upon irradiated solids. The investigation has demonstrated that silicates are preferentially pushed into an optically thick disk. The subsequent planetesimal production at the outward-moving edge renders metal-rich platenesimals close to the star and metal-depleted ones farther out in the nebula. After this very early separation of metals and silicates, the further development of Mercury goes gradually, with a slow emergence of a core-mantle boundary.

 A massive impact seems to provide a magic resolution to many of these difficulties. \citet{wcllr2012} suggest that the distribution of craters indicates that Mercury  spent some time in the synchronous state whence it was driven by an impact, eventually reaching the current 3:2 resonance.  \footnote{~\citet{wcllr2012} hypothesise that Mercury
 could have evolved into synchronism from an originally retrograde spin state. The probability of this step is reduced considerably by the likely entrapment into the 1:2 resonance, escape wherefrom would require an extra collision. At the same time, \citet{bwn2013} argue that the primordial orbit of Mercury could have been more circular than previously thought. This finiding, if correct, may give the second breath to the scenario with synchronism. If the eccentricity keeps a very low value for several millions of years, a swift despinning straight into synchronism is not  impossible.} We reconsidered the claim of inhomogeneous distribution of impact craters on the surface of Mercury, using more up-to-date and complete data and a robust method of statistical hypothesis testing. No evidence is found for a preferential location of craters on the opposite caps in the sub-solar and anti-solar directions, which is the principal prediction from the synchronous-state hypothesis. Still, the secondary prediction of an East-West asymmetry in the distribution of craters is confirmed.

  A consistent tidal theory based on rheology of minerals is, by itself, insufficient to disprove the scenario by Wieczorek et al. (2012). The scenario, however, comprises a number of propositions which are not impossible when considered separately, but which together combine into an unlikely course of events. Within this scenario, an initially retrograde spin slowed down and became prograde. Then
  Mercury would likely get into the first sub-synchronous resonance 1:2, an event almost inevitable for inviscid planets with a significant liquid core. Mercury would hardly be able to leave this state -- hence the need for a heavy impact in the right direction. The probability that a single impact is directed in the prograde sense is 1/2. A counter strike could reset the spin back to smaller values, whereafter Mercury would fall in the 1:2 state again. Once externally accelerated toward the 1:1 resonance, Mercury would reach it in a matter of millions of years and get inevitably captured there, irrespective of its eccentricity. Only most energetic impacts (again, correctly directed and almost tangential to the surface in the equatorial zone) can drive Mercury all the way to spin rates above the 3:2 resonance. There are few craters that could be left by such massive impactors, so this option appears to be unlikely. It is much easier to jolt Mercury out of the 1:1 state to spin rates just outside the resonance, especially if it already possesses a large liquid core. Still, if the eccentricity at the moment is smaller than 0.3, Mercury would quickly descend back into synchronism. Again, any counter-directed strike would only enable a short excursion outside the main resonance. For Mercury to get to the 3:2 state, being driven by the tides only, the eccentricity should be higher than 0.3-0.2 for the whole duration
  of spin-up.

 The scenario of initially prograde Mercury with a core meets its own set of difficulties in our analysis.  Given the short despinning time, the emergence of a core prior to capture is unlikely. (Tidal damping in Mercury is too weak to account for a high degree of segregation \citet{me2014}.) Besides, a decoupled core makes capture into the 2:1 resonance almost certain, and, as our simulations show, permanent. Hence we have to resort to the agency of an external impact to explain the current 3:2 resonance. This is not entirely impossible, simply because it is much easier to drive Mercury down from the 2:1 resonance than up from the 1:1 resonance, with small and moderate eccentricities. From the energy-balance viewpoint, it is more economic to descend down a bumpy slope than to be hoisted out of a hole at the bottom. Altogether, though, this scenario requires too many unlikely assumptions.

 To draw to a close, we advocate a capture of an initially prograde, yet undifferentiated Mercury. It requires less than
 20 Myr, the 3:2 resonance being the most probable end-state.

 \section*{Acknowledgements}

 Beno\^it Noyelles is an F.R.S.-$\,$FNRS postdoctoral research fellow, and he thanks this institution for having funded two of his visits to the USNO.
 This research has made use of the resources of the Interuniversity Scientific Computing Facility located at the University of Namur and supported by the F.R.S.-$\,$FNRS
 under Convention No 2.4617.07. Julien Frouard acknowledges support by NSF grant AST-1109776.

 On various occasions, the topics addressed in the paper were discussed by the authors with Sylvio Ferraz Mello, Val\'ery Lainey, Francis Nimmo, Stanton Peale,
 James G. Williams, Jack Wisdom and other colleagues, to all of whom the authors are profoundly grateful. The authors are thankful to Sebastiano Padovan for a
consultation on the rescaling method in \citep{pmhms2014}.

 The authors are indebted to Val\'ery Lainey and Christophe Le Poncin-Lafitte who made this collaboration possible by putting the authors in touch with one another.


 \vspace{3mm}

 \appendix
 \begin{center}
 \underline{\textbf{\Large{Appendix}}}~~~
 \end{center}

  \section{How long it takes Mercury to develop a liquid core\label{sec:protacalepsis}}


 Our modeling shows that, for a realistic rheology, capture happens extremely swiftly: almost certainly within 20 Myr,  more probably within 10 Myr. By that age, the liquid core of Mercury had not yet formed. Indeed, over a time span $\,\Delta t\,$, the temperature increases by
 \ba
 \Delta T ~=~ \frac{H}{C_p}~\Delta t~~,
 \label{heat}
 \ea
 where we neglected the diffusion, and denoted with $\,H\,$ and $\,C_p\,$ the radiogenic heating and specific heat capacity, correspondingly. Using the present-day values $\,H\,=\,3.5\,\times\,10^{-12}\,$ W/kg and $\,C_p\,=\,1200\,$ J/(K kg) borrowed from Turcotte \& Schubert (2002), we observe that by the age of $\,\Delta t\,=\,20\,$ Myr
 the radiogenic temperature increase of Mercury was $\,\Delta T\,=\,1.84\,$ K only. In the early Solar System, the value of $\,H\,$ could have been up to four times higher,
 but this would be still too low to melt the mantle of Mercury over such a short time. A likely age of melting is hundreds of millions to a billion of years. \footnote{~The authors are grateful to Francis Nimmo for the help he provided with this estimate.}

 In Section \ref{sec:core} we explore trapping of a Mercury with a liquid core. Our semi-analytical model (confirmed by limited numerical simulation) shows that even a weak laminar friction between the mantle and the core would result in a capture into the 2:1 or a higher resonance.

 These two circumstances indicate that Mercury was homogeneous when trapped.

 Even for a homogeneous Mercury, a decrease of the Maxwell time from hundreds to dozens of years (which corresponds, very roughly, to a 100 K temperature increase) makes the 2:1 resonance a more probable destination than 3:2 (see the subsection \ref{subsec:newmaxwell}). A further decrease of the Maxwell time makes the 3:2 final resonance even
 less probable. On all accounts, the warmer an early Mercury the less chances it has to end up where it is now.

  \section{\large{Expression for the tidal torque~~~~~~~~~~~}}\label{A}

 \subsection{Fourier tidal modes
 }

 The additional tidal potential $\,U\,$ of a tidally disturbed homogeneous sphere can be expanded into a Fourier series over the modes
 \ba
 \omega_{\textstyle{_{lmpq}}}\;\equiv\;({\it l}-2p)\;\dot{\omega}\,+\,({\it l}-2p+q)\;{\bf{\dot{\cal{M}}}}\,+\,m\;(\dot{\Omega}\,-\,\dot{\theta})
 \,\approx\,(l-2p+q)\,n\,-\,m\,\dot{\theta}\,~.~~~
 \label{L9}
 \label{A1}
 \ea
 where the notations $\,{\theta}\,$ and $\,\dot{\theta}\,$ stand for the sidereal angle and spin rate of the perturbed spherical body; while $\,\omega\,$, $\,\Omega\,$, $\,n\,$,
 and $\,{\cal{M}}\,$ are the pericentre, the node, the mean motion, and the mean anomaly of the perturber as seen from the perturbed body. While the tidal
 modes $\,\omega_{\textstyle{_{lmpq}}}\,$ can assume either sign, the physical forcing frequencies
 \ba
 \chi_{lmpq}\,=~|\,\omega_{lmpq}\,|~\approx~|~(l-2p+q)~n\,-\,m~\dot{\theta}~|\,~,
 \label{fr}
 \label{A2}
 \ea
 at which the tidal strain and stress are evolving, are positive-definite.

 Derivation of a partial sum of the series was pioneered by Darwin (1879),$\,$\footnote{~In the modern notation, Darwin's treatment is discussed by Ferraz-Mello et al. (2008).}
 $\,$the full series worked out in Kaula (1964). Extended presentation of this formalism, with explanation of some details omitted by Kaula (1964), is offered in Efroimsky \& Makarov (2013).

 \subsection{Expansion of the tidal torque}

 A Fourier expansion of the tidally generated torque was written down, without proof, by Goldreich \& Peale (1966) who took into account only the secular part of the
 torque. Later, a schematic derivation of their formula was offered by \citet{d2007}. A comprehensive derivation of the polar component of the torque can be found
 in \citet{e2012} where it is demonstrated that the torque contains both a secular and a rapidly oscillating part. The overall expression reads as
 \ba
 \nonumber
 {\cal T}_z^{^{\,(TIDE)}}= ~-\,\sum_{l=2}^{\infty}\,\left(\frac{R}{a}\right)^{\textstyle{^{2l+1}}}\frac{{G}\,M_{star}^{\textstyle{^{\,2}}}}{a}
 ~\sum_{m=1}^{l}\frac{(l - m)!}{(l + m)!} \;2\;m\;\sum_{p=0}^{l}F_{lmp}(i)
 \sum_{q=-\infty}^{\infty}G_{lpq}(e)~~~~~~~~~~~~~~~\\
                                   \nonumber\\
 \sum_{h=0}^{l}F_{lmh}(\inc)\sum_{j=-\infty}^{\infty} G_{lhj}(e)
 ~k_{\textstyle{_{l}}}~\sin\left[\,v_{lmpq}\,-\;v_{lmhj}\,-\;\epsilon_{l}\,\right]
   ~~_{\textstyle{_{\textstyle ,}}}~~~~~~~~~~~~~~~~~~~~~~
 \label{A7}
 \ea
 where $\,\epsilon_{\textstyle{_{l}}}\,$ are the phase lags, $\,G\,$ stands for the Newton gravitational constant, while $\,F_{lmp}(\inc)\,$ and $\,G_{lhj}(e)\,$ denote
 the inclination functions and eccentricity polynomials. The auxiliary quantities $\,v_{lmpq}\,$ are rendered by
 \ba
 v_{lmpq}\;\equiv\;(l-2p)~\omega\,+\,(l-2p+q)~{\cal M}\,+\,m\,\Omega~~~,
 \label{A8}
 \ea
 where $\,a,\,e,\,\inc,\,\omega,\,\Omega,\,{\cal M}\,$ stand for the orbital elements of the perturbing secondary, as seen in the reference frame associated with the
 equator of the perturbed primary. While the frame may precess with the primary's equator, it does {\it{not}} corotate with it.

 In our situation, the role of the primary is played by the planet, its host star being regarded as the secondary. This means that the apparent inclination $\,\inc\,$ of the
 secondary (the star) on the equator of the primary (the planet) will in fact be the planet's obliquity. This also means that the mass of the secondary, $\,M_{star}\,$, is to
 be the mass of the star.

 The expression (\ref{A7}) furnishes the tidal torque's projection onto the spin axis of the perturbed body. Taking into account this component solely, we neglect the
 planet's precession. This approach is acceptable when the inclination (in our case, obliquity) $\,\inc\,$ is small.

 As explained in detail in Efroimsky (2012a,b), expression (\ref{A7}) is obtained from the Darwin-Kaula expansion of the tidal potential through differentiation thereof with
 respect to the negative sidereal angle $\,\theta\,$, which enters the $\,lmpq\,$ term of the expansion as $\,m\theta\,$. Differentiation inserts the multiplier $\,m\,$ into
 each term, as wee see in (\ref{A8}). This way, while in the expansion for the potential the summation over $\,m\,$ goes from $\,m=0\,$, in the expansion for torque we sum
 beginning with $\,m=1\,$.

 \subsection{The tidal lags}

 The Fourier tidal modes $\,\omega_{\textstyle{_{lmpq}}}\,$ emerging in the Darwin-Kaula theory are given by (\ref{A1}), while the positive-definite actual frequencies
 $\,\chi_{\textstyle{_{lmpq}}}\,$ excited by the tides are rendered by (\ref{A2}). The corresponding positive-definite time delays $\,\Delta t_l(\chi_{\textstyle{_{lmpq}}})\,$
 depend on this physical frequency, the functional forms of this dependence being different for different materials.

 As demonstrated, e.g., in \citet{em2013}, within the Darwin-Kaula theory the phase lags emerge as the products of the Fourier modes $\,\omega_{\textstyle{_{lmpq}}}\,$ by the
 time lags which themselves are frequency-dependent: \footnote{~Just as the Love numbers $\,k_{\textstyle{_l}}(\omega_{\textstyle{_{lmpq}}})\,$, so the lags are written down
 as $\,\epsilon_{\textstyle{_l}}(\omega_{lmpq})\,$ and $\,\Delta t_{\textstyle{_l}}(\chi_{\textstyle{_{lmpq}}})~$ -- and not as $\,\epsilon_{\textstyle{_{lmpq}}}\,$
 and $\,\Delta t_{\textstyle{_{lmpq}}}\,$. This uniformity of nomenclature is needed to emphasise that for a homogeneous
 spherical body the functional forms of the lags and Love numbers (as functions of the Fourier mode) are determined only by the degree $\,l\,$. The dependence
 of $\, k_{\textstyle{_l}}(\omega_{\textstyle{_{lmpq}}})\,$, $\,\epsilon_{\textstyle{_l}}(\omega_{\textstyle{_{lmpq}}})\,$ and $\,\Delta t_{\textstyle{_l}}(\chi_{\textstyle{_{lmpq}}})\,$ upon $\,m,\,p,\,q\,$ comes into being only due to the $\,m,p,q-$dependence of the arguments $\,\omega_{\textstyle{_{lmpq}}}\,$ and $\,\chi_{\textstyle{_{lmpq}}}\,$.\\
 $~~~~~$ This however applies to spherical homogeneous objects only. For nonspherical bodies, the functional form of the lags and Love numbers acquires dependence
 also on integers other than the degree. For a symmetrical oblate body, the lags and Love numbers will depend also upon the order $\,m\,$. Accordingly, our
 notations $\,k_{\textstyle{_l}}(\omega_{\textstyle{_{lmpq}}})\,$, $\,
 \epsilon_{\textstyle{_l}}(\omega_{\textstyle{_{lmpq}}})\,$, $\,\Delta t_{\textstyle{_l}}(\chi_{\textstyle{_{lmpq}}})\,$ will have to be changed to
 $\,k_{\textstyle{_{lm}}}(\omega_{\textstyle{_{lmpq}}})\,$, $\,\epsilon_{\textstyle{_{lm}}}(\omega_{\textstyle{_{lmpq}}})\,$,
 $\,\Delta t_{\textstyle{_{lm}}}(\chi_{\textstyle{_{lmpq}}})\,$. For a slightly non-spherical body, the Love numbers and lags differ from the Love numbers and lags of the spherical
 reference body by terms of the order of the flattening, so a small non-sphericity can be neglected.
 }
 \begin{subequations}
 \ba
 \epsilon_l(\omega_{lmpq})\,=\,\omega_{lmpq}~\,\Delta t_l(\chi_{lmpq})~=~\pm~\chi_{lmpq}\,\Delta t_{lmpq}\,~,
 \label{lags}
 \label{A9a}
 \ea
 Owing to causality, time lags $\,\Delta t_{\textstyle{_l}}(\chi_{\textstyle{_{lmpq}}})\,$ are positive-definite, so the lags (\ref{lags}) become
 \ba
 \epsilon_l(\omega_{lmpq})\,=\,\chi_{lmpq}~\,\Delta t_l(\chi_{lmpq})~\,\mbox{Sgn}\,(\,\omega_{lmpq}\,)\,~,
 \label{lags_2}
 \label{A9b}
 \ea
 \label{A9}
 \end{subequations}
 where $\,\chi_{lmpq}\,$ are the forcing frequencies (\ref{A2}).

 In neglect of the apsidal and nodal precession
   and also of $\,\dot{\cal{M}}_0\,$, the expressions for the frequencies and lags can be approximated with: \footnote{~The perturber's mean anomaly is
   $\,{\cal{M}}={\cal{M}}_0+\int^t_{t_0}n(t)\,dt\,$, $\,$where
   $\,n(t)\equiv\,\sqrt{G(M_{star}+M_{\textstyle{_{planet}}})\,a^{-3}(t)\,}$. So
   $\,\stackrel{\bf\centerdot}{\cal{M}\,}=\,\stackrel{\bf\centerdot}{{\cal{M}}_0}+\,n\,$, with
   $\,\stackrel{\bf\centerdot}{{\cal{M}}_0}\,$ showing up when the orbital motion is
   perturbed.
   In our study, however, we use a secular value of $\,n\,$ (see Table \ref{table2}) which already takes into account the
   exterior perturbations. \label{f1}}
 \ba
 \omega_{lmpq}\,=\,({\it l}-2p+q)\,n\,-\,m\,\dot{\theta}~~~,~~
 \label{A10}
 \ea
 \ba
 \chi_{lmpq}\,\equiv\,|\,\omega_{lmpq}\,|\,=\,|\,({\it l}-2p+q)\,n\,-\,m\,\dot{\theta}\;|~~~,~~
 \label{A11}
 \ea
 and
 \ba
 \epsilon_{l}(\omega_{\textstyle{_{lmpq}}})~=~\chi_{lmpq}~\,\Delta t_l(\chi_{lmpq})~\,\mbox{Sgn}\,\left[\,(l-2p+q)\,n\,-\,m\,\dot{\theta}\,\right]~~~,~~
 \label{A12}
 \ea
 $\dot{\theta}\,$ being the sidereal spin rate of the perturbed body.

 Importantly, the Darwin-Kaula formalism and the ensuing expression for the torque impose no {\emph{a priori}} constraint on the form of frequency-dependence of the lags.
 \footnote{~The MacDonald (1964) theory imposes implicitly a particular dissipation law (Williams \& Efroimsky 2012, Efroimsky \& Makarov 2013). MacDonald (1964) did not
 notice this, and erroneously set the dissipation quality factor to be frequency-independent, which was in contradiction with the pre-imposed
 dissipation law. Later, his theory was corrected by Singer (1968) and Mignard (1979, 1980). Although compact and elegant, the so-corrected version of MacDonald's theory
 still has no practical application, because the dissipation law, whereon it relies, differs considerably from the rheology of realistic minerals (Efroimsky \& Lainey 2007).}

 \subsection{Simplification 1: ~algebraic development}\label{App4}

 Numerical approximation of the series (\ref{A7}) is not easy because of the summation over six indices. Fortunately, much of the work can be performed analytically, the
 numerical part of the work thus being reduced to the necessary minimum. In this subsection, we shall simplify the expansion analytically, and in the subsequent subsections shall
 truncate the series to a partial sum.

 From (\ref{A8}), we see that
 \ba
 \nonumber
 v_{{\it l}mpq}\,-\,v_{{\it l}mhj}&=&\left[\,({l}-2p)\omega\,+\,({\it l}-2p+q){\cal M}\,+\,m\,\Omega\,\right]\,
 -\,\left[\,({l}-2h)\omega\,+\,({\it l}-2h+j){\cal M}\,+\,m\,\Omega\,\right]\\
 \nonumber\\
 &=& 2~(h\,-\,p)~(\omega\,+\,{\cal M})~+~(q\,-\,j)~{\cal M}~~~,
 \label{32}
 \ea
 insertion whereof into (\ref{A7}) makes the latter look as
 \ba
 \nonumber
 {\cal T}_z^{^{\,(TIDE)}}=~-\,\sum_{l=2}^{\infty}\,\left(\frac{R}{a}\right)^{\textstyle{^{2l+1}}}\frac{G\,M_{star}^{\textstyle{^{\,2}}}}{a}
 ~\sum_{m=1}^{\it l}\frac{({\it l} - m)!}{({\it l} + m)!} \;2\;m\;\sum_{p=0}^{\it l}F_{{\it l}mp}(\inc)
 \sum_{q=-\infty}^{\infty}G_{{\it l}pq}(e)~~~~~~~~~~~~~~~~~~~~\\
                                   \nonumber\\
 \sum_{h=0}^{\it l}F_{{\it l}mh}(\inc)\sum_{j=-\infty}^{\infty} G_{{\it l}hj}(e)
 ~k_{\textstyle{_{l}}}~\sin\left[\,
 2~(h\,-\,p)~(\omega\,+\,{\cal M})~+~(q\,-\,j)~{\cal M}
 \,-\;\epsilon_{\textstyle{_l}}\,\right]
   ~~_{\textstyle{_{\textstyle .}}}~~~~~~~
 \label{A14}
 \ea
 If we expand each sine as
 \ba
 \sin\left[\,2\,(h-p)\,(\omega+{\cal M})\,+\,\left(q-j\right)\,{\cal{M}}\,\right]\,\cos\epsilon_{\textstyle{_{l}}}\,-\,
   \cos\left[\,2\,(h-p)~(\omega+{\cal M})\,+\,\left(q-j\right)\,{\cal{M}}\,\right]\,\sin\epsilon_{\textstyle{_{l}}}~~~~~~
 \label{A15}
 \ea
 and recall that $\cos\epsilon\approx 1+O(\epsilon^2)$, all the terms with $\,\sin\left[\,2\,(h-p)\,(\omega+{\cal M})\,+\,\left(q-j\right)\,{\cal{M}}\,\right]\,$ will cancel, in the order of $\,O(\epsilon^2)\,$, after the summation over $\,h,\,j,\,p,\,q\,$ is performed.$\,$\footnote{~Mind that the product
 $~F_{lmp}(i)\,G_{lpq}(e)\,F_{lmh}(\inc)\,G_{lhj}(e)~\sin\left[\,2\,(h-p)\,(\omega+{\cal M})\,+\,\left(q-j\right)\,{\cal{M}}\,\right]~$ changes its sign under the simultaneous swap $~p\rightleftharpoons h~, ~j\rightleftharpoons q~$. So the sum of all such terms vanishes.
 } This will leave us with
 \ba
 \nonumber
 {\cal T}_z^{^{\,(TIDE)}}=~\sum_{l=2}^{\infty}\,\left(\frac{R}{a}\right)^{\textstyle{^{2l+1}}}\frac{G\,M_{star}^{\textstyle{^{\,2}}}}{a}
 ~\sum_{m=1}^{\it l}\frac{({\it l} - m)!}{({\it l} + m)!} \;2\;m\;\sum_{p=0}^{\it l}F_{{\it l}mp}(\inc)
 \sum_{q=-\infty}^{\infty}G_{{\it l}pq}(e)~~~~~~~~~~~~~~~~~~~~~~~~\\
                                   \nonumber\\
 \sum_{h=0}^{\it l}F_{lmh}(\inc)\sum_{j=-\infty}^{\infty} G_{{\it l}hj}(e)
 ~k_{\textstyle{_{l}}}~\sin\epsilon_{l}~\cos\left[\,
 2~(h\,-\,p)~(\omega\,+\,{\cal M})~+~(q\,-\,j)~{\cal M}
 \,\right]
   ~~_{\textstyle{_{\textstyle .}}}~~~~\qquad
 \label{A16}
 \ea
 In the literature, it is customary to consider only the terms with $\,h=p\,,~q=j\,$, i.e., with $~\cos\left[\,2~(h\,-\,p)~(\omega\,+\,{\cal M})~+~(q\,-\,j)~{\cal M}\,\right]\,=\,1~$.
 The reason for this is that for $\,h\neq p\,$ or/and $\,q\neq j\,$ the cosine averages out to nil,$\,$\footnote{~The averaging period should be the period of apparent
 rotation of the perturber about the perturbed body. In our case, it is the synodal day of the planet.} wherefore the oscillating terms vanish {\it{in
 average}}. This leaves one with the secular part of the torque.

 As explained in Efroimsky (2012), there exists a reason why the oscillating part of the torque may influence the process of entrapment into resonances. The
 frequencies $\,n(2h-2p+q-j)\,$, which show up in the oscillating part of the torque, are integers of $\,n\,$, and thus become commensurate with the spin
 rate $\,\stackrel{\bf{\centerdot}}{\theta\,}\,$ near resonances. It indeed was discovered, via numerical modeling, that the presence of the oscillating part
 changes histories. Whether a history stemming from a particular set of initial conditions leads to entrapment or to transition -- that is influenced by the
 presence or absence of the oscillating part of the torque. The statistics, though, remain virtually unchanged (Makarov et al. 2012).

 \subsection{Simplification 2: ~approximation of the series with its quadrupole part}

 We shall approximate the polar torque with the $\,l=2\,$ input:
 \ba
 {\cal T}_z^{^{\,(TIDE)}}=~^{\textstyle{^{{\textstyle{_{(l=2)\,}}}}}}{\cal T}_z^{^{\,(TIDE)}}\,+~O\left(\,(R/a)^{7}\,\epsilon\,\right)
                         =~^{\textstyle{^{{\textstyle{_{(lm=21)\,}}}}}}{\cal T}_z^{^{\,(TIDE)}}\,+~
                            ^{\textstyle{^{{\textstyle{_{(lm=22)\,}}}}}}{\cal T}_z^{^{\,(TIDE)}}
                         ~+~O\left(\,(R/a)^{7}\,\epsilon\,\right)\,~,\qquad
 \label{A17}
 \ea
 where the $\,l=2\,$ input is of the order of $\,(R/a)^{5}\,\epsilon\,$, while the $\,l=3,\,4,\,.\,.\,.\,$ inputs constitute $\,O\left(\,(R/a)^{7}\,\epsilon\,\right)\,$. As
 ever, $\,R\,$ and $\,a\,$ denote the planet's radius and semimajor axis, while $\,\epsilon\,$ stands for the phase lag. The index $\,m\,$ runs through the values
 \footnote{~Recall that, while the $\,m=0\,$ terms enter the potential, they add nothing to the torque, because $\,m\,$ enters (\ref{A14}) as a multiplier.} from $\,1\,$ through $\,l=2\,$.

 In a comprehensive form, (\ref{A17}) reads as
 \ba
 \nonumber
 {\cal T}_z^{^{\,(TIDE)}}= ~\left(\frac{R}{a}\right)^{\textstyle{^{5}}}\frac{G\,M_{star}^{\textstyle{^{\,2}}}}{a}
 ~\sum_{m=1}^{2}\frac{(2 - m)!}{(2 + m)!} \;2\;m\;\sum_{p=0}^{2}F_{2mp}(i)
 \sum_{q=-\infty}^{\infty}G_{2pq}(e)~\sum_{h=0}^{2}F_{2mh}(i)~~~~~~~~~~~~~~~~~~~~~~\\
                                   \nonumber\\
 \sum_{j=-\infty}^{\infty} G_{2hj}(e)
 ~k_{\textstyle{_{2}}}~\sin\epsilon_{\textstyle{_2}}~\cos\left[\,
 2~(h\,-\,p)~(\omega\,+\,{\cal M})~+~(q\,-\,j)~{\cal M}
 \,\right]\,+~O\left(\,\epsilon\,(R/a)^{7}\,\right)
   ~~_{\textstyle{_{\textstyle ,}}}~~~~~~~~
 \label{A18}
 \ea

 \subsection{Simplification 3: ~diagonalisation over the indices $\,p\,$ and $\,h\,$}\label{AA3}

 Above we mentioned that our formulae for the tidal torque furnish the torque's component along the spin axis of the perturbed body. This component is
 sufficient when the inclination of the perturber, $\,\inc\,$, is small. When the role of a tidally disturbed primary is played by a planet and the role of the
 perturber is played by the host star, the apparent inclination coincides with the planet's obliquity which, for Mercury, is $~i\,=\,2.04\,\pm\,0.08\,$ arcmin$\,$
 \citep{mpshgjygpc2012}. This enables us to make a further simplification of the formula for the torque, truncating away the terms of the order of $\,O(\inc^2)\,$.

 To this end, we should notice that, among the terms with $\,l=2\,$, we should keep only those containing the inclination functions
 with subscripts $\,({\emph{l}}mp)\,=\,(220)\,,\,(210)\,,\,(211)\;$:
 \ba
 F_{220}(\inc)\,=\,3\,+\,O(\inc^2)~~~,~~~~~F_{210}(\inc)\,=\,\frac{3}{2}\;\sin\inc\,+\,
 O(\inc^2)\;\;\;,~~~~~F_{211}(\inc)\,=\;-\;\frac{3}{2}\;\sin\inc\,+\,O(\inc^2)~~~,~~~~~
 \label{A19}
 \ea
 all the other $\,F_{2mp}(\inc)\,$ being of order $\,O(\inc^2)\,$ or higher. As the inclination functions enter the expansion in
 combinations $~F_{{\it l}mp}(\inc)\,F_{{\it l}mh}(\inc)~$,  we see that it is sufficient, in the $\,O(\inc^2)\,$ approximation, to
 take into account only the terms with $\,F_{220}^2(\inc)\,$, ignoring those containing $\,F_{210}^2(\inc)\,$, $\,F_{211}^2 (\inc)\,$,
 or $\,F_{210}(\inc)\,F_{211}(\inc)\,$. Thus the $\,l=2\,$ input into the torque gets simplified to:
 \ba
 \nonumber
 {\cal T}_z^{^{\,(TIDE)}}~=~
 ^{\textstyle{^{{\textstyle{_{(l=2)\,}}}}}}{\cal T}_z^{^{\,(TIDE)}}+O\left(\,(R/a)^{7}\epsilon\right)~=~^{\textstyle{^{{\textstyle{_{(lmph=2200)\,}}}}}}{\cal T}_z^{^{\,(TIDE)}}+O(i^2\epsilon)+O\left(\,(R/a)^{7}\epsilon\right)~=
 ~\qquad~~
 ~\\ \label{A20}\\  \nonumber
 \frac{\textstyle 3}{\textstyle2}\,\frac{G\,M_{star}^{\textstyle{^{\,2}}}}{a}\left(\frac{R}{a}\right)^5
 \sum_{q=-\infty}^{\infty}G_{\textstyle{_{\textstyle{_{20\mbox{\it{q}}}}}}}(e)
 \sum_{j=-\infty}^{\infty}G_{\textstyle{_{\textstyle{_{20\mbox{\it{j}}}}}}}(e)\,
 k_{\textstyle{_2}}\;\sin\epsilon_{\textstyle{_{
 2}}}
 \,\cos\left[\,\left(q- j\right)\,{\cal M}\,\right]
 +O(i^2\epsilon)+O\left(\,(R/a)^{7}\epsilon\right)
   ~~_{\textstyle{_{\textstyle ,}}}
 \ea
 where the absolute error $\,O(i^2\,\epsilon)\,$ comes into being after the neglect of terms with $\,p,h\,\geq\,1\,$.

 \subsection{Simplification 4: truncation over the indices $\,q\,$ and $\,j~$}\label{App7}

 In our numerical simulations, the eccentricity of Mercury never exceeded $\,e=0.45\,$. Moreover, it exceeded $\,e=0.4\,$ in several isolated cases and for
 only short time spans. In neglect of these unique episodes, we may assume that the eccentricity stays lower than $\,0.4\,$.

 How many terms of the series (\ref{A20}) should be taken into consideration? The issue is nontrivial, because the higher the eccentricity the slower the
 convergence of the series. On the one hand, $\,|G_{lpq}(e)|\,=\,O(e^{|q|})\,$, which means that $\,|G_{lpq}(e)|\,=\,A_{lpq}\,e^{|q|}\,+
 \,O(e^{|q|+1})\,$, with $\,A_{lpq}\,$ being real numbers. On the other hand, the numbers $\,A_{lpq}\,$ are growing rapidly with the increase of $\,|q|\,$, so
 the product $\,A_{lpq}\,e^{|q|}\,$ does not go immediately into decrease with an increasing $\,|q|\,$. For example, $\,G_{206}(0.4)=0.25\,$, a significantly
 large value. Thence, for higher values of the eccentricity $\,e\,$, one needs to go to larger $\,|q|\,$, to make sure that the factor $\,e^{|q|}\,$ is small
 enough to keep the value of $\,A_{lpq}\,e^{|q|}\,$ lower than the needed precision level.

 Apart from the eccentricity polynomials, a $\,2mpq\,$ term of the series (\ref{A20}) includes the multiplier $\,k_{\textstyle{_2}}\;\sin\epsilon_{\textstyle{_2}}\,$
 which is a function of the Fourier tidal mode $\,\omega_{\textstyle{_{2mpq}}}\,$. Therefore, to decide on how many terms should be kept in the series (\ref{A20}),
 we should know the shape and magnitude of the so-called {\it{quality function}}
  %
  %
 \bs
 \ba
 k_{\textstyle{_2}}\;\sin\epsilon_{\textstyle{_2}}\,=\,k_{\textstyle{_2}}(\omega_{\textstyle{_{2mpq}}})\;\sin\epsilon_{\textstyle{_2}}(\omega_{\textstyle{_{2mpq}}})
 ~=~k_{\textstyle{_2}}(\omega_{\textstyle{_{2mpq}}})\;\sin|\epsilon_{\textstyle{_2}}(\omega_{\textstyle{_{2mpq}}})|~\mbox{Sgn}\,(\omega_{\textstyle{_{2mpq}}})\,~.
 \label{A21a}
 \ea
 For an arbitrary degree $\,l\,$, the appropriate quality functions would, of course, look:
 \ba
 k_{\textstyle{_l}}\;\sin\epsilon_{\textstyle{_l}}\,=\,k_{\textstyle{_l}}(\omega_{\textstyle{_{lmpq}}})\;\sin\epsilon_{\textstyle{_l}}(\omega_{\textstyle{_{lmpq}}})
 ~=~k_{\textstyle{_l}}(\omega_{\textstyle{_{lmpq}}})\;\sin|\epsilon_{\textstyle{_l}}(\omega_{\textstyle{_{lmpq}}})|~\mbox{Sgn}\,(\omega_{\textstyle{_{lmpq}}})\,~.
 \label{A21b}
 \ea
 \label{A21}
 \es
 As will be explained in Appendix \ref{B} below, for realistic terrestrial bodies the quality function has the shape of a kink centered at the
 point $\,\omega_{\textstyle{_{lmpq}}}\,=\,0\,$, see Figure \ref{Fig1}.

 Inserting into (\ref{A21}) the expression (\ref{A1}) for the Fourier mode, we present the multiplier $\,k_{\textstyle{_2}}\;\sin\epsilon_{\textstyle{_2}}\,$ as a
 function of the spin rate $\,\stackrel{\bf\centerdot}{\theta\,}$ of the tidally perturbed body:
 \bs
 \ba
 k_{\textstyle{_2}}\;\sin\epsilon_{\textstyle{_2}}\,=\,k_{\textstyle{_2}}(\,(2-2p+q)\,n\,-\,m\,\dot{\theta}\,)\;\sin\epsilon_{\textstyle{_2}}(\,(2-2p+q)\,n\,-\,m\,\dot{\theta}\,)\,~.
 \label{A22a}
 \ea
 For an arbitrary degree $\,l\,$, that would be:
 \ba
 k_{\textstyle{_l}}\;\sin\epsilon_{\textstyle{_l}}\,=\,k_{\textstyle{_l}}(\,(l-2p+q)\,n\,-\,m\,\dot{\theta}\,)\;\sin\epsilon_{\textstyle{_l}}(\,(l-2p+q)\,n\,-\,m\,\dot{\theta}\,)
 \,~.
 \label{A22b}
 \ea
 \label{A22}
 \es
 Here $\,(\,(l-2p+q)\,n\,-\,m\,\dot{\theta}\,)\,$ denotes functional dependence on the argument $\,(l-2p+q)\,n\,-\,m\,\dot{\theta}\,$, not multiplication by a factor
 of $\,(l-2p+q)\,n\,-\,m\,\dot{\theta}\,$. The resulting functions of $\,\stackrel{\bf\centerdot}{\theta\,}$ will have a kink shape too, though the kink will now
 be centered around the point $\,\dot{\theta}\,=\,n\,(l-2p+q)/m\,$ corresponding to $\,\omega_{\textstyle{_{lmpq}}}\,=\,0\,$. Summation over the indices will yield
 a superposition of kinks, as in Figure \ref{Fig2}. We now say {\it{functions}} and not {\it{function}}, because for different sets of the integers $\,m,\,p,\,q\,$ we
 indeed obtain different functional dependencies of $\,k_{\textstyle{_l}}\;\sin\epsilon_{\textstyle{_l}}\,$ upon $\,\stackrel{\bf\centerdot}{\theta\,}$. The
 numerical values of these functions are small everywhere except in close vicinities of resonances, i.e., of the afore mentioned points $\,\dot{\theta}\,=\,n\,(l-2p+q)/m\,$.

 We limit ourselves to the degree $\,l=2\,$. Moreover, as was demonstrated in subsection \ref{AA3}, in the order $\,O(i^2)\,$ it is sufficient to take
 into account only the terms with $\,\{mp\}\,=\,\{20\}\,$ in the expression (\ref{A20}) for the solar torque. So we shall be interested only in the dependencies
 \ba
 k_{\textstyle{_2}}\;\sin\epsilon_{\textstyle{_2}}\,=\,k_{\textstyle{_2}}(\omega_{\textstyle{_{220q}}})\;\sin\epsilon_{\textstyle{_2}}(\omega_{\textstyle{_{220q}}})\,=\,
 k_{\textstyle{_2}}(\,(2+q)\,n\,-\,2\,\dot{\theta}\,)\;\sin\epsilon_{\textstyle{_2}}(\,(2+q)\,n\,-\,2\,\dot{\theta}\,)\,~.\qquad
 \label{A23}
 \ea
 It would not hurt to reiterate that $\,(\,(2-2p+q)\,n\,-\,m\,\dot{\theta}\,)\,$ denotes functional dependence on the argument $\,(2-2p+q)\,n\,-\,m\,\dot{\theta}\,$, not
 multiplication by a factor of $\,(2-2p+q)\,n\,-\,m\,\dot{\theta}\,$.

 The functions (\ref{A23}) will be peaked at $\,\dot{\theta}\,=\,n\,(l-2p+q)/m\,=\,n\,(2-q)/2\,$. For example, the terms with $\,q=\,6\,$ will peak at the 4:1
 resonance, i.e., when $\,\dot\theta/n=4\,$.

\begin{figure}[h]
   \begin{center}
        \epsfxsize=147mm
        \epsfbox{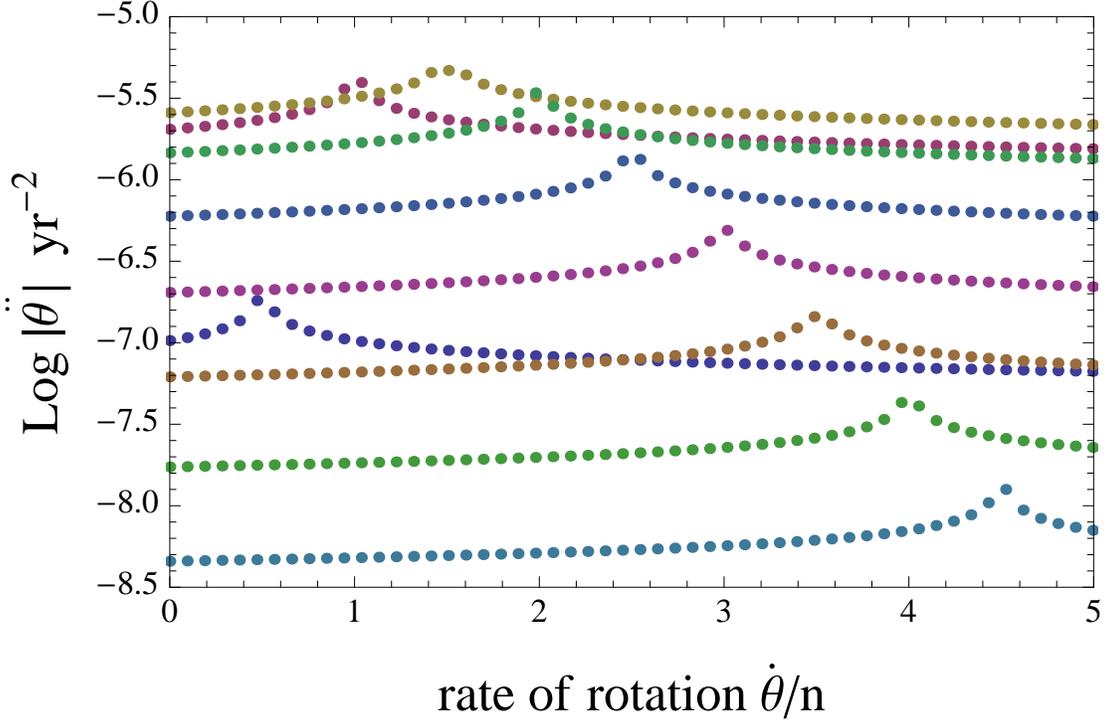}
   \end{center}
 \caption{\small{~Inputs into the angular acceleration caused by the secular part of the tidal torque. Curves on the plot depict the inputs
 from terms with $\,q=-1,\,0,\,\ldots\,,\,7\,$. A curve corresponding to the $\,q^{th}\,$ term can be identified by its peak located at $\,\dot\theta/n=(2+q)/2\,$. The plot is
 built for $\,e=0.32\,$, which is close to the upper limit on eccentricity in our simulation. For this value of eccentricity, the $\,q=0\,$ term is
 superseded by two of the higher-order terms (those with $\,q=1,\,2\,$).\label{contrib.fig}}}
 \end{figure}

 Figure \ref{contrib.fig} demonstrates several contributions into the secular ($\,j=q\,$) part of the torque (\ref{A20}). These are the contributions from the terms
 with $\,\{lmp\}\,=\,\{220\}\,$, $\,j\,=\,q\,$, calculated for the eccentricity value $\,e=0.32\,$, for several values of $\,q\,$. Each contribution is rendered by
 the decimal logarithm of the absolute value of the angular acceleration caused by the appropriate input into the secular part of the torque.

 It can be understood from these plots that, unless we are interested in the passage through higher spin-orbit resonances, the terms with $\,|q|,|j|\,>\,7\,$ can be
 omitted. Indeed, their input into the secular tidal torque is significantly smaller that that of the terms with $\,0\leq |q| \leq 7\,$, even for the high value $\,0.32\,$
 of Mercury's eccentricity, which is close to the upper boundary of our range. Thus the simulations presented in this paper included the terms with $\,-1\leq q \leq 7~$. Our
 choice limits the Taylor expansions of eccentricity-dependent functions to terms of the orders up to $\,e^7\,$, inclusive. In our simulations, this truncation also
 precludes the possibility of capture into a resonance higher than 9:2. We found that a capture into the 4:1 resonance took place so rarely, that the omission of the
 remote possibility of capture into the 4:1, or higher, resonances does not alter our results appreciably. The resulting relative precision of our calculations is in most cases
significantly better than 0.1\%$\,$, because the largest
   omitted term, the one with $\,q=8\,$, is 0.00055 times the largest included terms with $\,\dot\theta\,=\,\pi\,n\,$.
For eccentricity values lower than the one used to generated these plots ($\,e=0.32\,$), the relative precision is much better, and it does degrade rapidly
towards the upper-limit eccentricities.

 We would finally mention that the terms with $\,q=-2\,$ and $\,j=-2\,$ fall out, because the numerical factors accompanying these terms happen to vanish
 identically. Another interesting observation is that the terms with $\,q,\,j\,=\,-\,7\,$ through $\,q,\,j\,=\,-\,3\,$ are accompanied with extremely small numerical
 factors and are inessential. Thus we arrive at
 \ba
 \nonumber
 {\cal T}_z^{^{\,(TIDE)}}~=
 \qquad~\qquad~\qquad~\qquad~\qquad~\qquad~\qquad\qquad\qquad~\qquad\qquad\qquad~\qquad\qquad\qquad~\qquad\,\quad\quad
 ~\\ \label{A24}\\  \nonumber
 \frac{\textstyle 3}{\textstyle2}\,\frac{GM_{star}^{\textstyle{^{\,2}}}}{a}\left(\frac{R}{a}\right)^5 \sum_{q,\,j=-1}^{7}G_{\textstyle{_{\textstyle{_{20\mbox{\it{q}}}}}}}(e)
 \,G_{\textstyle{_{\textstyle{_{20\mbox{\it{j}}}}}}}(e)\,
 k_2\,\sin\epsilon_{\textstyle{_{
 2}}}
 \,\cos\left[\left(q- j\right){\cal M}\right]
 +O(e^8\epsilon)+O(i^2\epsilon)+O\left(\,(R/a)^{7}\epsilon\right)
   ~_{\textstyle{_{\textstyle ,}}}\,~
  \ea
 the absolute error $\,O\left(\,e^{8}\epsilon\right)\,$ originating from our neglect of terms with $\,|q|,|j|>7\,.$
 In the sum (\ref{A24}), the terms with $\,q=j\,$ constitute the secular part, the other terms being oscillating.

 \subsection{The secular part of the torque}\label{App8}

 While the secular part of the total torque (\ref{A14}) reads as
 \bs
 \ba
 \langle{\cal T}_z^{^{\,(TIDE)}}\rangle=\,\sum_{l=2}^{\infty}\,\left(\frac{R}{a}\right)^{\textstyle{^{2l+1}}}\frac{G\,M_{star}^{\textstyle{^{\,2}}}}{a}
 ~\sum_{m=1}^{\it l}\frac{({\it l} - m)!}{({\it l} + m)!} \;2\;m\;\sum_{p=0}^{\it l}F^2_{lmp}(\inc) \sum_{q=-\infty}^{\infty}G^2_{lpq}(e)
 ~k_{\textstyle{_{l}}}~\sin\epsilon_{\textstyle{_l}}~~_{\textstyle{_{\textstyle ,}}}~~\qquad~
 \label{A25a}
 \ea
 the secular part of the truncated version (\ref{A24}) assumes the form of
 \ba
 \langle\,{\cal{T}}_z^{\rm{_{\,(TIDE)}}}\rangle_{\textstyle{_{\textstyle_{\textstyle{_{l=2}}}}}}~=~
 \frac{3}{2}~G\,M_{star}^{\textstyle{^{\,2}}}\,R^5\,a^{-6}\sum_{q=-1}^{7}\,
 G^{\,2}_{\textstyle{_{\textstyle{_{20\mbox{\it{q}}}}}}}(e)~k_2
 ~\sin\epsilon_2
 +O(e^8\,\epsilon)+O(\inc^2\,\epsilon)
 ~~.~\quad~
 \label{A25b}
 \ea
 \label{A25}
 \es
  It is sufficient to employ the latter expression, with the oscillating terms excluded. The neglect of those leaves the resulting statistics virtually unaltered (Makarov et al. 2012).

 \section{\large{How rheology and self-gravitation define the shape\\ of the quality functions $~k_l(\omega_{\textstyle{_{lmpq}}})~\sin\epsilon_l(\omega_{\textstyle{_{lmpq}}})~$}}\label{B}

 An $\,lmpq\,$ term in the expansion of the tidal torque is proportional to a factor $~k_l\,\sin\epsilon_l\,=\,k_l(\omega_{\textstyle{_{lmpq}}})~\sin\epsilon_l
 (\omega_{\textstyle{_{lmpq}}})~$ which is a function of the Fourier mode $\,\omega_{\textstyle{_{lmpq}}}\,$ and goes under the name of {{a}}$\,$\footnote{~We say ``a",
 because the functional forms of these dependencies depend upon $\,l\,$.} quality function. The shapes of the quality functions are determined by the interplay of the
 planet's rheology and self-gravitation. So one should expect that these functional dependencies will include, as parameters, the rheological constants of the mantle and also
 the density and radius of the planet.

 In our previous works (Efroimsky 2012a,b), the topic was explored in detail for a homogeneous near-spherical body. It was demonstrated how the quality functions are
 expressed through the real and imaginary parts of the complex compliance of the mantle material and the planet's radius and mass. The quality functions turned out to be
 odd. Although slightly different for different values of the degree $\,l\,$, the quality functions demonstrate similarity of shape and always look like a kink -- see
 Figure \ref{Fig1}.$\,$\footnote{~In a hypothetical situation of a planet despinning through a resonance at a constant rate, the appropriate Fourier mode is linear in
 time. Then the tidal torque, expressed as a function of time, will acquire a similar kink shape. This situation is described in Ferraz-Mello (2013, Figure 7b). Any
 physically acceptable rheological model must lead to this or similar kind of tidal torque behaviour near a resonance.}
 Each function
 $~k_l(\omega_{\textstyle{_{lmpq}}})~\sin\epsilon_l(\omega_{\textstyle{_{lmpq}}})~$ looks as a kink centered around the resonant point $\,\omega_{\textstyle{_{lmpq}}}=0\,$.
 This way, an $\,lmpq\,$ term of the torque changes its sign when the appropriate commensurability is transcended.

  Being odd, the quality functions can be rewritten as $~k_l(\omega_{\textstyle{_{lmpq}}})~\sin|\,\epsilon_l(\omega_{\textstyle{_{lmpq}}})\,|~\,\mbox{Sgn}\,(\,\omega_{
 \textstyle{_{lmpq}}}\,)~$, where the products $~k_l(\omega_{\textstyle{_{lmpq}}})~\sin|\,\epsilon_l(\omega_{\textstyle{_{lmpq}}})\,|~$ are even functions of the
 Fourier modes and therefore can be regarded as functions not of the tidal mode $\,\omega_{\textstyle{_{lmpq}}}\,$ but of its absolute value $\,\chi_{\textstyle{_{lmpq}}}\,=\,|\,\omega_{\textstyle{_{lmpq}}}\,|\,$, which is the actual physical frequency of the tidally induced stress:
 \ba
 \nonumber
 k_l(\omega_{\textstyle{_{lmpq}}})~\sin\epsilon_l(\omega_{\textstyle{_{lmpq}}})&=&k_l(\omega_{\textstyle{_{lmpq}}})~\sin|\,\epsilon_l
 (\omega_{\textstyle{_{lmpq}}})\,|~\,\mbox{Sgn}\,(\,\omega_{\textstyle{_{lmpq}}}\,)\\
 \nonumber\\
 &=&k_l(\chi_{\textstyle{_{lmpq}}})~\sin|\,\epsilon_l
 (\chi_{\textstyle{_{lmpq}}})\,|~\,\mbox{Sgn}\,(\,\omega_{\textstyle{_{lmpq}}}\,)\,~.
 \label{eq:quality}
 \ea
 As was shown in {\it{Ibid.}}, the quality functions have the form
 \ba
 k_l(\omega_{\textstyle{_{lmpq}}})\;\sin\epsilon_l(\omega_{\textstyle{_{lmpq}}})\,=\;\frac{3}{2\,({\it l}\,-\,1)}\;\,\frac{-\;A_l\;J\;{\cal{I}}{\it{m}}\left[\bar{J}(\chi)\right]}{
 \left(\;{\cal{R}}{\it{e}}\left[\bar{J}(\chi)\right]\;+\;A_l\;J\;\right)^2\;+\;\left(\;{\cal{I}}{\it{m}}\left[\bar{J}(\chi)\right]\;
 \right)^2} ~\,\mbox{Sgn}\,(\,\omega_{\textstyle{_{lmpq}}}\,)~~~,~~~~
 \label{L39b}
 \ea
 where $\,\chi\,$ is a shortened notation for the frequency $\,\chi_{\textstyle{_{lmpq}}}\,$. The coefficients $\,A_l\,$ are given by
 \ba
 A_{l}\,
 \equiv\;\frac{\textstyle{(2\,{\it{l}}^{\,2}\,+\,4\,{\it{l}}\,+\,3)\,{\mu}}}{\textstyle{{\it{l}}\,\mbox{g}\,
 \rho\,R}}\;=\;\frac{\textstyle{3\;(2\,{\it{l}}^{\,2}\,+\,4\,{\it{l}}\,+\,3)\,{\mu}}}{\textstyle{4\;{\it{l}}
 \,\pi\,G\,\rho^2\,R^2}}\;\;\;,~~~~~~~
 \label{eq:amplitude}
 \ea
 where $\,R\,$, $\,\rho\,$, $\,\mu\,$, and g  are the radius, mean density, unrelaxed rigidity, and surface gravity of the planet, while
 $\,G\,$ is the Newton gravity constant.

 With $\,\Gamma\,$ denoting the Gamma function, the functions
 \ba
 {\cal R}{\it e} [ \bar{J}(\chi)]\;=\;J\;+\;J\,(\chi\tau_{_A})^{-\alpha}\;\cos\left(\,\frac{\alpha\,\pi}{2}\,\right)
 \;\Gamma(\alpha\,+\,1)~~~~\quad\quad\quad\quad\quad~\quad\quad\quad\quad\quad\quad\quad
 \label{A4ccc}
 \ea
 and
 \ba
 {\cal I}{\it m} [ \bar{J}(\chi)]\;=\;-\;J~(\chi\tau_{_M})^{-1}\;-\;J\,(\chi\tau_{_A})^{-\alpha}\;\sin\left(
 \,\frac{\alpha\,\pi}{2}\,\right)\;\Gamma(\alpha\,+\,1)~~~,~~~~~~~~~~~~~~\quad\quad\quad
 \label{A3ccc}
 \ea
 are the real and imaginary parts of the complex compliance $\,\bar{J}(\chi)\,$ of the planet. They contain several rheological parameters. Specifically, the notation
 $\,J\,$ stands for the unrelaxed compliance of the mantle, which is inverse to the unrelaxed rigidity $\,\mu\,$. The Andrade parameter $\,\alpha\,$ is known to
 be about $\,0.3\,$ for solid silicates and about $\,0.14 - 0.2\,$ for partial melts. (We used $\,\alpha=0.2\,$.)

 As ever, the notation $\tau_{_M}$ stands for the Maxwell time which is the ratio of the mantle's viscosity $\,\eta\,$ and rigidity $\,\mu\,$. For silicate mantles, the
 values of $\,\tau_{_M}\,$ may vary greatly, given the exponential dependence of the viscosity on the temperature. For the Earth, $\tau_{_M}\approx$ 500 yr

 The so-called Andrade time $\,\tau_{_A}\,$ was introduced in Efroimsky (2012a,b) as a substitute to the previously used Andrade parameter $\,\beta\,$. The necessity for
 that substitution stemmed from the commonly employed $\,\beta\,$ having fractional dimensions and thus lacking a clear physical interpretation. Referring the reader to
 {\it{Ibid.}} for details, we would only recall here that beneath some threshold frequency the friction in the mantle becomes predominantly viscous, while the
 inelastic processes seize to play a major role. Therefore, at frequencies below that threshold, the mantle's response becomes close to that of a Maxwell body. Mathematically, this
 means that below the threshold the parameter $\,\tau_{_A}\,$ increases rapidly as the frequency goes down. As a result of this, only the first term in
 (\ref{A4c}) and the first term in (\ref{A3c}) remain relevant, and the resulting complex compliance coincides with that of a Maxwell material.

 At frequencies near and above the threshold, both the viscoelastic and inelastic mechanisms are present. This implies that the parameters $\,\tau_{_M}\,$ and $\,\tau_{_A}\,$ should have comparable values, as can be seen from the expressions (\ref{A4c}) and (\ref{A3c}) for the complex compliance. In that frequency band, the mantle behaves as an Andrade body. With increase of the frequency, the inelastic reaction takes over the viscoelastic reaction, as defect-unpinning becomes a dominating mechanism of friction.

 In our computations, we treated $\,\tau_{_A}\,$ in the same way as in Makarov et al. (2012) and Makarov (2012): we set $\,\tau_{_A}\,=\,\tau_{_M}\,$ at the frequencies above
 the threshold (chosen to be 1 yr$^{-1}\,$, like for the solid Earth). Below the threshold, we set $\,\tau_{_A}\,$ to increase exponentially with the decrease of the frequency,
 so at low frequencies the rheology approached that of Maxwell. Numerical runs have shown that the resulting capture probabilities are not very sensitive to how
 quickly the switch from the Andrade to Maxwell model is performed (Makarov 2012).

 Programming the quality functions, it is practical to divide both the numerator and denominator of (\ref{L39b}) by $\,J^{\,2}\,$:
 \ba
 k_l(\,\omega_{\textstyle{_{lmpq}}}\,)\;\sin\epsilon_l(\,\omega_{\textstyle{_{lmpq}}}\,)\;=\;\frac{3}{2\,({\it l}\,-\,1)}\;\,\frac{-\;A_l\;{\cal{I}}}{\left(\;{\cal{R}}
 \;+\;A_l\;\right)^2\;+\;{\cal{I}}^{\textstyle{^{\,2}}}}~\,\mbox{Sgn}\,(\,\omega_{\textstyle{_{lmpq}}}\,) ~~~,~~~~~
 \label{L39}
 \ea
 ${\cal R}\,$ and $\,{\cal I}\,$ being the {\it{dimensionless}} real and imaginary parts of the complex compliance:
 \ba
 {\cal R}\;=\;1\;+\;(\chi\tau_{_A})^{-\alpha}\;\cos\left(\,\frac{\alpha\,\pi}{2}\,\right)
 \;\Gamma(\alpha\,+\,1)~~~,\quad\quad\quad\quad\quad~\quad\quad\quad\quad\quad\quad\quad
 \label{A4c}
 \ea
 \ba
 {\cal I}\;=\;-\;(\chi\tau_{_M})^{-1}\;-\;(\chi\tau_{_A})^{-\alpha}\;\sin\left(
 \,\frac{\alpha\,\pi}{2}\,\right)\;\Gamma(\alpha\,+\,1)~~~.~~~~~~~~~~~~~~\quad\quad\quad\quad\quad
 \label{A3c}
 \ea
 These dependencies were used by Makarov et al (2012), to study the spin-orbit dynamics of the planet GJ581d, and in Makarov (2012), to explore the dynamics of a Mercury-like planet

 \section{\large{The separatrix dividing circulation from libration.\\
 Derivation of the equation (\ref{separ.eq})}}\label{D}

 In neglect of tidal and core-mantle friction, the law of motion near the $\,q\,'\,$ resonance can be written as
 \ba
 C_{\rm m}~\ddot{\gamma} ~+~\frac{3}{2}~(B_{\rm m}\,-\,A_{\rm m})~\frac{M_{star}}{M_{star}\,+\,M_{planet}}~n^2~G_{20\mbox{\small{q}}\,'}(e)~\sin 2\gamma~=~0\,~.
 \label{*}
 \label{655}
 \ea
 As was pointed out by Goldreich (1966), the problem is identical to a simple pendulum. Indeed, in terms of $\,\beta=2\gamma\,$,
 equation (\ref{655}) becomes $~\ddot{\beta}+2\omega_0^2\,\sin\beta=0~$, with a constant positive $\,\omega_0^2\,$ and with $\,\beta\,$ playing the role of the pendulum angle.
 \footnote{~Approximation of $\,\sin 2 \gamma\,$ with $\,2\gamma\,$ in the equation (\ref{*}) indicates that the quantity
 \ba
 \sqrt{2}~\omega_0~=~ n~\left[\,3~\frac{B_{\rm m}-A_{\rm m}}{C_{\rm m}}~\,\frac{M_{star}}{\,M_{star}\,+\,M_{planet}\,}~G_{20\mbox{\small{q}}\,'}(e)\,\right]^{1/2}
 \nonumber
 \ea
 is the frequency of $\,${\it{small}}$\,$ librations. For planets (but not necessarily for small bodies), the condition $\,\frac{\textstyle B-A}{\textstyle C}~\frac{\textstyle M_{star}}{\textstyle M_{star}\,+\,M_{planet}}\,\ll\,1~$ is obeyed, whence $\,\sqrt{2}\,\omega_0\ll n\,$. Larger amplitudes increase the period and yield a smaller frequency, so $\,\sqrt{2}\,\omega_0\,$ is the {\it{largest}}$\,$ frequency of libration. (Being unharmonic, large-amplitude librations still have a period. Accordingly, they may be attributed a frequency, albeit an unharmonic one.)\\
 \vspace{1mm}
 $\quad$ We denote this frequency with $\,\sqrt{2}\,\omega_0\,$ to conform to the notation from \citet{pb1977,pb}.
 }

 Multiplying the equation (\ref{655}) by $\,\dot{\gamma}\,$ and integrating over time $\,t\,$, we arrive at the first integral of motion,
 \ba
 \frac{1}{2}~C_{\rm m}~{{\dot{\gamma}}^{\,2}}~-~\frac{3}{4}~(B_{\rm m}\,-\,A_{\rm m})~\frac{M_{star}}{M_{star}\,+\,M_{planet}}~n^2~G_{20\mbox{\small{q}}\,'}(e)~\cos 2\gamma
 ~=~E~~~,
 \label{**}
 \label{656}
 \ea
 whose value is defined by the initial conditions. Just as in the pendulum case, there exists a critical value $\,E\,=\,E_b\,$ serving as a boundary separating
 circulation from libration. To find $\,E_b\,$, we should set simultaneously $\,\gamma=\pi/2\,$ and $\,\dot{\gamma}=0~$ in the formula (\ref{656}). This furnishes:
 \ba
 E_b ~=~ \frac{3}{4}~(B_{\rm m}\,-\,A_{\rm m})~\frac{M_{star}}{M_{star}\,+\,M_{planet}}~G_{20\mbox{\small{q}}\,'}(e)~n^2~~,
 \label{657}
 \ea
 insertion whereof into the formula (\ref{656}) renders us the equation of the separatrix dividing circulation from libration:
 \footnote{~Our equations (\ref{655} - \ref{656}) and (\ref{sepa}) coincide with those in Goldreich \& Peale (1968), but differ from similar equations in Makarov (2012) and
 Makarov et al. (2012). A small difference in numerical factors stems from the fact that the $\,\gamma\,$ used in Goldreich \& Peale (1968) and in the current paper differs by a factor of 1/2
 from the $\,\gamma\,$ used in Makarov (2012) and Makarov et al. (2012).}
 \be
 \dot\gamma~=~n\,\left[\,\frac{3~(B_{\rm m}\,-\,A_{\rm m})}{C_{\rm m}}~\,\frac{M_{star}}{M_{star}\,+\,M_{planet}}
 ~\,G_{20\mbox{\small{q}\,}'}(e)\,\right]^{{1}/{2}}\cos{\gamma}\,~.
 \label{sepa}
 \ee



\begin{thebibliography}{}


\bibitem[Archinal et al.(2011)]{archinal} Archinal, B. A.; A'Hearn, M. F.; Bowell, E.; Conrad, A.; Consolmagno, G. J.; Courtin, R.; Fukushima, T.; Hestroffer, D.;
Hilton, J. L.; Krasinsky, G. A.; Neumann, G.; Oberst, J.; Seidelmann, P. K.; Stooke, P.; Tholen, D. J.; Thomas, P. C.; and Williams, I. P. 2011.
    ``Report of the IAU Working Group on Cartographic Coordinates and Rotational Elements: 2009."
    {\it{Celestial Mechanics and Dynamical Astronomy}}, Vol. {\bf{109}}, pp. 101 - 135.

  \bibitem[Batygin \& Laughlin(2008)]{bl2008} Batygin, K., and Laughlin, G. 2008. ``On the dynamical stability of the Solar System." \apj, Vol. {\bf{683}}, pp. 1207 - 1216

\bibitem[Benz et al.(1988)]{bsc1988} Benz, W.; Slattery, W. L., and Cameron, A. G. W. 1988. ``Collisional stripping of Mercury's mantle." \icarus, Vol.{\bf{74}}, pp. 516 - 528

\bibitem[Benz et al.(2007)]{bahw2007} Benz, W.; Anic, A.; Horner, J., and Whitby, J.A. 2007. ``The origin of Mercury." \ssr, Vol. {\bf{132}}, pp. 189 - 202

 \bibitem[Bou{\'e} et al.(2012)]{blf2012} Bou{\'e}, G.; Laskar, J.; and Farago, F. 2012. ``A simple model of the chaotic eccentricity of Mercury." \aap, Vol. {\bf{548}}, p. A43

 \bibitem[Brasser et al.(2013)]{bwn2013} Brasser, R.; Walsh, K.J., and Nesvorn\'y, D.
 2013. ``Constraining the primordial orbits of the terrestrial planets." \mnras, Vol. {\bf{433}}, pp. 3417 - 3427

 \bibitem[Bretagnon(1974)]{b1974} Bretagnon, P. 1974. ``Termes \`a longues p\'eriodes dans le syst\`eme solaire." \aap, Vol. {\bf{30}}, pp. 141 - 154

 \bibitem[Bretagnon(1982)]{b1982} Bretagnon, P. 1982. ``Theory for the motion of all the planets - The VSOP82 solution." \aap, Vol. {\bf{114}}, pp. 278 - 288

\bibitem[Cameron(1985)]{c1985} Cameron, A. G. W. 1985. ``The partial volatilization of Mercury." \icarus, Vol. {\bf{64}}, pp. 285 - 294

\bibitem[Castillo-Rogez et al.(2011)]{cel2011} Castillo-Rogez, J.C.; Efroimsky, M.; and Lainey, V. 2011. ``The tidal history of Iapetus:
Spin dynamics in the light of a refined dissipation model." \jgr, Vol. {\bf{116}}, E09008

 \bibitem[Colombo(1965)]{c1965} Colombo, G. 1965. ``Rotational period of the planet Mercury." \nat, Vol. {\bf{208}}, p. 575

\bibitem[Correia \& Laskar(2004)]{cl2004} Correia, A.C.M., and Laskar, J. 2004. ``Mercury's capture into the 3/2 spin-orbit resonance as a result of
its chaotic dynamics." \nat, Vol. {\bf{429}}, pp. 848 - 850

\bibitem[Correia \& Laskar(2009)]{cl2009} Correia, A.C.M., and Laskar, J. 2009. ``Mercury's capture into the 3/2 spin-orbit resonance including the effect of
core-mantle friction." \icarus, Vol. {\bf{201}}, pp. 1 - 11

\bibitem[Correia \& Laskar(2010)]{cl2010} Correia, A.C.M., and Laskar, J. 2010. ``Long-term evolution of the spin of Mercury I. Effect of the obliquity and
core-mantle friction." \icarus, Vol. {\bf{205}}, pp. 338 - 355

\bibitem[Correia \& Laskar(2012)]{cl2012} Correia, A.C.M., and Laskar, J. 2012. ``Impact cratering on Mercury: Consequences for the spin evolution." \apj, 751:L43

\bibitem[Counselman(1969)]{c1969} Counselman III C.C., 1969, ``Spin-orbit resonance of Mercury'', PhD dissertation, Massachusetts Institute of Technology

\bibitem[Dahlquist \& Bj\"orck(2008)]{db2008} Dahlquist G. and Bj\"orck {\AA}. 2008. Numerical Methods in Scientific Computing, Vol. 1. SIAM, Philadelphia, pp. 351-520 (Chapter 4)

\bibitem[Danby(1962)]{danb} Danby, J.M.A. 1962. Fundamentals of Celestial Mechanics. MacMillan, New York

\bibitem[Darwin(1879)]{darwin} Darwin, G. H. 1879. ``On the precession of a viscous spheroid and on the remote history of the Earth."
{\it{Philosophical Transactions of the Royal Society of London}}, Vol. {\bf{170}}, pp. 447 - 530

\bibitem[Denevi et al.(2009)]{drsmbdmehwc2009} Denevi, B.W.; Robinson, M.S.; Solomon, S.C.; Murchie, S.L.;, Blewett, D.T.; Domingue, D.L.; McCoy, T.J.;
Ernst, C.M.; Head, J.W.; Watters, T.R.; and Chabot, N.L., 2009, ``The evolution of Mercury's crust: A global perspective from MESSENGER'', {\it{Science}},
Vol.{\bf{324}}, pp. 613 - 618

\bibitem[Dobrovolskis(2007)]{d2007} Dobrovolskis, A.R. 2007. ``Spin states and climates of eccentric exoplanets." \icarus, Vol. {\bf{192}}, pp. 1 - 23

\bibitem[Dones \& Tremaine(1993)]{dt1993} Dones, L., \& Tremaine, S., 1993 ``On the origin of planetary spins'', \icarus, Vol. {\bf{103}}, pp. 67 - 92

\bibitem[Duriez(2002)]{d2002} Duriez L. 2002. Cours de m\'ecanique c\'eleste classique, \verb=http://lal.univ-lille1.fr/mecanique_celeste.html=

\bibitem[Efroimsky \& Lainey(2007)]{el2007} Efroimsky, M., and Lainey, V. 2007. ``Physics of bodily tides in terrestrial planets and the appropriate scales of
dynamical evolution." \jgr, 112, E12003

 \bibitem[Efroimsky(2012a)]{e2012} Efroimsky, M. 2012a. ``Bodily tides near spin-orbit resonances." {\it{Celestial Mechanics and Dynamical Astronomy}}, Vol. {\bf{112}}, pp. 283 - 330\\
~~Extended version: ~http://arxiv.org/abs/1105.6086

\bibitem[Efroimsky(2012b)]{e2012b} Efroimsky, M. 2012b. ``Tidal dissipation compared to seismic dissipation: in small bodies, earths, and superearths."
                                        {\it{The Astrophysical Journal}}, Vol. {\bf{746}}, ~id. 150\\ doi:10.1088/0004-637X/746/2/150 ~~~~http://arxiv.org/abs/1105.3936\\
                                        ERRATA: {\it{ApJ}}, Vol. {\bf{763}}, ~id. $\,$150 (2013)

 \bibitem[Efroimsky \& Makarov(2013)]{em2013} Efroimsky, M., and Makarov, V.V. 2013. ``Tidal friction and tidal lagging. Applicability limitations of a popular
 formula for the tidal torque." \apj, 764:26

 \bibitem[Eggleton et al.(1998)]{ekh1998} Eggleton, P. P.; Kiseleva, L. G.; and Hut, P. 1998.
 ``The equilibrium tide model for tidal friction." {\it{The Astrophysical Journal}}, Vol. {\bf{499}}, pp. 853 - 870

 \bibitem[Elkins-Tanton \& Hager(2005)]{eh2005} Elkins-Tanton, L.T.; and Hager, B.H.; 2005, ``Giant meteoroid impacts can cause volcanism.'',
 {\it{Earth and Planetary Science Letters}}, Vol. {\bf{239}}, pp. 219 - 232

 \bibitem[Fassett et al.(2012)]{fhbzsnskscppop2012} Fassett, C.I., Head, J.W., Baker, D.M.H., Zuber, M.T., Smith, D.E., Neumann, G.A.,
 Solomon, S.C., Klimczak, C., Strom, R.G., Chapman, C.R., Prockter, L.M., Phillips, R.J., Oberst, J., and Preusker, F., 2012,
 ``Large impact basins on Mercury: Global distribution, characteristics, and modification history from MESSENGER orbital data'', \jgr, Vol.{\bf{117}}, E00L08

\bibitem[Fegley \& Cameron(1987)]{fc1987} Fegley Jr., B., and Cameron, A. G. W. 1987. ``A vaporization model for iron / silicate fractionation in the Mercury
protoplanet." {\it{Earth and Planetary Science Letters}}, Vol.{\bf{82}}, pp. 207 - 222

 \bibitem[Ferraz-Mello et al.(2008)]{frh2008} Ferraz-Mello, S.; Rodr\'iguez, A.; and Hussmann, H. 2008. ``Tidal friction in close-in satellites and exoplanets: The Darwin theory re-visited." {\it{Celestial Mechanics and Dynamical Astronomy}}, Vol. {\bf{101}}, pp. 171 - 201

 \bibitem[Ferraz-Mello(2013)]{ferraz} Ferraz-Mello, S. 2013. ``Tidal synchronization of close-in satellites and exoplanets. A rheophysical approach."
 {\it{Celestial Mechanics and Dynamical Astronomy}}, Vol. {\bf{116}}, pp. 109 - 140

 \bibitem[Galassi et al.(2009)]{gdtgjabr2009} Galassi, M.; Davies, J.; Theiler, J.; Gough, B.; Jungman, G.; Alken, P.; Booth, M.; and Rossi, F. 2009.
 {\it{GNU Scientific Library Reference Manual}} (v1.12), 3rd ed. Network Theory Ltd., Bristol

 \bibitem[Goldreich(1966)]{g1966} Goldreich, P. 1966. ``Final spin states of planets and satellites." {\it{The Astronomical Journal}}. Vol. {\bf{71}}, pp. 1 - 7

 \bibitem[Goldreich \& Peale(1966)]{gp1966} Goldreich, P., and Peale, S. 1966. ``Spin-orbit coupling in the Solar System." {\it{The Astronomical Journal}}, Vol. {\bf{71}}, pp. 425 - 438

 \bibitem[Goldreich \& Peale(1967)]{gp1967} Goldreich, P., and Peale, S. 1967. ``Spin-orbit coupling in the Solar System. II. The resonant
 rotation of Venus" {\it{The Astronomical Journal}}, Vol. {\bf{72}}, pp. 662 - 668

 \bibitem[Goldreich \& Peale(1968)]{gp1968} Goldreich, P., and Peale, S.J. 1968.
  ``The Dynamics of Planetary Rotations." {\it{Annual Review of Astronomy and Astrophysics}}, Vol. {\bf{6}}, pp. 287 - 320

  \bibitem[Gomes et al.(2005)]{gltm2005} Gomes, R.; Levison, H.F.; Tsiganis, K.; and Morbidelli, A.; 2005, ``Origin of the cataclysmic Late Heavy
  Bombardment period of the terrestrial planets.'', \nat, Vol. {\bf{435}}, pp. 466 - 469

 \bibitem[Guzzo(2005)]{guzzo2005} Guzzo, M. 2005. ``The web of three-planet resonances in the outer Solar System." \icarus, Vol. {\bf{174}}, pp. 273 - 284

  \bibitem[Guzzo(2006)]{guzzo2006} Guzzo, M. 2006. ``The web of three-planet resonances in the outer Solar System. II. A source of orbital instability for Uranus
  and Neptune." \icarus, Vol. {\bf{181}}, pp. 475 - 485

 \bibitem[Hairer et al.(1987)]{hnw1987} Hairer E., N{\o}rsett S.P. and Wanner G., 1987, Solving Ordinary Differential Equations. I: Nonstiff Problems,
 Springer Series in computational Mathematics, Springer-Verlag, Berlin

 \bibitem[Hayes(2007)]{hayes} Hayes, W. B. 2007. ``Is the outer Solar System chaotic?" {\it{Nature Physics}}, Vol. {\bf{3}}, No 10, pp. 689 - 691

 \bibitem[Hut(1981)]{hut} Hut, P. 1981. ``Tidal evolution in close binary systems." {\it{Astronomy \& Astrophysics}}, Vol. {\bf{99}}, pp. 126 - 140


\bibitem[Kaula(1964)]{k1964} Kaula, W.M. 1964. ``Tidal dissipation by solid friction and the resulting orbital evolution." Reviews of Geophysics, 2, 661 - 685

\bibitem[Kokubo \& Ida(2007)]{ki2007} Kokubo, E., and Ida, S., 2007, ``Formation of terrestrial planets from protoplanets -- II. Statistics of planetary spin.'', \apj,
Vol. {\bf{671}}, pp. 2082 - 2090

\bibitem[Laskar(1986)]{l1986} Laskar J., 1986, Secular terms of classical planetary theories using the results of general theory, \aap, 157, 59-70

  \bibitem[Laskar(1988)]{l1988} Laskar, J. 1988. ``Secular evolution of the solar system over 10 million years." \aap, 198, 341 - 362

\bibitem[Laskar(1989)]{l1989} Laskar J., 1989, A numerical experiment on the chaotic behaviour of the solar system, \nat, 338, 237

   \bibitem[Laskar et al.(2004a)]{lrjgcl2004a} Laskar, J.; Robutel, P.; Joutel, F.; Gastineau, M.; Correia, A.~C.~M. and Levrard, B. 2004a.
   ``A long-term numerical solution for the insolation quantities of the Earth." \aap, Vol. {\bf{428}}, pp. 261 - 285

   \bibitem[Laskar et al.(2004b)]{lcgjlr2004b} Laskar, J.; Correia, A.~C.~M.; Gastineau, M.; Joutel, F.; Levrard, B. and Robutel, P. 2004b.
  ``Long term evolution and chaotic diffusion of the insolation quantities of Mars." \icarus, Vol. {\bf{170}}, pp. 343 - 364

 \bibitem[Laskar(2008)]{l2008} Laskar, J. 2008. ``Chaotic diffusion in the Solar System." \icarus, 196, 1 - 15

 \bibitem[Laskar \& Gastineau(2009)]{lg2009} Laskar, J., and Gastineau, M. 2009. ``Existence of collisional trajectories of Mercury, Mars and Venus with the Earth."
 \nat, 459, 817 - 819

 \bibitem[Laskar et al.(2011)]{lfgm2011} Laskar, J.; Fienga, A.; Gastineau, M.; and Manche, H. 2011. ``La2010: a new orbital solution for the long-term motion of the Earth."
 \aap, Vol. {\bf{532}}, p. A89


\bibitem[Le Feuvre \& Wieczorek(2011)]{lefe} Le Feuvre, M., \& Wieczorek, M.A. 2011. ``Nonuniform cratering of the Moon and a revised crater chronology of the inner Solar System." {\it{Icarus}}, Vol. {\bf{214}}, pp. 1 - 20


\bibitem[MacDonald(1964)]{m1964} MacDonald, G.J.F. 1964. ``Tidal friction." {\it{Reviews of Geophysics}}, Vol. {\bf{2}}, pp. 467 - 541

 \bibitem[Makarov(2012)]{m2012} Makarov, V.V. 2012. ``Conditions of passage and entrapment of terrestrial planets in spin-orbit resonances." \apj, 752:73

 \bibitem[Makarov et al.(2012)]{makarovetal2012} Makarov, V.V.; Berghea, C.; and Efroimsky, M. 2012. ``Dynamical evolution and spin-orbit resonances of potentially
                                                 habitable exoplanets. The case of GJ 581d." {\it{ApJ}}, {{761}}:83

 \bibitem[Makarov \& Efroimsky(2013)]{me2013} Makarov, V.V., and Efroimsky, M. 2013. ``No pseudosynchronous rotation for terrestrial planets and moons." \apj, 764:27

 \bibitem[Makarov \& Efroimsky(2014)]{me2014} Makarov, V.V., and Efroimsky, M. 2014, ``Tidal dissipation in a homogeneous spherical body -- II. Examples: Io, Mercury, and Kepler-10$\,$b."
Submitted to: ~{\it{The Astrophysical J.}} ~\\ arXiv:1406.2352

 \bibitem[Marchi et al.(2013)]{mcfhbs2013} Marchi, S.; Chapman, C.R.; Fassett, C.I.; Head,  J.W.; Bottke, W.F.; and Strom, R.G.; 2013, ``Global resurfacing of Mercury
 4.0-4.1 billion years ago by heavy bombardment and volcanism'', \nat, Vol. {\bf{499}}, pp. 59 - 61

 \bibitem[Margot et al.(2007)]{mpjsh2007} Margot, J.-L.; Peale, S.J.; Jurgens, R.F.; Slade, M.A.; and Holin, I.V. 2007. ``Large longitude libration of Mercury reveals a
                                          molten core." {\it{Science.}} Vol. {\bf{316}}, pp. 710 - 714

 \bibitem[Margot et al.(2012)]{mpshgjygpc2012} Margot, J.-L.; Peale, S.J.; Solomon, S.C.; Hauck II, S.A.; Ghigo, F.D.; Jurgens, R.F.; Yseboodt, M.; Giorgini, J.D.;
                                               Padovan, S.; and Campbell, D.B. 2012. ``Mercury's moment of inertia from spin and gravity data." \jgr, Vol. {\bf{117}}, E00L09

 \bibitem[Meriggiola \& Iess(2012)]{mi2012} Meriggiola R. \& Iess L., 2012, ``A new rotational model of Titan from Cassini SAR data'', European Planetary Science Congress 2012, id. EPSC2012-593

 \bibitem[Mignard(1979)]{mignard1979} Mignard, F. 1979. ``The Evolution of the Lunar Orbit Revisited. I."
           {\it{The Moon and the Planets.}} Vol. {\bf{20}}, pp. 301 - 315

 \bibitem[Mignard(1979)]{mignard1980} Mignard, F. 1980. ``The Evolution of the Lunar Orbit Revisited. II."
           {\it{The Moon and the Planets.}} Vol. {\bf{23}}, pp. 185 - 201.

 \bibitem[Morbidelli \& Vokrouhlick\'y(2003)]{mv2003} Morbidelli A. \& Vokrouhlick\'y D., 2003, ``The Yarkovsky-driven origin of near-earth asteroids.'', \icarus,
 Vol. {\bf{163}}, pp. 120 - 134.

 \bibitem[Ness et al.(1974)]{nblws1974} Ness, N.F., Behannon, K.W., Lepping, R.P., Whang, Y.C., \& Schatten,K.H., 1974,
 ``Magnetic field observations near Mercury: Preliminary results from Mariner 10'', {\it{Science}}, {\bf{185}}, 151-160

 \bibitem[Ness et al.(1975)]{nblw1975} Ness, N.F., Behannon, K.W., Lepping, R.P., \& Whang, Y.C., 1975, ``The magnetic field of Mercury, 1'', \jgr., {\bf{80}}, 2708-2716

 \bibitem[Nobili et al.(1989)]{nmc1989} Nobili A.M., Milani A., Carpino M. 1989. ``Fundamental frequencies and small divisors in the orbits of the outer planets."
                                        \aap, Vol. {\bf{210}}, pp. 313 - 336

 \bibitem[Noyelles \& Lhotka(2013)]{NL} Noyelles, B., and Lhotka, C. 2013. ``The influence of orbital dynamics, shape and tides on the obliquity of Mercury.",
 {\it Advances in Space Research}, Vol. {\bf{52}}, pp. 2085 - 2101

\bibitem[Padovan et al.(2014)]{pmhms2014} Padovan, S., Margot, J.-L., Hauck II, S.A., Moore, W.B., and Solomon, S.C., 2014, ``The tides of Mercury and possible implications for its interior structure.", {\it Journal of Geophysical Research: Planets}, Vol.{\bf{119}}, 850-866

 \bibitem[Peale \& Boss(1977a)]{pb1977} Peale, S.J., and Boss, A.P. 1977a. ``A spin-orbit constraint on the viscosity of a Mercurian liquid core." \jgr, Vol. {\bf{82}}, pp. 743 - 749

 \bibitem[Peale \& Boss(1977b)]{pb} Peale, S.J., and Boss, A.P. 1977b. ``Mercury's Core: The Effect of Obliquity on the Spin-Orbit Constraints." \jgr, Vol. {\bf{82}}, pp.
  3423 - 3429

 \bibitem[Peale(2005)]{peale2005} Peale, S.J. 2005. ``The free precession and libration of Mercury." {\it{Icarus}}, Vol. {\bf{178}}, pp. 4 - 18

 \bibitem[Peplowski et al.(2011)]{pehmbgeghlmnsrsss2011} Peplowski, P.N.; Evans, L.G.; Hauck II, S.A.; McCoy, T.J.; Boynton, W.V.; Gillis-Davis, J.J.;
 Ebel, D.S.; Goldsten, J.O.; Hamara, D.K.; Lawrence, D.J.; McNutt Jr., R.L.; Nittler, L.R.; Solomon, S.C.; Rhodes, E.A.; Sprague, A.L.; Starr, R.D.; and
 Stockstill-Cahill K.R. 2011. ``Radioactive elements on Mercury's surface from MESSENGER: Implications for the planet's formation and evolution", {\it Science}, Vol. {\bf {333}}, pp. 1850 - 1852

 \bibitem[Pettengill \& Dyce(1965)]{pd1965} Pettengill, G.H., and Dyce, R.B. 1965. ``A radar determination of the rotation of the planet Mercury." \nat, Vol. {\bf{206}}, p. 1240

 \bibitem[Press et al.(1992)]{ptvf1992} Press, W.H.; Teutolski, S.A.; Vetterling, W.T.; and Flannery, B.P. 1992. {\it{Numerical Recipes}}. Cambridge University Press

  \bibitem[Quinn et al.(1991)]{qtd1991} Quinn, T.R.; Tremaine, S.; and Duncan, M. 1991. ``A three million year integration of the earth's orbit." \aj, Vol. {\bf{101}},
  pp. 2287 - 2305

 \bibitem[Rasio et al.(1996)]{ras} Rasio, F.A.; Tout, C. A.; Lubow, S. H.; and Livio, M. 1996. ``Tidal Decay of Close Planetary Orbits."
                                   {\it{The Astrophysical Journal}}, Vol. {\bf{470}}, pp. 1187 - 1191

 \bibitem[Singer(1968)]{singer1968} Singer, S. F. 1968. ``The Origin of the Moon and Geophysical Consequences." {\it{The
            Geophysical Journal of the Royal Astronomical Society}}, Vol. {\bf{15}}, pp. 205 - 226.

 \bibitem[Smith et al.(2012)]{szpshlmnpmjtprght2012}  Smith, D.E.; Zuber, M.T.; Phillips, R.J; Solomon, S.C.; Hauck II, S.A.; Lemoine, F.G.; Mazarico, E.; Neumann, G.A.;
 Peale, S.J.; Margot, J.-L.; Johnson, C.L.; Torrence, M.H.; Perry, M.E.; Rowlands, D.D.; Goossens, S.; Head, J.W., and Taylor, A.H.; 2012,
 ``Gravity field and internal structure of Mercury from MESSENGER'', {\it{Science}}, Vol. {\bf{336}}, pp. 214 - 217

 \bibitem[Sonneveld(1969)]{s1969} Sonneveld, P. 1969. ``Errors in cubic spline interpolation." {\it{Journal of Engineering Mathematics}}, Vol. {\bf{3}}, pp. 107-117

 \bibitem[Strom et al.(2005)]{smiyk2005} Strom, R.G.; Malhotra, R.; Ito, T.; Yoshida, F.; and Kring, D.A.; 2005, ``The origin of planetary impactors in the Inner Solar System'',
 {\it{Science}}, Vol. {\bf{309}}, pp. 1847 - 1850.

 \bibitem[Strom et al.(2008)]{scmsh2008} Strom, R.G.; Chapman, C.R.; Merline, W.J.; Solomon, S.C.; and Head III, J.W.; 2008, ``Mercury cratering record viewed from MESSENGER's first flyby.'',
 {\it{Science}}, Vol. {\bf{321}}, pp. 79 - 81.

 \bibitem[Strom et al.(2011)]{sbcffhmps2011} Strom, R.G.; Banks, M.E.; Chapman, C.R.; Fassett, C.I.; Forde, J.A.; Head III, J.W.; Merline, W.J.; Prockter, L.M.; and Solomon, S.C.;
 2011, ``Mercury crater statistics from MESSENGER flybys: Implications for stratigraphy and resurfacing history.'', \planss, Vol. {\bf{59}}, pp. 1960 - 1967

 \bibitem[Sussman \& Wisdom(1992)]{sw1992} Sussman, G.J., and Wisdom, J. 1992. ``Chaotic evolution of the solar system." {\it{Science}}, Vol. {\bf{257}}, pp. 56 - 62

 \bibitem[Turcotte \& Schubert(2002)]{ts2002} Turcotte, D. L., \& Schubert, G. 2002. Geodynamics, 2nd edition, Cambridge University Press.

 \bibitem[Weidenschilling(1978)]{w1978} Weidenschilling, S. J. 1978. ``Iron / silicate fractionation of the origin of Mercury", {\it{Icarus}}, Vol. {\bf{35}}, pp. 99 - 111


 \bibitem[Wieczorek et al.(2012)]{wcllr2012} Wieczorek, M.A.; Correia, A.C.M.; Le Feuvre, M.; Laskar, J.; and Rambaux, N. 2012. ``Mercury's spin-orbit resonance explained
                                             by initial retrograde and subsequent synchronous rotation." {\it{Nature Geoscience}}, Vol. {\bf{5}}, pp. 18 - 21

 \bibitem[Williams \& Efroimsky(2012)]{we2012} Williams, J.G., and Efroimsky, M. 2012. ``Bodily tides near the 1:1 spin-orbit resonance: correction to Goldreich's dynamical model."
 {\it{Celestial Mechanics and Dynamical Astronomy}}, Vol. {\bf{114}}, pp. 387 - 414


 \bibitem[Wisdom(1987)]{wisdom} Wisdom, J. 1987. ``Chaotic Behaviour in the Solar System." {\it{Proceedings of the Royal Astronomical Society of London}},
  Series A, Vol. {\bf{413}}, pp. 109 - 129

 \bibitem[Wurm et al.(2013)]{wurm} Wurm, G.; Trieloff, M.; and Rauer, H. 2013. ``Photophoretic Separation of Metals and Silicates: The
 Formation of Mercury-like Planets and Metal Depletion in Chondrites." {\it{The Astrophysical Journal}}, Vol. {\bf{769}}, article id. 78

\end{thebibliography}
\end{document}